\documentclass[aps,prb,a4paper,twocolumn,superscriptaddress,floatfix,nofootinbib,showpacs]{revtex4-1}
\pdfoutput=1
\usepackage{graphicx} 
\usepackage{amsmath} \usepackage{array} \usepackage{amssymb}
\usepackage{amsfonts} \usepackage{color} \usepackage{enumitem}
\usepackage{bm} \usepackage{verbatim} \usepackage{hyperref}
\usepackage{morefloats} \usepackage{dcolumn} \usepackage{xspace}
\usepackage{relsize} \usepackage{footnote} \usepackage{lipsum}

\hypersetup{ citecolor=blue, colorlinks=true, urlcolor=blue }

 \newcommand{\vektor}[1]{{\bf
    #1}} 
 
\newcommand{\la}{\langle} \newcommand{\ra}{\rangle}
 
\newcommand{\be}{\begin{equation*}} \newcommand{\ee}{\end{equation*}}
\newcommand{\bea}{\begin{eqnarray*}}
  \newcommand{\eea}{\end{eqnarray*}}

\newcommand{\eref}[1]{Eq.~(\ref{#1})}
\newcommand{\fref}[1]{Fig.~\ref{#1}}
\newcommand{\sref}[1]{Sec.~\ref{#1}}  \newcommand{\sector}[1]{\mbox{$\hat{\mathrm{#1}}$}}
\newcommand{\etal}{ \emph{et al.}}

\makeatletter \def\blfootnote{\xdef\@thefnmark{}\@footnotetext}
\makeatother

\begin{document}

\title{Detection and characterization of symmetry-broken long-range
  orders in the spin-$\frac{1}{2}$ triangular Heisenberg model}

\author{S.~N.~Saadatmand} \email{s.saadatmand@uq.edu.au}
\affiliation{ARC Centre of Excellence for Engineered Quantum Systems,
  School of Mathematics and Physics, The University of Queensland, St
  Lucia, QLD 4072, Australia} 
\author{I.~P.~McCulloch}
\affiliation{ARC Centre of Excellence for Engineered Quantum Systems,
  School of Mathematics and Physics, The University of Queensland, St
  Lucia, QLD 4072, Australia}

\begin{abstract}

  We present new numerical tools to analyze symmetry-broken phases in
  the context of $SU(2)$-symmetric translation-invariant matrix product
  states (MPS) and density-matrix renormalization-group (DMRG) methods
  for infinite cylinders, and determine the phase diagram of the
  geometrically-frustrated triangular Heisenberg model with nearest
  and next-nearest neighbor (NN and NNN) interactions. The appearance
  of Nambu-Goldstone modes in the excitation spectrum is
  characterized by ``tower of states'' levels in the
  momentum-resolved entanglement spectrum. Symmetry-breaking
  phase transitions are detected by a combination of the correlation
  lengths and second and fourth cumulants of the magnetic order
  parameters (which we call the Binder ratio), even though symmetry
  implies that the order parameter itself is strictly zero. Using
  this approach, we have identified $120^{\circ}$ order, a columnar
  order, and an algebraic spin liquid (specific to width-6 systems),
  alongside the previously studied topological spin liquid phase. For
  the latter, we also demonstrate robustness against chiral
  perturbations.
  
\end{abstract}


\date{\today}

\maketitle

\section{Introduction}
\label{sec:intro}

The groundstate of the one-dimensional nearest-neighbor Heisenberg model (originally
determined in an important work by
Heisenberg~\cite{Heisenberg28_original}), $H_{NN} = J \sum_{i}
\vektor{S}_i \cdot \vektor{S}_{i+1}$, for $J<0$, exhibits
long-range ferromagnetic (FM) order, which breaks the spins' rotational
symmetry (the $SU(2)$ group) and elementary excitations are spin-waves
(also known as Nambu-Goldstone bosons or Magnons; see for example
\onlinecite{Schollwock04_book,Auerbach94_book,Sachdev11_book}).
The Bethe Ansatz~\cite{Bethe31_original}
can be employed (e.g.~see~\onlinecite{Karbach98}) to study the
antiferromagnet (AFM) spin-$\frac{1}{2}$ Heisenberg model, $J>0$,
which demonstrates the absence of magnetic ordering in clear contrast
to the FM case. Today, we know there exist \textit{no}
continuous-symmetry-broken long-range order (LRO) in any
one-dimensional system. In fact, magnetism in 1D and few-leg ladders
is peculiarly different to higher dimensions (where LROs exist; see
below), since the magnetic ordering at zero temperature is suppressed
by quantum fluctuations due to the same
mechanism as described by Mermin-Wagner-Hohenberg theorem for
finite-temperatures~\cite{MerminWagner66_original,Hohenberg67_original}
(i.e.~due to the low cost of creating quantum long-range
fluctuations, which increases the entropy).  In contrast to 1D,
long range magnetic ordering is possible in 2D Heisenberg-type
Hamiltonians; early examples arose from studying
anisotropy~\citep{Frohlich78},
the AFM Heisenberg model with $S=\frac{1}{2}$~\citep{Anderson52,Oitmaa78,Huse88,Huse88_PRB,Kennedy88_JSP,Reger88},
$S\geq \frac{3}{2}$~\citep{Neves86}, large-$S$ values~\citep{Anderson52} and for the antiferromagnetic
XY~\citep{Oitmaa78,Kennedy88_PRL} and XXZ~\citep{Kubo88,Okabe88}
models for all spin magnitudes.  For the majority of two-dimensional
magnetic materials, if there exist \textit{no} frustration, the
groundstate exhibit~\citep{Lhuillier05,Sachdev11_book,Farnell14}
either ferromagnetism or antiferromagnetism (i.e.~the
well-known bipartite N\'eel order~\cite{Neel48_original}). It is
widely believed that the celebrated Landau symmetry-breaking
theorem~\citep{Landau37,Landau50} explains the physics behind
all such conventional magnetic ordering: Hamiltonians such as $H_{NN}$
contain a set of symmetries which are absent in the groundstate, a
feature known as spontaneous symmetry breaking (SSB). As a result
of symmetry-breaking, a well-defined order parameter exists in the
model that can be used to characterize the magnetic ordering,
unambiguously. After uncovering the key mechanisms behind the
conventional ordering (in particular, ferromagnetism and bipartite antiferromagnetism), the
field of low-dimensional quantum magnetism enjoyed a new boost of
attention aimed at understanding exotic phases of quantum matter
that appear in frustrated 
one-dimensional~\citep{Tsvelik07_book,Gu09,Auerbach94_book,PollmannTurner12,Pollmann10,Pollmann12,Turner11}
and two-dimensional~\citep{Lhuillier05,Balents10,Chen10,Powell11,Farnell14}
systems.  This happened partly due to the rise of the
geometrically-frustrated antiferromagnets on non-bipartite Archimedean
lattices~\citep{Lhuillier05,Schollwock04_book,Diep04_book,Lacroix11_book,Farnell14,Sachdev11_book}.
Interestingly, the existence of geometrical frustration is enough by itself
to often lead to the `melting' of the magnetic ordering,
stabilizing a family of nonmagnetic phases, collectively classified as
\emph{spin liquids} (also known as paramagnetic states)
\citep{Kalmeyer87,Wen89,Wen91,Read91,Balents10,Chen10,Wen02}. Such
quantum liquids preserve all Hamiltonian symmetries and, consequently,
their existence cannot be understood through Landau's
symmetry-breaking paradigm.  The search for new, hidden order
parameters has been challenging theorists for the last 20 years,
and has led to the discovery of even more intriguing phases of the
quantum matter. A canonical example is the discovery of the
\emph{topological order}~\cite{Wen02,Wen07_book,Chen10}, such as
symmetry-protected topological (SPT) ordering (including the Haldane
phase and the closely relevant Affleck-Kennedy-Lieb-Tasaki
groundstate~\citep{AKLT87,AKLT88}) and the intrinsic topological
states~\citep{Wen91,Chen10} (including the $Z_2$-gauge
groundstate of the toric
code~\cite{Kitaev03_original,Kitaev06,Kitaev09}), which can only exist
in $D \ge 2$.

Magnetic ordering is often identified using the scaling behavior of a
static two-point (or higher order) correlation function. For a spin
system, a two-point correlator can be written in terms of the
(principal) correlation length, $\xi$, as $|G_2(i,i')| = \la
\vektor{S}_i \cdot \vektor{S}_{i^\prime} \ra \sim
C + e^{-\frac{r_{ii'}}{\xi}} + \cdots$, where $C$ stands for a constant (which can be zero),
$r_{ii'}$ is the distance
between sites, and ellipses represent faster decaying terms. 
Different type of ordering can be defined as follows. For magnetically
disordered states, with no conventional order parameter (i.e.~no broken symmetry), 
the correlation function decays to zero
exponentially fast, $C$ is zero, $\xi$ is finite, and there is a bulk gap in the
excitation spectrum. In this case, instead 
of symmetry breaking, we have \emph{symmetry protection}, giving rise
to the SPT order.
Such an exponential drop is observed in
the Haldane phase (as an example, see the original
calculations by White and Huse\cite{White93}).
For true LROs, such as
N\'{e}el-type AFMs and the FM state with a conventional order
parameter, the correlation function tends to a constant at large
distances, $\xi\rightarrow\infty$.  There exists another distinct
long-range phase of the quantum matter, which is referred to as a
quantum critical state (or a quasi-LRO). In such phases, the
correlation function decays with a \emph{power-law} with distance.
Power-law decaying correlation functions can be approximated as the
sum of many exponential functions, as occurs in the MPS ansatz~\citep{Rommer97,Ostlund95_original},
which again translates to having diverging $\xi$, consistent with
the Bethe Ansatz' prediction for the spin-$\frac{1}{2}$ Heisenberg
chain.  Critical states are common in 1D quantum magnetism and
appear at a transition between
two gapped disordered phases with different
symmetries, when the gap necessarily closes; however, they can also
stabilize in an extended region, as in the \textit{XY-phase} of the
anisotropic Heisenberg chain~\citep{Schollwock04_book,Sachdev11_book}.

High-accuracy numerical methods, such as exact diagonalization (ED),
quantum Monte Carlo (QMC), (see~\onlinecite{Sandvik10} for a review), and
coupled cluster~\citep{Farnell14} methods,  are often used
for low-temperature frustrated magnets,
modeled as strongly-interacting spin Hamiltonians exhibiting
many-sublattice groundstates. 
In this paper, we employ and expand the
functionality of the finite DMRG
(fDMRG)~\cite{White92_original,White93_original,McCulloch07,Schollwock11},
and the state-of-the-art infinite DMRG
(iDMRG)~\cite{McCulloch08,Schollwock11} methods to characterize LROs
of a geometrically-frustrated system, when the many-body states are
constructed through the $SU(2)$-symmetric (non-Abelian) MPS and infinite MPS
(iMPS) ans\"atze~\cite{Perez-Garcia07,McCulloch07,McCulloch08,Schollwock11}, respectively. The latter is a
translation-invariant MPS that allows the calculation of many useful quantities
directly in the thermodynamic limit via transfer matrix methods.
Currently, there exist few well-established numerical tools, in the
context of non-symmetric DMRG, to identify LROs. In finite-system MPS studies,
SSB needs to be treated carefully because in exact calculations
SSB does not occur at all\cite{Koma1994,Lhuillier05}, as
finite size effects induce a gap between states that would be degenerate
in the thermodynamic limit. In practice, with finite-precision arithmetic
symmetry breaking can occur when the finite-size gap is smaller than the characteristic
energy scale set by the accuracy of the numerics 
(in MPS calculations, this is set by the energy scale associated with the basis truncation).
This can be difficult to control, as symmetry breaking might occur as a side-effect of 
the numerical algorithm or it might require an additional perturbation. 
Infinite-size MPS (or very large finite MPS) are better
behaved in this respect, where there are a variety of techniques;
one can look at the scaling of the correlation length of the groundstate against
MPS number of states, $m$, which distinguishes gapped and gapless 
states~\cite{Tagliacozzo08,Stojevic15,McCulloch08}, direct
measurement of local magnetization order parameters, the
entanglement entropy\cite{Eisert10_colloquium}, and
the static spin structure factor (SSF -- see below).  However, when
the Hamiltonian symmetries are preserved explicitly, the order parameter
is zero by construction and a robust set of
numerical tools for characterizing magnetic ordering is not readily
available. Here, we introduce and verify the accuracy of two new
numerical tools, in the context of $SU(2)$-symmetric iMPS/iDMRG, to
characterize and locate phase transitions incorporating LROs in the
triangular Heisenberg model (THM) on infinite cylinders. New tools
include study of the cumulants (cf.~\onlinecite{Stuart09_book} for the
definitions and relevant discussions on the non-central moments and
the cumulants in the context of the probability theory) and a Binder
ratio of magnetization order parameters, and further developments on tower-of-states (TOS) 
level patterns in the
momentum-resolved entanglement spectra (ES)\cite{Li08_original,Kolley13,Cincio13}.

The triangular lattice has the highest geometrical frustration in the
Archimedean crystal family with a coordination number of $Z_c\!=\!6$.
Anderson and Fazekas\cite{Anderson73,Fazekas74} argued that the high
frustration of the triangular lattice might be enough to melt the
long-range magnetic ordering observed for the Heisenberg model on the
square lattice
(e.g.~see~\onlinecite{Lhuillier05,Sachdev11_book,Farnell14}).  In the
first work, Anderson conjectured that the spin-$\frac{1}{2}$ THM with
antiferromagnetic NN bonds should stabilize a resonating-valance-bond
(RVB) groundstate (i.e.~the equally-weighted superposition of all
possible arrangements of the singlet dimers on the lattice; RVBs are
the building blocks of the quantum liquids).  The failure of robust
analytical and numerical studies to find an RVB groundstate motivates
the search for a minimal extension to $H_{NN}$ that increases the
frustration. The obvious choice is frustration through the addition of
a NNN coupling term, which frustrates a LRO $120^{\circ}$-ordered arrangements of sublattices (see below and
\onlinecite{Huse88,Deutscher93,Capriotti99,Jolicoeur89,Jolicoeur90,
Chubukov92,Bernu94,White07,Kaneko14,Farnell14,Li15,Zhu15,Hu15,Iqbal16}).
This led to the introduction of the $J_1$-$J_2$
THM, for which the Hamiltonian is defined as
\begin{equation}
  \label{eq:J1J2-Ham}
  H_{J_2} = J_1 \sum_{\langle i,j \rangle} \vektor{S}_i \cdot {\bf S}_j + 
  J_2 \sum_{\langle\langle i,j \rangle\rangle} {\bf S}_i \cdot {\bf S}_j \; ,
\end{equation}
where $\langle i,j \rangle$ ($\langle\langle i,j \rangle\rangle$)
indicates that the sum goes over all NN (NNN) couplings. The
$SU(2)$-symmetry of $H_{J_2}$ can be simply realized by noticing
$[H_{J_2},\vektor{S}]=0$ ($\vektor{S}$ stands for the total spin vector),
which means that eigenvalues of $\vektor{S}^2$ are good quantum numbers and can
be used to label groundstate symmetry sectors.  Geometrical
frustration forbids the bipartite N\'{e}el order as a stable groundstate
of the antiferromagnetic NN model ($J_1>0$ and
$J_2=0$).  Consequently, one expects the groundstate, for the majority
of the phase diagram of the antiferromagnetic $H_{J_2}$, to be a
compromise, such as a $120^{\circ}$-ordered
arrangement~\cite{Saadatmand15,Lhuillier05,Farnell14}.
By now, it is well-known that the groundstate of the nearest-neighbor
THM does \emph{not} exhibit an RVB, but is instead a
quasi-classical LRO $120^{\circ}$ state,
which is less
stable~\citep{Jolicoeur89,Jolicoeur90,Bernu94,Kaneko14,Li15} than the
N\'{e}el order on the square lattice, since the sublattice
magnetization of the triangular lattice is significantly reduced
compared to its classical value. Because of this reduced stability,
inherent to the triangular lattice, upon perturbing the Hamiltonian
one may expect to see a variety of new phases.  There have been some
historically important semi-classical spin-wave theory (SWT) and ED
studies\citep{Jolicoeur90,Dagotto89,Deutscher93,Hirsch89,Chubukov92}
for the model. However, such studies did not cover the physics of
the whole phase diagram and were not able to capture the detailed
properties of the groundstates.  Previously, we
elucidated\cite{Saadatmand15} the complete phase diagram of the
$J_1$-$J_2$ THM on three-leg finite- and infinite-length cylinders to
understand the crossover of 1D and 2D physics in the model.  Moreover,
other precise numerical
approaches~\citep{Manuel99,Mishmash13,Kaneko14,Zhu15,Hu15,Li15,Iqbal16,Wietek16,Saadatmand16}
demonstrate the existence of a spin-liquid (SL) state that stabilizes in a
region ranging from $J_2^{\text{low}} \approx 0.05$~\citep{Mishmash13}
up to $J_2^{\text{high}} \approx 0.19$~\citep{Manuel99}. Some 
numerical studies discovered magnetic orders outside this
approximate SL region (see for
example~\onlinecite{Jolicoeur90,Chubukov92,Bernu94,Farnell14,Li15,Wietek16}).
However, we suggest the detailed properties of the magnetic
groundstates are still unclear in comparison to the well-understood
counterparts in the (semi-)classical THM~\cite{Jolicoeur90,Chubukov92}
and the quantum model on the three-leg cylinders. In particular, for
the finite-size lattices, the largest system sizes for which the
magnetic ordering of the $J_1$-$J_2$ THM was thoroughly studied is an
$18 \times 18$ torus~\citep{Kaneko14} and a $30 \times 3$
cylinder~\citep{Saadatmand15}.

It is noteworthy that the $J_1$-$J_2$ THM can describe some
low-temperature properties of quasi-2D organic lattices, such as
$\kappa$-(BEDT-TTF)$_2$Cu$_2$(CN)$_3$ and Me$_3$EtP[Pd(dmit)$_2$]$_2$,
and inorganic materials, such as Cs$_2$CuCl$_4$, Cs$_2$CuBr$_4$, and
RbFe(MoO$_4$)$_2$ (see~\onlinecite{Powell11,Svistov03,Balents10} for
details).

In this work, we establish the phase diagram for the $J_1$-$J_2$ 
THM on infinite-length cylinders with
width up to 12 sites, and show that the model contains an LRO coplanar
three-sublattice $120^\circ$ order ($J_2 \leq 0.105(5)$), a four-fold
degenerate toric-code-type $\text{Z}_2$-gauge spin liquid ($0.105(5)
\leq J_2 \leq 0.140(5)$; see also \onlinecite{Saadatmand16}), an
LRO collinear two-sublattice columnar order ($0.140(5) \leq J_2
\leq 0.170(5)$ and $0.200(5) \leq J_2 \leq 0.5 $ for width-6
cylinders, and $0.140(5) \leq J_2 \leq 0.5$ for larger cylinders), and
an algebraic spin liquid\cite{Wen02} (ASL) groundstate ($0.170(5) \leq
J_2 \leq 0.200(5)$, only for width-6 cylinders).
The new tools for the
cumulants and Binder ratios of the order parameter allowed us to
locate the aforementioned phase transitions accurately, while the
patterns of the TOS levels in the momentum-resolved ES revealed the
structure of the magnetic order (or its absence). We
also consider the explicit breaking of the time-reversal symmetry in
the $J_1$-$J_2$ THM and the possibility of the stabilization of a
chiral LRO. We note that there were recent, indecisive
discussions~\cite{Hu16,Wietek16,Saadatmand16} on the robustness of the
topological spin liquids against perturbing $H_{J_2}$ with a chiral
term. Here, we confirm the non-chiral nature of such groundstates and
the existence of a continuous phase transition toward the chiral spin
liquid~\cite{Wen89,Kalmeyer87,Wen02} (CSL) phase through the study of
a scalar chiral order parameter on width-8 cylinders.

The rest of the paper is organized as follows. In \sref{sec:methods},
we provide the details of the employed $SU(2)$-symmetric MPS and DMRG
methods (in particular, how we construct order parameter operators)
and the geometry of the cylindrical lattices (in particular, the MPS
mapping). In \sref{sec:overview}, we present an overview of each of the
phases, a schematic phase diagram for the model, and a 
more quantitative diagram in the
form of short-range correlations. In \sref{sec:FiniteMag}, we directly
measure the magnetization order parameters on some small-width
($L_y=3,4,5,6$) finite-length cylinders using MPS/fDMRG algorithms, to
benchmark our calculations with another algorithm.
Afterward, we focus on
presenting our more precise iMPS/iDMRG results on infinite cylinders
(having widths up to $L^{max}_y=12$).  In \sref{sec:CorrLengths}, we
investigate the scaling behaviors of the correlation lengths against
$m$, deep in each phase region.  In \sref{sec:EE}, in order to better
understand the entanglement entropy of the symmetry-broken LROs on cylinders, we study
the entropy in the columnar magnetically-ordered phase. Details of our
numerical tools are presented in \sref{sec:cumulants}
and \sref{sec:ES}, for cumulants and Binder ratios of the
magnetization order parameters, and for TOS levels in the
momentum-resolved ES, respectively.
In \sref{sec:chirality}, we test the robustness of a
topological SL groundstate against chirality perturbations of the
Hamiltonian to investigate the formation of long-range chiral orders,
before some concluding remarks in \sref{sec:conclusion}.

\section{Methods}
\label{sec:methods}

\begin{figure}
  \begin{center}
    \includegraphics[width=0.99\linewidth]{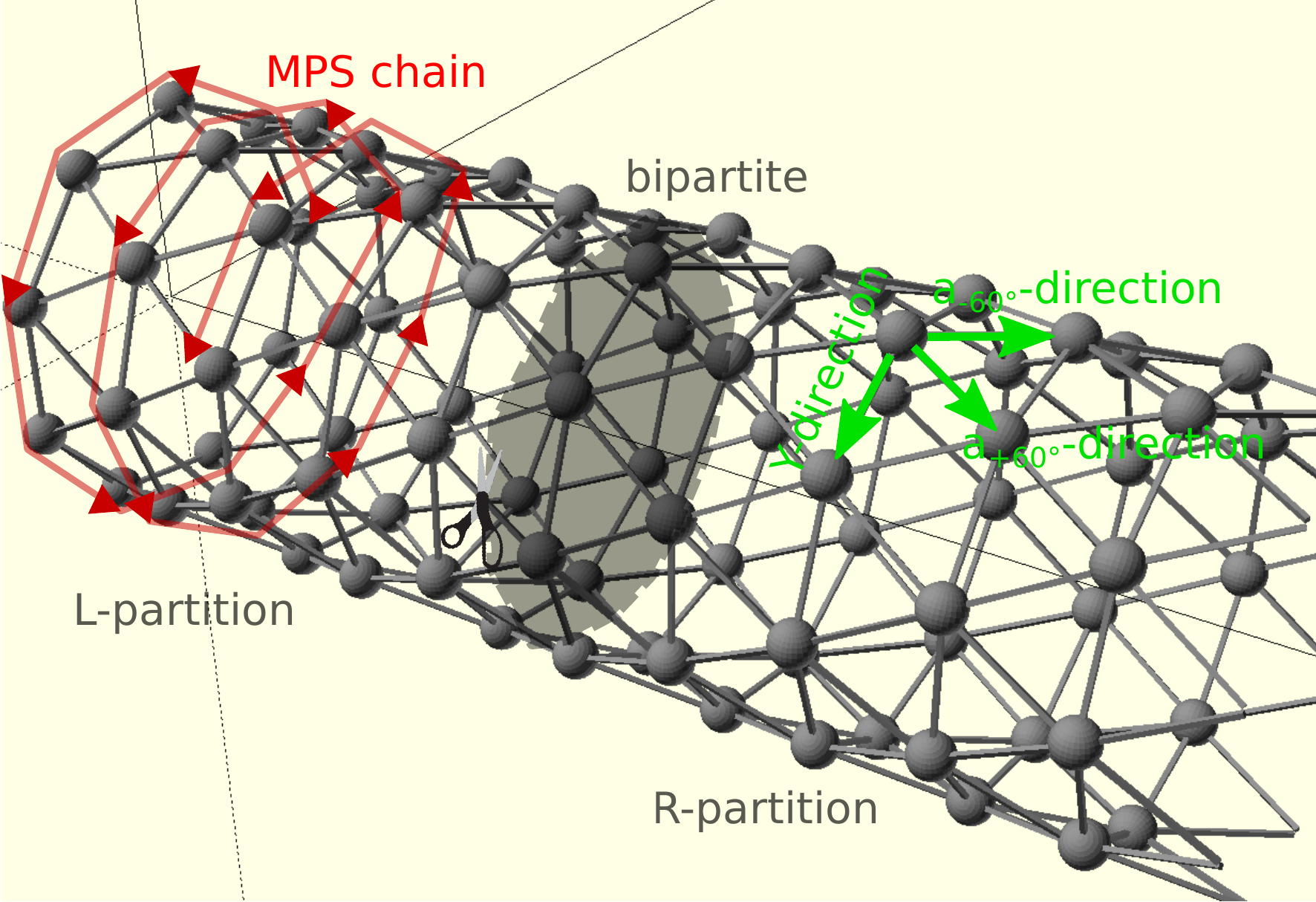}
    \caption{(Color online)
    Cartoon visualization of a triangular lattice on a YC cylinder. Spins
    sit on spheres. An `efficient' mapping of the MPS chain is shown using the red spiral.
    The green arrows represent the unit vectors on three principal lattice directions.
    The transparent gray plane corresponds to the bipartite cut that creates partitions $L$ 
    and $R$, without crossing any $Y$-direction bond. 
    \label{fig:TypicalLattice}}
\end{center}
\end{figure}

To obtain the variational groundstate of the THM for a wide range of
FM and AFM $J_2$-values in \eref{eq:J1J2-Ham}, we set $J_1=1$ as the
unit of the energy, and employ the single-site DMRG
algorithm (incorporating density-matrix mixing\cite{White05} with 
subspace expansion\cite{Hubig15}
and $SU(2)$ symmetry\cite{McCulloch02,McCulloch07}).
In addition, we construct the operators using the efficient
formalism of matrix product
operators\citep{McCulloch07,McCulloch08,Schollwock11,Hubig17} (MPOs),
which represents an operator analogous to an ordinary MPS matrix.  The MPO
structure provides a formulation of any polynomial operator
(with an expectation value that scales polynomially with the size of the lattice)
in a Schur form (an upper- or lower-triangular matrix) for infinite systems\cite{McCulloch08,Michel10}
(see below for an example and also \onlinecite{SaadatmandThesis} for an overview),
which allows the calculation of the asymptotic limit of the expectation value per site.
We keep up to $m=2,000$ number of states
(approximately equivalent to 6,000 states of an Abelian $U(1)$-symmetric
basis) in MPS/fDMRG, and up to $m=3,000$ number of states
(approximately 9,000 $U(1)$-states) in iMPS/iDMRG calculations. Due to
inherent 1D nature of the MPS, a mapping between the ansatz
wavefunction and the triangular lattice is necessary. For the mapping
purposes, we wrap the lattice in a way to create a long (or
infinite-length) $L_x\!\times\!L_y$-site cylinder ($L_x$ can go to
infinity; we also set $L=L_x\!\times\!L_y$) as in
\fref{fig:TypicalLattice}.
We will employ a standard notation, previously presented in
\onlinecite{Saadatmand15} (originally developed for single-wall carbon
nano-tubes\cite{Wilder98}), to specify the wrapping vectors of the
cylinders, $\vektor{C}_0$, in terms of principal lattice directions
using a notation of
$(\hat{\vektor{a}}_{+60^\circ},\hat{\vektor{a}}_{-60^\circ})$.  For
the majority of the calculations, we choose the so-called YC wrapping,
$\vektor{C}_0[\text{YC}]=(L_y,-L_y)$ (we shall use the shorthand
notations of YC$L_y$ and YC$L_x\!\times\!L_y$ to specify different YC
lattice sizes).
The YC structure is the only wrapping method with a circumference that
equals to $L_y$ ($Y$-axis now coincides with the lattice
short-direction and the $X$-axis coincides with the lattice
long-direction) and is the best choice for the momentum-resolved ES (see below).
However, in general, the choice of $L_y$ and
$\vektor{C}_0$ should respect sublattice ordering (if any) of the
target state to avoid frustrating the groundstate.
Consequently, depending on the desired width, the YC structure cannot be always used.
Therefore, in finite-$L_x$ fDMRG calculations, we use a YC6 structure
in all regions (allowing the stabilization of up to
tripartite-symmetric groundstates), and YC3,
$\vektor{C}_0[L_y=4]=(4,-2)$, and $\vektor{C}_0[L_y=5]=(5,-4)$
cylinders only in the $120^\circ$ and the SL phase regions. We also
consider a YC4 structure in the columnar and the SL phase regions
(occasionally, the YC3 and $\vektor{C}_0[L_y=4]=(5,-4)$ systems are
employed in the columnar phase region, however they are frustrating
some forms of the collinear ordering -- see below). For finite-length cylinders, we fix $L_x$ to
a value that after which, an increase of the cylinder's length would
not change the average bond energy in the bulk of the system up to
numerical uncertainties coming from the DMRG systematic
errors\cite{Schollwock11,SaadatmandThesis}. In $L_x\!=\!\infty$ iDMRG
calculations, we use YC6 and YC12 structures in all regions, reserving
YC9 only for the $120^\circ$ region, plus YC8 and YC10 in the columnar phase. 
We always set an efficient mapping for the infinite cylinders
that minimizes the one-dimensional range of NN and NNN interactions,
as shown in \fref{fig:TypicalLattice}. Finally to calculate bipartite
quantities, such as reduced density matrix, $\tilde{\rho}$, and entropy of
the DMRG wavefunctions\cite{Schollwock11}, we make a cut that does not cross any
$Y$-direction bond and creates partitions $L$ and $R$, as shown in the
figure.

\begin{figure*}
  \begin{center}
    \includegraphics[width=0.74\linewidth]{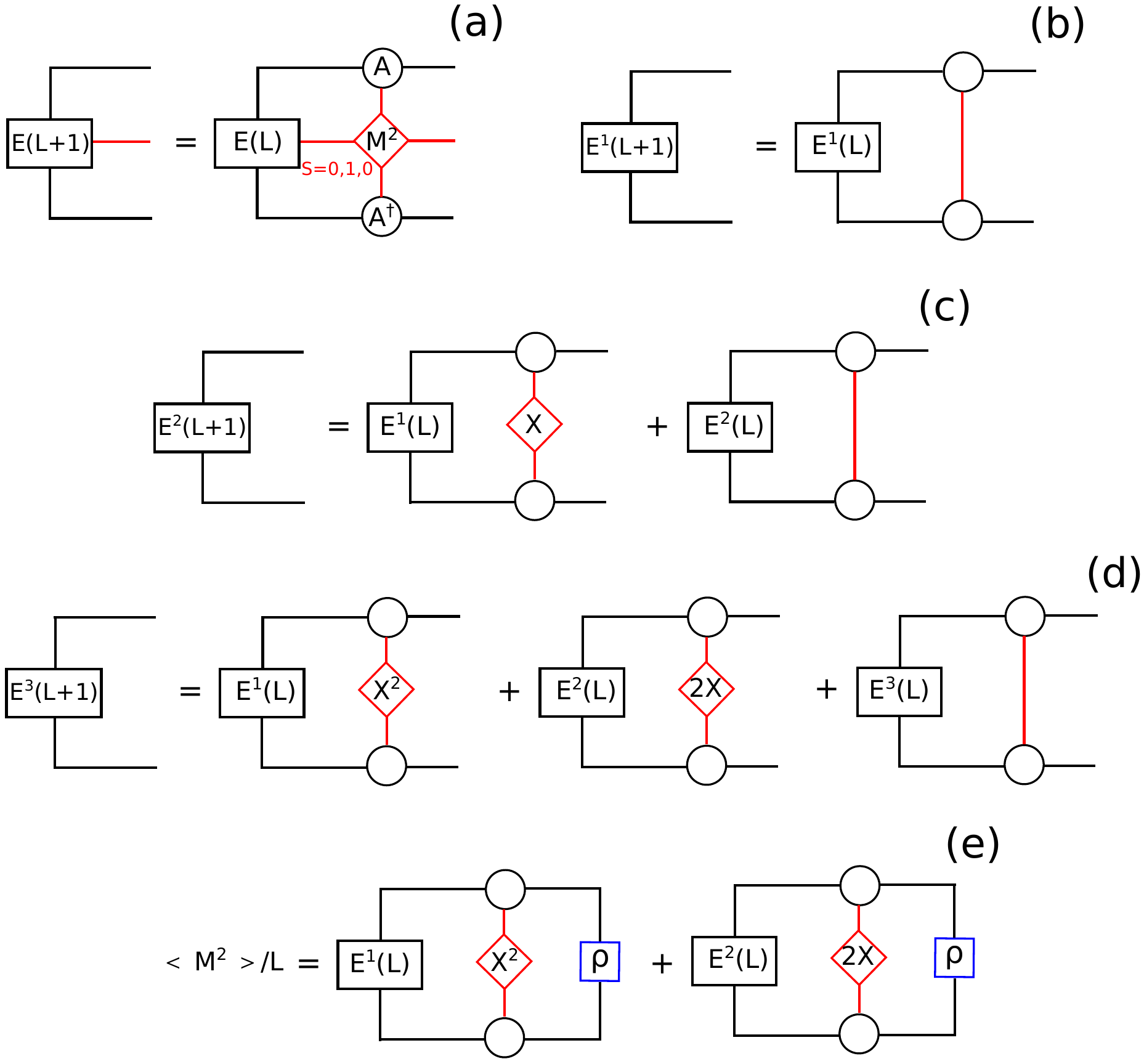}
    \caption{(Color online)
    (a) MPS diagram for the fixed-point equation of $E^a$-matrices
    of the second moment of $\vektor{M}$. MPS diagrams for the (b) first, $S=0$, (c) second, $S=1$, and
    (d) third, $S=0$, columns of $\vektor{M}^2$, \eref{eq:M2-MPO_ch2}. (e) The MPS recipe to calculate
    the final expectation value of the second moment.
    \label{fig:MPS-moments}}
\end{center}
\end{figure*}

We now present an overview of how to calculate higher moments of a
(possibly non-local) observable in a
translation-invariant infinite-size system. This is required for the
measurements of the cumulants and Binder ratios of the magnetization
order parameters (see below). For symmetry broken (or symmetry protected) states
the Binder cumulant of the (string) order parameter can be evaluated directly in the 
thermodynamic limit\cite{kolley15_binder}. However in this case, because
we preserve $SU(2)$ symmetry
the magnetic order parameter is strictly zero and the Binder cumulant is not well
defined. However, as we show below, the moments can still be used to detect
the signature of magnetic ordering.
Suppose we are interested in calculating
the matrix elements of the moments of an order parameter MPO,
$\vektor{M}^{[k]}$, of dimension $\tilde{m}$,
that transforms under $SU(2)$ as a rank $k$ tensor. The explicit
preservation of $SU(2)$ symmetry leads to the vanishing
of the order parameter, but the even moments can be non-zero.  In this case,
the measurement of the expectation values of the higher-order
magnetic moments, $\la M^n \ra$ (of order $n$), is of
interest. These can be done using the method of the
\emph{transfer operator}~\cite{Michel10}.  The generalized transfer
(super-)operator, $\mathbb{T}_X$, associated with some operator of finite
support (acting on a unit cell of an iMPS), $\hat{X}$, is defined as
\begin{equation}
  \mathbb{T}_X(E^a) = \sum_{s^\prime s} \la s^\prime | \hat{X} | s \ra A^{s^\prime \dagger} E^a A^s,
\end{equation}
where $A^s$ are ordinary MPS matrices and $E^a$ denote the so
called $\vektor{E}$-matrices. In this context, $E^a$ is essentially
an eigenmatrix, however, the $\vektor{E}$-matrices are more familiar for
their role in the expectation value of an MPO $\la \mathcal{A} | \hat{M} | \mathcal{A}
\ra$ (see~\onlinecite{McCulloch07,McCulloch08} for full
details). $E^a$ is in principle extensive, and on a $n$-site system, 
can be defined recursively as
\begin{align}
  E^{a^\prime}(n) \equiv \sum_{s^\prime, s, a} \mathcal{A}^{s^\prime_n
    \dagger} M_{a^\prime a}^{s^\prime s} \mathcal{A}^{s_n} E^{a}(n-1).
  \label{eq:MPS-E}
\end{align}
In the following example, for the sake of the simplicity we assume a
one-site unit cell, although in practice for a magnetically ordered system
the unit cell will be at least as large as the number of sublattices;
The generalization for larger unit-cell sizes is straightforward.
We give an example here for the second moment, the higher moments can
be obtained recursively\cite{Michel10}.
To calculate the asymptotic limit of $\la M^2
\ra$, one only needs to solve the diagrammatic fixed-point equation
shown in \fref{fig:MPS-moments}(a), where the $E^a$-matrices are
connected according to
\begin{align}
  E^i(L+1) = \mathbb{T}_{M_{ii}}(E^i(L)) + \sum_{j>i}
  \mathbb{T}_{M_{ij}}(E^j(L)) \; ,
  \label{eq:E-recursive_ch2}
\end{align}
which can be solved sequentially, from $E^1$, $E^2$, $\ldots$, $E^{\tilde{m}}$.
In practice, \fref{fig:MPS-moments}(a) shows the fixed point at which the
addition of an extra site (or unit cell) to $E^a$-matrices will
leave the system unchanged. The MPO form of the order parameter on a
unit cell can be written\citep{McCulloch07,McCulloch08} as a
super-matrix (a matrix where elements are local operators acting on a single site
or unit cell of the lattice):
\begin{equation}
  M = \begin{pmatrix} I & X \\ & I \end{pmatrix} \; .
  \label{eq:M-MPO}
\end{equation}
and we can attach $SU(2)$ quantum numbers $S=0,1$ to the rows/columns.
The operator for the second moment has the form
\begin{align}
  M^2 = \begin{pmatrix} I & X \\ & I \end{pmatrix} \otimes
  \begin{pmatrix} I & X \\ & I \end{pmatrix} =
  \begin{pmatrix} I & X  & X & X^2 \\ & I & 0 & X \\  & & I & X \\  & & & I \end{pmatrix} \notag \\
  \Rightarrow \begin{pmatrix} I & X & X^2 \\ & I & 2X \\ & & I
  \end{pmatrix},
  \label{eq:M2-MPO_ch2}
\end{align}
where in the last step, we have collapsed the middle rows to create a
$3 \times 3$ matrix with new quantum numbers of $S=0,1,0$ labeling the
rows (assuming here that we want to calculate only the scalar
part of $X^2$ -- for the calculation of higher moments we need the other
spin projections too). We can now write the fixed point of the last MPO in the form of
recursive equations for the $E^a$-matrices, as shown in
\fref{fig:MPS-moments}(b),(c), and (d). We note that the objects that
appear on the right hand sides of the figures are nothing other than
the generalized transfer operators. Translating the graphical notation into 
equations, for example, \fref{fig:MPS-moments}(b) can be written as
\begin{equation}
  E^1(L+1) = \mathbb{T}_I( E^1(L) ),
  \label{eq:MPS-from-MPO_0}
\end{equation}
which means $E^1(L)$ is an eigenmatrix of the transfer operator, which,
for a properly orthogonalized MPS is just the identity matrix, so
$E^1(L)=I$. As a result, equations for
\fref{fig:MPS-moments}(c) and (d) can be written as
\begin{align}
  E^2(L+1) &= \mathbb{T}_X( I ) + \mathbb{T}_I( E^1(L) ) \notag \\
  &= C_X + \mathbb{T}_I( E^1(L) ),
  \label{eq:MPO1_ch2}
\end{align}
where $C_X = \mathbb{T}_X( I )$ is a constant
matrix, and
\begin{equation}
  E^3(L+1) = \mathbb{T}_{X^2}( I ) + 2\mathbb{T}_X( E^2(L) ) + \mathbb{T}_I( E^3(L) )~.
  \label{eq:MPO2_ch2}
\end{equation}
The desired expectation value is encoded in the final matrix,
i.e.~$\la M^2 \ra = Tr(E^3\tilde{\rho})$. However, importantly, most
of the matrix elements of $E^3$ do \emph{not} contribute to the expectation
value of the second moment per site, we need only the component of
$E^3$ that has non-zero overlap with $\tilde{\rho}$.
Note that
$\tilde{\rho}$ is the right eigenmatrix of $\mathbb{T}_I$ with the
unity eigenvalue, hence the component of $E^3$ that gives the
expectation value is the component in the direction of the
corresponding left eigenmatrix of $\mathbb{T}_I$.

The calculation of the matrix elements of $E^2$ can be done efficiently
using a linear solver. To see how this works,
consider the eigenmatrix expansion for the transfer
operator, $\mathbb{T}_I = \sum_{n=1}^{m^2} \eta_n | \eta_n )
( \eta_n |$, to obtain the eigenvalues $\eta_n$ and eigenvectors
$|\eta_n)$. If we write $C_X$ and $E^2$
matrices in this $\{ |\eta_n) \}$ basis with expansion coefficients
$c^{(2)}_{n}$ and $e^{(2)}_{n}(L)$,
\begin{align}
  C_X &= \sum_{n=1}^{m^2} c^{(2)}_{n} |\eta_n) \;, \notag \\
  E^2(L) &= \sum_{n=1}^{m^2} e^{(2)}_{n}(L) |\eta_n) \; ,
  \label{eq:MPO1-basis_ch2}
\end{align}
then \eref{eq:MPO1_ch2} is, for each component,
\begin{equation}
  e^{(2)}_{n}(L+1) = c^{(2)}_{n} + \eta_n e^{(2)}_{n}(L) \;.
  \label{eq:MPO1-n_ch2}
\end{equation}
Following \onlinecite{Michel10}, we further decompose the
coefficients into a component parallel and components perpendicular
to the identity matrix, $I$ (i.e.~the left eigenmatrix of
$\mathbb{T}_I$, which has the largest eigenvalue of $\eta_1=1$
due to the MPS orthogonalization condition), and define
\begin{align}
  \tilde{C}_X &= \sum_{n=2}^{m^2} c^{(2)}_{n} |\eta_n) \;, \notag \\
  \tilde{E}^2(L) &= \sum_{n=2}^{m^2} e^{(2)}_{n}(L) |\eta_n) \; ,
  \label{eq:I-decomposition_ch2}
\end{align}
so that $C_X = \tilde{C}_X + c^{(2)}_1 I$ and $E^2 = \tilde{E}^2 + e^{(2)}_1 I$.
The reason for this is that the component in the direction of the identity
$e^{(2)}_1$ diverges in the summation,
whereas the other components that are perpendicular to the identity do
not. Hence, we need to find the fixed points of these parts separately.

Solving \eref{eq:MPO1-n_ch2} for the parallel components reveal the
local expectation value of $X$ per site, which is an straightforward
calculation,
\begin{equation}
  e_1^{(2)}(L+1) = e_1^{(2)}(L) + c^{(2)}_1 \; ,
\end{equation}
where $c^{(2)}_1$ is just the expectation value of the order parameter on one site. 
Hence $e_1^{(2)}(L+1) = \sum_{i=1}^L \langle X_i \rangle$, which is zero because of the 
$SU(2)$ symmetry (indeed, $c^{(2)}_1=0$ by construction, since it is in the wrong
quantum number sector for the identity eigenvector of the transfer operator). The
perpendicular components lead to
\begin{equation}
  \tilde{E}^2_{(n)}(L+1) = \tilde{C}_{(n)} + \eta_n \tilde{E}^2_{(n)}(L)~,
  \label{eq:MPO1-vertical_ch2}
\end{equation}
where now $n \geq 2$, and the eigenvalues $|\eta_n|<1$. Thus,
\eref{eq:MPO1-vertical_ch2} is of the form of the sum of a convergent geometric series.
Upon taking the limit $L\rightarrow\infty$ and writing back the projection operators
as the original matrices, \eref{eq:MPO1-vertical_ch2} converges to a
fixed-point:
\begin{equation}
  (1-\mathbb{T}_I) \tilde{E}^2(\infty) = \tilde{C}_X,
  \label{eq:FixedPoint-E1}
\end{equation}
which is a rather simple system of linear equations, and is numerically
stable because the condition number of $1-\mathbb{T}_I$ is simply
related to the leading correlation length,
$1/(1-|\eta_2|) \simeq \xi$. In practice, generalized minimal residual method (GMRES) is a good choice of linear solver
for \eref{eq:FixedPoint-E1}.
Upon obtaining the matrix elements of
$\tilde{E}^2$, we can proceed to calculate the final expectation value
as shown in \fref{fig:MPS-moments}(e). Note that this does not require
all of the matrix elements of $E^{3}$, since we only require the overlap
between $E^{3}$ and the density matrix (the right eigenvector of $\mathbb{T}_I$
with eigenvalue 1). This
means that $\la M^2 \ra = L \times \text{Tr}(
\tilde{\rho}\mathbb{T}_{X^2}( I ) + 2\tilde{\rho}\mathbb{T}_X( E^2(L)
) )$, which is demonstrated in the MPS diagrammatic equation of
\fref{fig:MPS-moments}(e).  I.e.~the only unknown is the $E^2$-matrix.
This is a useful optimization and rather general -- in calculating
the expectation value of a triangular MPO of dimension $\tilde{m}$, only the
matrix elements up to $E^{\tilde{m}-1}$ are required.

For calculating the 4th moment of a magnetization order parameter using $SU(2)$ symmetry,
$X^4$ decomposes as
\begin{equation}
  X^4 = (\vektor{X} \cdot \vektor{X})^2 + (\vektor{X} \otimes \vektor{X}) \cdot (\vektor{X} \otimes \vektor{X}) \; ,
\end{equation}
where the dot product $\vektor{X} \cdot \vektor{X} = -\sqrt{3} [\vektor{X} \times \vektor{X}]^{[0]}$
and 
outer product $\vektor{X} \otimes \vektor{X} = \sqrt{6/5}[\vektor{X} \times \vektor{X}]^{[2]}$ are
proportional to the $S\!=\!0$ and $S\!=\!2$ projections of the operator product, respectively,
with an additional factor 
arising from the $SU(2)$ coupling coefficients.
In general, we would need to also include the cross-product term 
$(\vektor{X} \times \vektor{X}) \cdot (\vektor{X} \times \vektor{X})$ (proportional to the spin-1 projection), 
however, this vanishes due to antisymmetry under time reversal.

\section{Overview of the Phase Diagram}
\label{sec:overview}

\begin{figure*}
  \begin{center}
    \includegraphics[width=0.74\linewidth]{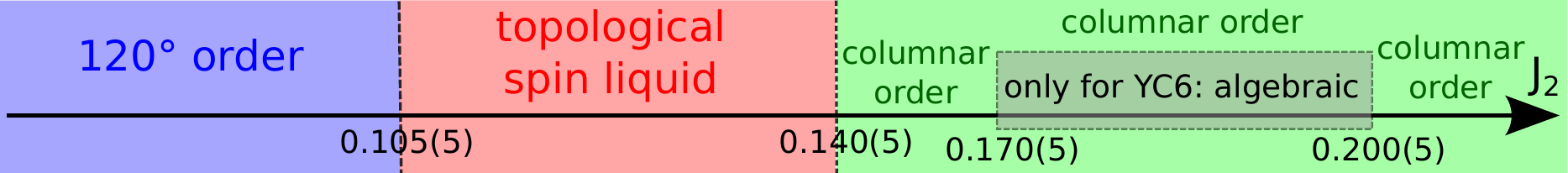}
    \caption{(Color online)
    Schematic phase diagram of the $J_1$-$J_2$ THM, \eref{eq:J1J2-Ham}, on infinite
    cylinders. Phase 
    transition boundaries are obtained from the Binder ratios of the magnetization 
    order parameter (see below).
    \label{fig:SemiPhaseDiagram}}
\end{center}
\end{figure*}

In this section, we present our findings for the phase boundaries
and properties of \eref{eq:J1J2-Ham}, for different $J_2 / J_1$ with
$J_1 > 0$, using iDMRG and some benchmark comparisons
using fDMRG. 

In \fref{fig:SemiPhaseDiagram}, we show the summary of the
phase diagram, with four distinct phases; two
phases with symmetry-broken magnetic order, a $Z_2$ spin liquid,
and (only for the $YC6$ geometry) an algebraic spin liquid.
For each of these phases, we present below visualizations of the
correlation functions, obtained from well-converged iDMRG groundstates. 
In these visualizations, we depict spin-spin correlations, $\la \vektor{S}_i \cdot
\vektor{S}_0 \ra$, with respect to a reference site $\vektor{S}_0$,
using the size and the color of some spheres, and the NN
correlators are depicted using the thickness and the color of some
bonds. The reference site is denoted with the gray sphere.
We also present the SSF up to the second
Brillouin zone. Using the discrete Fourier
transform of the real-space correlations to switch to the momentum
space, one can write
\begin{equation}
  \text{SSF}(\vektor{k}, N\!=\!\infty) = \lim_{N\rightarrow\infty} \frac{1}{N} \sum_{i,i'}^N \la \vektor{S}_{i} \cdot
  \vektor{S}_{i'} \ra e^{i \vektor{k} \cdot (\vektor{r}_{i} - \vektor{r}_{i'})},
  \label{eq:SSF_ch6}
\end{equation}
where $\vektor{r}_{i}$ denotes the position vector of a spin
$\vektor{S}_i$ in the \emph{planar} map of the lattice.  
The momentum vector,
$\vektor{k}$, will sweep the extended Brillouin zones. When the
momentum vector coincides with the lattice's wave vector, $\vektor{Q}$, the
occurrence of the condition $\lim_{N\rightarrow\infty}
\frac{\text{SSF}(N)}{N} = \text{Const.}$ guarantees the existence of a
true LRO. Plotting the SSF in the $(k_x,k_y)$ plane will reveal occurrence
of strong FM correlations as Bragg peaks. 
However, for a fixed-$L_y$ infinite cylinder,
one can only estimate the sums appearing in \eref{eq:SSF_ch6} using a
finite length correlation. Therefore, we consider a large
enough \emph{cutoff} as an upper limit for $i$, namely $N_{c}$. 
We note that it is possible to obtain $SSF(\vektor{k}, N\!=\!\infty)$ directly
using the same method as described above for the moments (see also \onlinecite{Michel10}), however,
this is an expensive process and for calculating the entire $\vektor{k}$ space it
is much faster to calculate the real-space correlations and perform a Fourier transform.
Here, we truncate the real-space correlation at the first point where 
$|\la
\vektor{S}_0 \cdot \vektor{S}_{N_{c}} \ra| \leq 10^{-5}$ is met for
the nonmagnetic short-range correlated states (i.e.~spin liquids) and
the condition $|\la \vektor{S}_0 \cdot \vektor{S}_{N_{c}} \ra| \leq
10^{-3}$ is met for the symmetry-broken quasi-LROs
(i.e.~$120^\circ$ and columnar states). The obtained phases are:
\begin{enumerate}
\item[1.] $J_2 \rightarrow -\infty$:
  In this limit, one can readily show that the lattice decouples
  into \emph{three} sublattices, each of which is a NN triangular
  lattice with bond strength $J_2$. In the case of vanishing
  interactions between sublattices ($J_2/J_1 \rightarrow -\infty$), the
  groundstate for each sublattice is trivially a fully-saturated ferromagnet
  (see also \onlinecite{Saadatmand15}) with total spin
  magnetization of $S^{\text{total}}_{A,B,C} = \frac{L_{u}}{2}$ per
  unit cell of each sublattice ($A$, $B$, or $C$). For a
  width-$L_y$ infinite-length YC structure, $L_{u}=L_y/3$ and
  $S^{\text{total}}_{A,B,C} = L_y/6$. The overall state can be any
  arbitrary mixture of three $\vektor{S}^{\text{total}}_A$,
  $\vektor{S}^{\text{total}}_B$, and $\vektor{S}^{\text{total}}_C$
  spin vectors, where they only have to follow the angular momentum
  summation rules. This will cause a large degeneracy for the overall
  groundstate, supporting total magnetization in a range of $0 \leq
  S^{\text{total}} \leq \frac{3L_{u}}{2}$. Perturbing the Hamiltonian with a positive $J_1$ would then
  break this degeneracy and impose a $120^\circ$-ordered groundstate.
  Similarly to the case of three-leg cylinders\cite{Saadatmand15}, we find no signs
  of a phase transition for any $J_2 < 0$.
\begin{figure}
 \begin{center}
    \includegraphics[width=0.91\columnwidth]{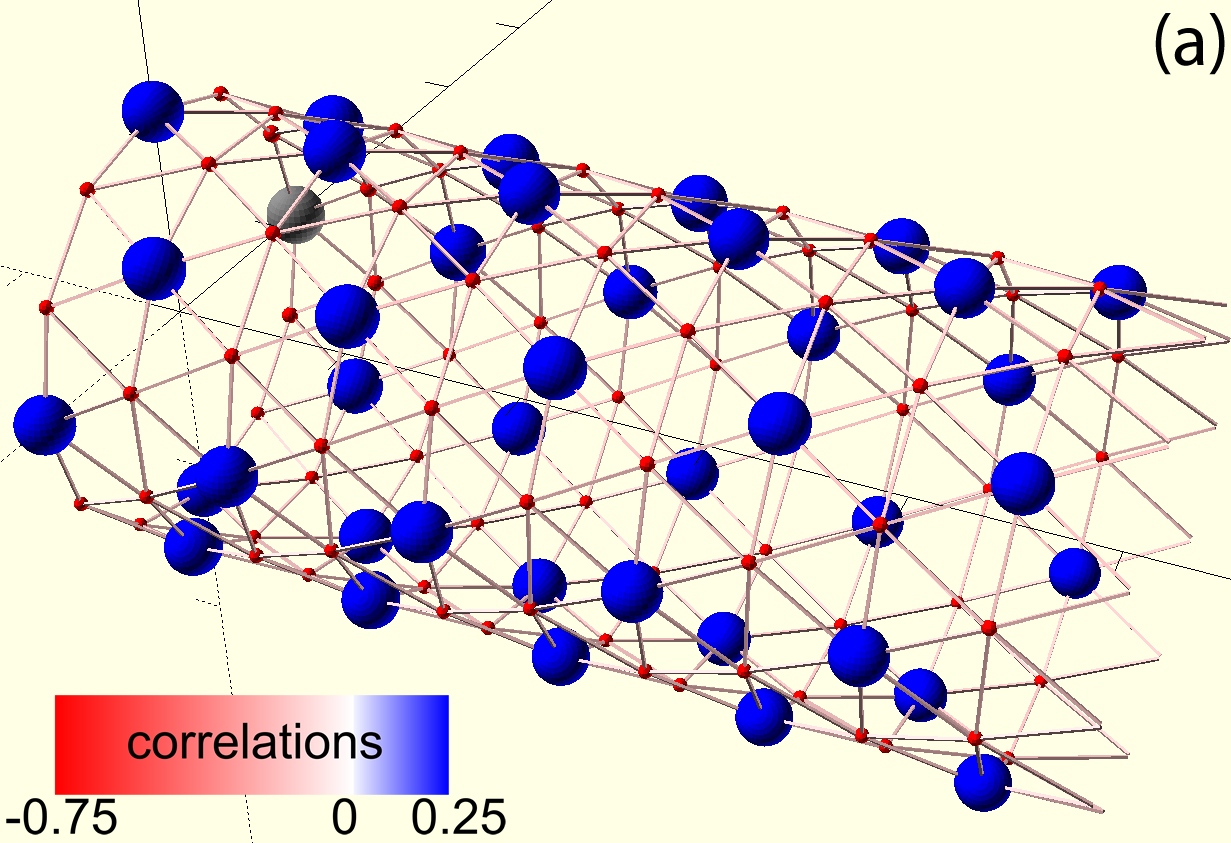}
    \includegraphics[width=0.99\columnwidth]{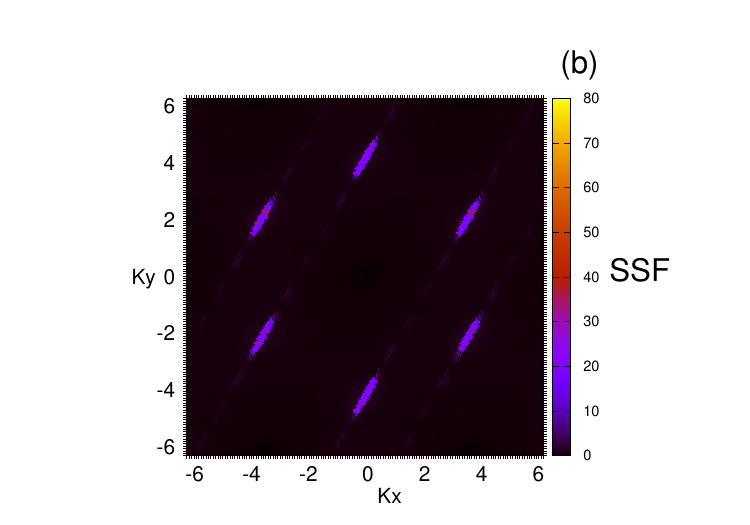}
    \caption{(Color online)
    Lattice visualizations for the iDMRG groundstates
    of the THM on infinite cylinders at $J_2=-1.0$ ($120^\circ$ order). (a) Correlation function for a YC12
    system. The size and the color of the spheres indicate the (long-range) spin-spin correlations 
    in respect to the principal (gray) site, and the thickness and the color of the 
    bonds indicate the strength of the NN correlations. (b) SSF for a
    YC6 system. Bragg peaks are presented up to the second Brillouin zone of the inverse 
    lattice.
    \label{fig:Visualizations_120order}}
 \end{center}
\end{figure}
\item[2.] $J_2 \leq 0.105(5)$:
  The groundstate is a coplanar quasi-classical $120^\circ$ order. 
  Our investigations on infinite YC6, YC9, and YC12 structures find
  a three-sublattice
  magnetic ordered state exhibiting SSB in the thermodynamic limit
  (cf.~\sref{sec:cumulants} and \sref{sec:ES}). By imposing $SU(2)$ symmetry,
  the low-lying Nambu-Goldstone modes are evident and viewing the infinite cylinder
  as a 1D system it appears as a 1D quantum critical gapless state
  (cf.~\sref{sec:CorrLengths}).
  In
  \fref{fig:Visualizations_120order}(a), we present the correlation function for a
  YC12 groundstate at $J_2=-1.0$. The appearance of
  $\frac{L_y}{3}=4$ blue (ferromagnet) spheres per ring exhibiting a
  roughly constant size (for short distances) and all-AFM (red) bonds
  (throughout the cylinder) are characteristics of the phase.  In
  \fref{fig:Visualizations_120order}(b), we present the SSF for a
  YC6 groundstate, deep in the $120^\circ$ phase.
  The formation of \emph{six} strong Bragg peaks
  on a slightly-distorted regular hexagon is another characteristic
  for the phase. Using this data, we predict a wave vector of
  $Q_{120^{\circ}}\approx(\pm3.64,\pm2.09)$, which is very close to
  our expected theoretical value of
  $Q_{120^{\circ}}^{\text{theory}}=(\pm\frac{2\pi}{\sqrt{3}},\pm\frac{2\pi}{3})
  \approx (\pm3.63,\pm2.09)$ for a $120^\circ$ product
  state\cite{Jolicoeur90,Chubukov92,Saadatmand15,Sachdev11_book}. We
  note that the correlation functions of YC6 and YC9, and SSFs of YC9 and YC12 structures in the $120^\circ$ phase
  are essentially identical to the results of \fref{fig:Visualizations_120order}.
\begin{figure}
 \begin{center}
    \includegraphics[width=0.91\columnwidth]{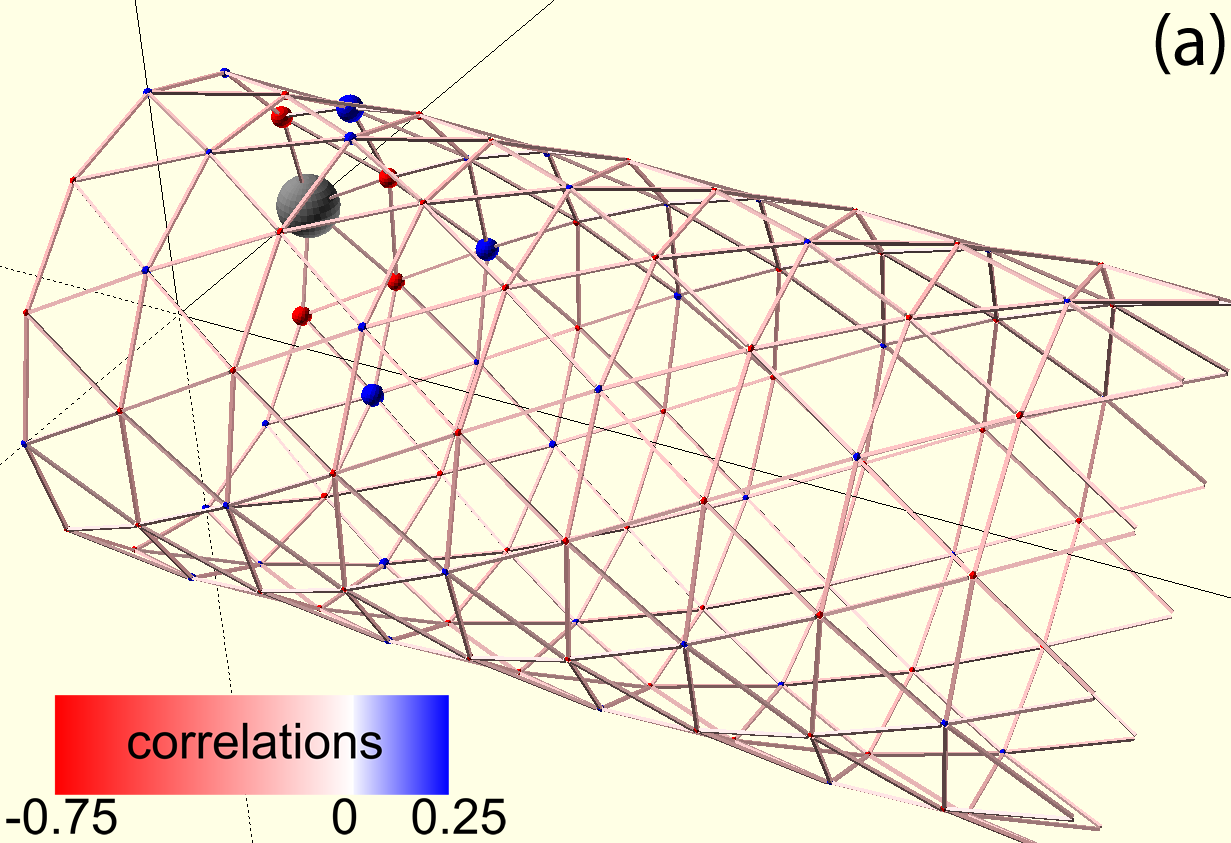}
    \includegraphics[width=0.91\columnwidth]{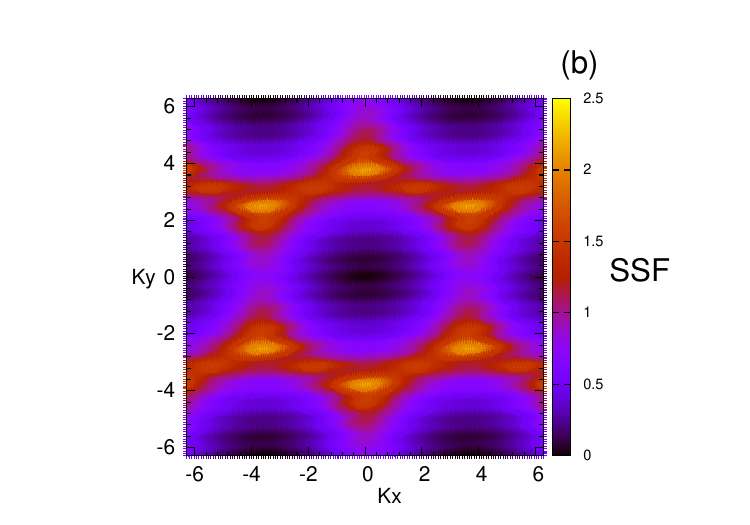}
    \caption{(Color online)
    Lattice visualizations for the iDMRG groundstates
    of the THM on infinite cylinders at $J_2=-0.125$ (topological spin liquid). 
    (a) Correlation function for a YC12-\sector{i}
    system. The size and the color of the spheres indicate the (long-range) spin-spin correlations 
    in respect to the principal (gray) site, and the thickness and the color of the 
    bonds indicate the strength of the NN correlations. (b) SSF for a
    YC10-\sector{b} system. Bragg peaks are presented up to the second Brillouin zone of the inverse
    lattice.
    \label{fig:Visualizations_TopoSLs}}
\end{center}
\end{figure}
\item[3.] $0.105(5) \leq J_2 \leq 0.140(5)$:
  The groundstate is a four-fold degenerate toric-code-type
  $\text{Z}_2$ topological spin liquid (denoted by
  YC$L_y$-\sector{a} for the anyonic sector \sector{a} $\in \{ \sector{i}, \sector{b},
  \sector{f}, \sector{v} \}$;
  see~\onlinecite{Saadatmand16} for full details).
  In \fref{fig:Visualizations_TopoSLs}(a), we present the correlation function
  for a YC12-\sector{i} groundstate at $J_2=0.125$. The
  appearance of spheres with rapidly decaying radii and relatively
  weak all-AFM (red) bonds throughout the cylinder, is
  characteristic of the SL states (such a behavior of the correlations
  is also observed for other topological
  sectors and system sizes, except there exist some weak bond
  anisotropies\cite{Saadatmand16,Hu15}).  In
  \fref{fig:Visualizations_TopoSLs}(b), we present the SSF for a
  YC10-\sector{b} groundstate at $J_2=0.125$. The
  spectral function is almost homogeneous, although being noisy and
  containing some weak diffusive peaks (compared to the strong Bragg
  peaks of magnetically-ordered states) reminiscent of gradual
  disappearance of the $120^{\circ}$ order. We notice that this
  overall pattern is virtually the same for all anyonic sectors and
  system sizes. Furthermore, our qualitative studies demonstrate that
  the homogeneity of the SSF is growing with increasing $L_y$ (not shown in the figures). 
  For the topological SL phase, we find the lower and upper
  phase boundaries of $J_2^{\text{low}}=0.105(5)$ and $J_2^{\text{high}}=0.140(5)$,
  respectively. Using fDMRG for rather small YC6 widths (see below) we obtain similar results,
  $0.101(4) \leq J_2 \leq 0.136(4)$. These phase boundaries are fairly close, but not
  identical, to those found by other 
  authors~\citep{Manuel99,Mishmash13,Kaneko14,Zhu15,Hu15,Li15,Iqbal16,Wietek16}.
\begin{figure}
 \begin{center}
    \includegraphics[width=0.91\columnwidth]{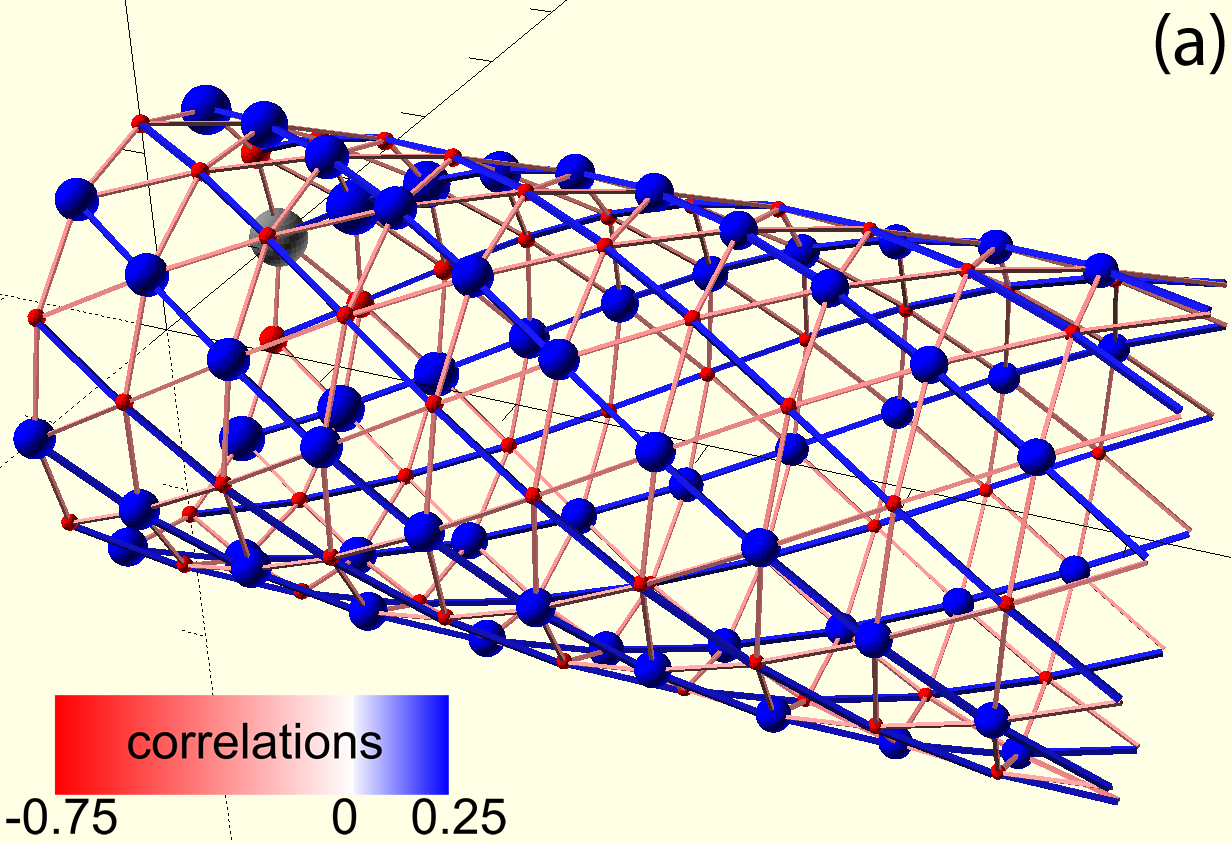}
    \includegraphics[width=0.99\columnwidth]{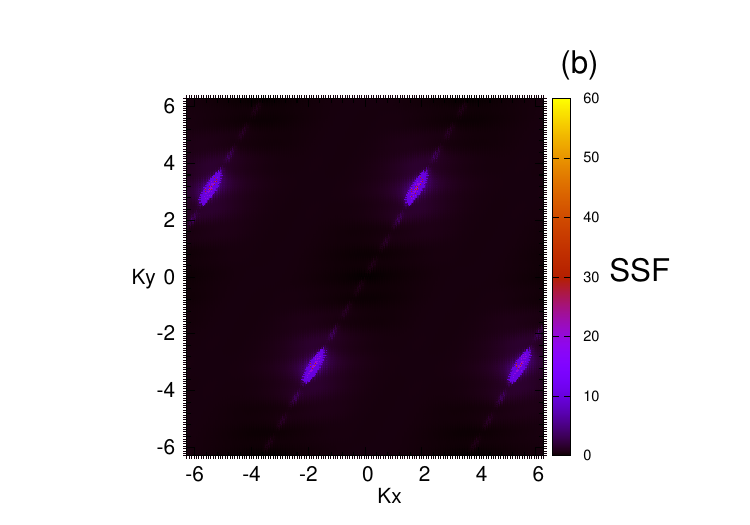}
    \caption{(Color online)
    Lattice visualizations for the iDMRG groundstates
    of the THM on infinite cylinders at $J_2=0.5$ (columnar order). (a) Correlation function for a YC12
    system. The size and the color of the spheres indicate the (long-range) spin-spin correlations 
    in respect to the principal (gray) site, and the thickness and the color of the 
    bonds indicate the strength of the NN correlations. (b) SSF for a
    YC8 system. Bragg peaks are presented up to the second Brillouin zone of the inverse
    lattice.
    \label{fig:Visualizations-ColumnarOrder}}
\end{center}
\end{figure}
\item[4.] $0.140(5) \leq J_2 \leq 0.5$, but excluding a region
\emph{only for YC6} of $0.170(5) \leq J_2 \leq 0.200(5)$:
  The groundstate is a quasi-classical collinear columnar (striped)
  order. Our investigations on infinite YC6, YC8, and YC10 structures (plus few
  more $J_2$ points on YC12 structures) show that that the columnar
  order is two-sublattice AFM state exhibiting SSB in the thermodynamic limit
  (cf.~\sref{sec:cumulants} and \sref{sec:ES}). Again, with $SU(2)$ symmetry
  the state appears on an infinite cylinder as 1D quantum critical.
  The correlation function for a
  YC12 groundstate at $J_2=0.5$ is presented in
  \fref{fig:Visualizations-ColumnarOrder}(a), where the appearance of
  robust FM stripes in the $\vektor{a}_{+60^{\circ}}$-direction is
  clearly recognizable. In fact, the columnar order on the triangular
  lattice has three possible arrangements~\citep{Lecheminant95} of FM
  stripes, each aligning with one of the three principal lattice
  directions, which are only degenerate in the thermodynamic limit.
  For the THM on three-leg (trivially) and four-leg cylinders (both
  finite and infinite-length cases), we found that the columnar order
  always has FM stripes in the lattice short (Y) direction, while for
  wider-width finite-length YC structures, FM stripes will be in
  either of $\vektor{a}_{+60^{\circ}}$ or
  $\vektor{a}_{-60^{\circ}}$-directions, producing only \emph{two}
  degenerate groundstates.
  We numerically confirmed that, upon choosing a suitable wavefunction
  unit cell, iDMRG states randomly converge to one of these two
  states. We present the SSF for a YC8
  groundstate at $J_2=0.5$ (with $\vektor{a}_{+60^{\circ}}$-direction
  FM stripes), in \fref{fig:Visualizations-ColumnarOrder}(b). The
  formation of \emph{four}, comparatively very strong Bragg peaks on a
  slightly-distorted regular parallelogram (with $60^{\circ}$ angles)
  is a characteristic of the phase.  A wave vector of
  $Q_{\text{striped}}\approx\pm(1.82,3.18)$ can be estimated for the
  SSF, which is close to our expected theoretical value of
  $Q_{\text{striped}}^{\text{theory}}=\pm(\frac{\pi}{\sqrt{3}},\pi)
  \approx \pm(1.81,3.14)$ for a columnar product
  state\cite{Jolicoeur90,Chubukov92,Saadatmand15}. We note that the SSFs of the
  columnar orders on YC6, YC10, and Y12 systems are rather similar
  to this result, however, the wave vector changes to
  $Q_{\text{striped}}^{\text{theory}}=\pm(\pi,-\frac{\pi}{\sqrt{3}})$,
  when the direction of FM stripes are switched. Our numerical
  calculations extend only to $J_2 = 0.5$. However we expect that
  there will be some additional geometry-dependent magnetically ordered phases
  for larger $J_2$ before reaching the large $J_2$ limit (see below).
\begin{figure}
 \begin{center}
    \includegraphics[width=0.91\columnwidth]{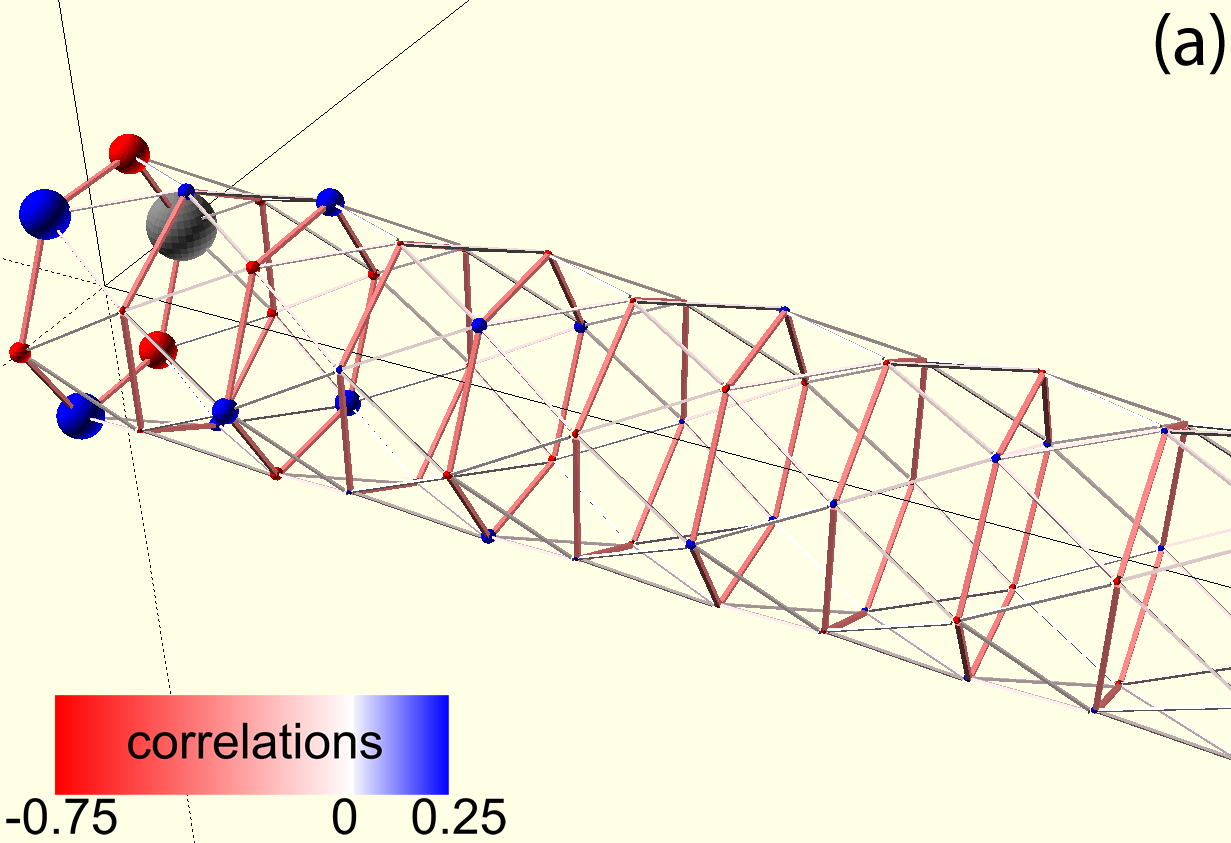}
    \includegraphics[width=0.99\columnwidth]{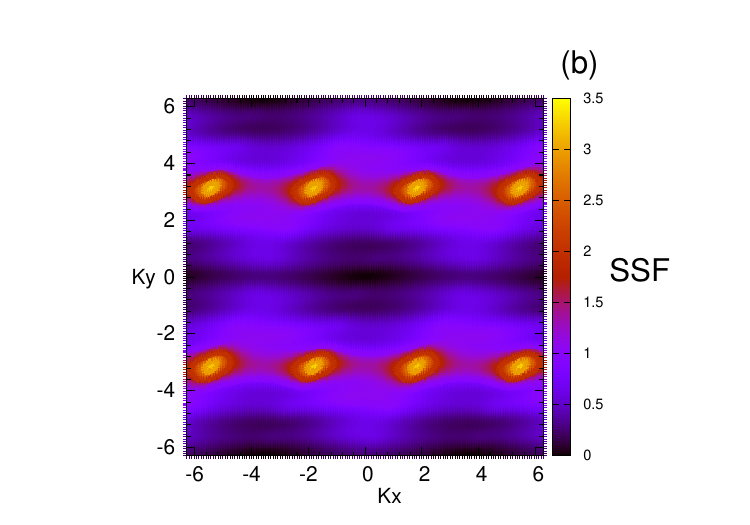}
    \caption{(Color online)
    Lattice visualizations for the iDMRG groundstate
    of the THM on an infinite YC6 system at $J_2=0.185$ (ASL phase). (a) Correlation function results, where
    the size and the color of the spheres indicate the (long-range) spin-spin correlations 
    in respect to the principal (gray) site, and the thickness and the color of the 
    bonds indicate the strength of the NN correlations. (b) SSF results, where the
    Bragg peaks are presented up to the second Brillouin zone of the inverse 
    lattice.
    \label{fig:Visualizations-AlgebraicSL}}
\end{center}
\end{figure}
\item[5.] $0.170(5) \leq J_2 \leq 0.200(5)$, only for YC6:
  The YC6 geometry appears special in that we find signatures of an algebraic
  spin liquid, rather distinct from any other phase that we have observed
  in the model. Our results suggest that this phase is a
  quantum critical, gapless state with power-law scaling of the
  correlation function (cf. \sref{sec:CorrLengths}) and
  \emph{no} magnetic order. In \fref{fig:Visualizations-AlgebraicSL}, the
  presented correlation function and SSF appear to be reminiscent of a columnar-like
  ordering, but there are subtle differences. The size of spheres, representing
  the two-point correlation function, 
  \fref{fig:Visualizations-AlgebraicSL}(a), decays faster
  than the columnar phase. In addition, the size of SSF peaks,
  \fref{fig:Visualizations-AlgebraicSL}(b), are considerably smaller
  than the typical size of the Bragg peaks in the columnar order
  having the same system width. In  \sref{sec:cumulants} and
  \ref{sec:ES}, below, we show that this phase has no signatures of magnetic
  ordering, which indicates that there are no broken symmetries and hence
  some kind of algebraic spin liquid.
\item[6.] $J_2 \rightarrow +\infty$:
  Following the arguments presented for the $J_2 \rightarrow -\infty$
  case, in the limit of $|J_2| \gg 1$, the physical lattice will
  transform to three decoupled sublattices with antiferromagnetic NN
  bonds of the strength $J_2$. For $J_2 \rightarrow +\infty$, the
  groundstate on each new sublattice is the same as the overall
  groundstate for $J_2=0$, i.e., the $120^\circ$
  order. However for few-leg ladder systems, other symmetry broken phases
  could appear due to the restricted geometry.
  As an example, for three-leg finite
  cylinders\cite{Saadatmand15} in $J_2\rightarrow\infty$, we found
  that the groundstate is
  three weakly-coupled copies of a NNN Majumdar-Ghosh state. Interestingly, we
  found a similar 
  dual Majumdar-Ghosh phase for four-leg finite cylinders\cite{SaadatmandThesis}).
  Consistent with the expected 2D limit, we did not observe any
  signature of such Majumdar-Ghosh-type phases for $L_y > 4$ ladders. In
  addition, semi-classical SWT studies~\citep{Jolicoeur90} confirms
  that the ``order from disorder'' mechanism would choose three-fold
  degenerate and decoupled states for $J_2 \gg 1$, which are
  energetically favorable to arrange according to $120^{\circ}$
  ordering. Hence, we expect that such exotic ordered phases are particular
  features of narrow cylinders.
\end{enumerate}

\begin{figure}
  \begin{center}
    \includegraphics[width=0.99\columnwidth]{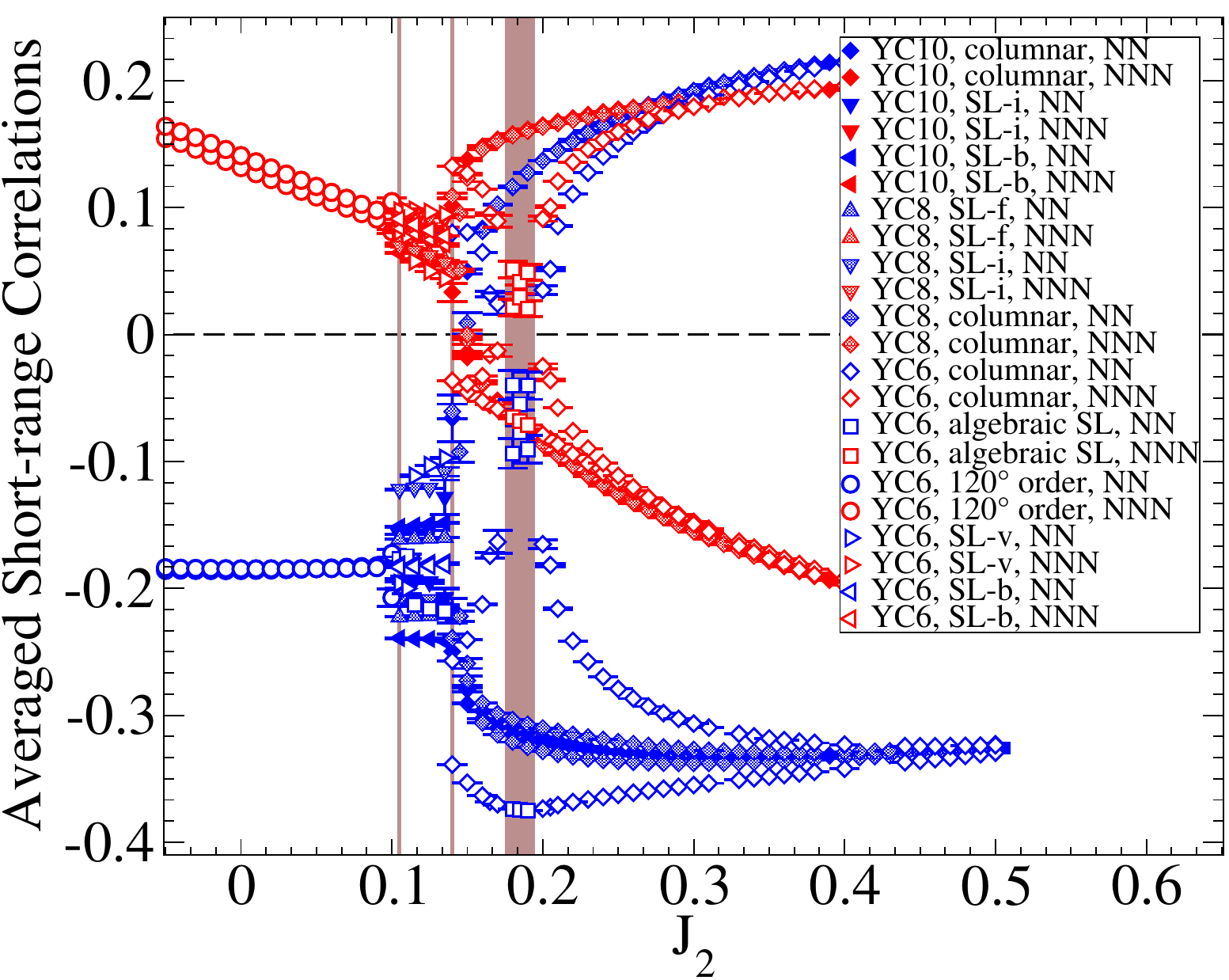}
    \caption{(Color online)
    Short-range correlation functions for the iDMRG groundstates of the 
    THM on YC6, YC8, and YC10 structures versus NNN coupling strength, $J_2$. Each
    correlator value is averaged over a wavefunction unit cell, then 
    extrapolated \emph{linearly} with iDMRG truncation errors toward the 
    thermodynamic limit of $m\rightarrow\infty$. For each $J_2$, red symbols represent NN bonds
    in principal $Y$, $\vektor{a}_{+60^\circ}$, and $\vektor{a}_{-60^\circ}$ directions. 
    Similarly, blue symbols represent NNN bonds in non-principal
    directions of $\frac{1}{\sqrt{2}}(1,1)$, $\frac{1}{\sqrt{5}}(2,-1)$, 
    and $\frac{1}{\sqrt{5}}(2,1)$. Narrow brown stripes indicate
    predicted phase transitions from the Binder ratio results. 
    Thick brown stripe shows a speculated region for the existence of the ASL phase on YC6 structures (see below). 
    \label{fig:ShortRangeCorr}}
\end{center}
\end{figure}

To get a better quantitative insight on the phase diagram of the THM,
we study the short-range (NN and NNN) spin-spin correlations, $\la
\vektor{S}_i \cdot \vektor{S}_j \ra$, \fref{fig:ShortRangeCorr}.
Short-distance correlators in a crystalline phase have a
repetitive pattern reflecting the bulk properties of
the groundstate. In \fref{fig:ShortRangeCorr}, we choose six reference
bonds, including three NN and three NNN correlators, to build up a
picture of the real-space correlations for different
system widths. In the $120^{\circ}$ phase region, correlators are very
nearly isotropic, where NN (NNN) bonds are all AFM (FM). On the other
hand, topological spin liquids on finite-width systems contain strong
anisotropies\cite{Hu15,Zhu15}, which is clearly seen in 
\fref{fig:ShortRangeCorr}. As we
showed previously\cite{Saadatmand16}, in the thermodynamic
limit anyonic sectors \sector{b} and \sector{f} are 
anisotropic on finite cylinders, while \sector{i} and \sector{v} are isotropic\cite{Slagle14,Lu16}.
The behavior of the 
correlation functions is distinct in the columnar phase,
where there are always \emph{two} FM bonds (one is a NN and another
one a NNN correlator) and \emph{four} AFM bonds (two are NN and other
two NNN correlators) out of the six reference bonds. The FM stripes of
the columnar order can, of course, choose either of
$\vektor{a}_{+60^\circ}$ or $\vektor{a}_{-60^\circ}$ directions, so
such data-points in this region are exchangeable.
Furthermore, curiously for YC6, in the ASL phase region ($ 0.170(5)
\leq J_2 \leq 0.200(5) $), the system temporarily
restore all symmetries, again, by crossing the
$\vektor{a}_{\pm60^\circ}$-direction bonds.

\section{Direct measurement of the order parameters on finite-length cylinders}
\label{sec:FiniteMag}

To provide a verification of the phase boundaries for comparison
against our iDMRG results, we calculated two magnetization order parameters
on $L_y \leq 6$ finite-length cylinders (small compare to the largest
width of our infinite-length YC systems) using an approach originally
suggested by White and Chernyshev\cite{White07}. Consider the
arbitrary magnetization vector order parameter of
$\vektor{M}(m)$ for a wavefunction with $m$ number of states (the
preservation of the SU(2)-symmetry causes the structural vanishing of
all projection components). Upon a suitable choice of the system size
and careful extrapolation toward the thermodynamic limit, non-zero
values for the \textit{second} moment of $\vektor{M}$ (which is
directly proportional to the spin susceptibility) can be derived. In
White and Chernyshev's method, one first extrapolates the order
parameter \emph{linearly} with the DMRG truncation errors,
$\varepsilon_m$, toward the thermodynamic limit of
$m\rightarrow\infty$ ($\varepsilon_m\rightarrow0$) to calculate $\la
\vektor{M}^2(\infty) \ra$. Then, using only fixed aspect-ratio
($\frac{L_y}{L_x} = \text{Const.}$) system sizes, $L_x$ and
$L_y$ should be simultaneously extrapolated toward the thermodynamic
limit of $L \rightarrow \infty$.  By employing a similar approach,
plus some simple dimensional analyses and numerical examination of
the magnetic moments, we suggest in the MPS constructions of the $SU(2)$
$S\!=\!0$-sector groundstates on fixed aspect-ratio cylinders ($L_x >
L_y$), the normalized order parameter, $\vektor{M}^2(\infty)$ per
site, scales as
\begin{equation}
  \la \bar{\vektor{M}}^2(\infty) \ra = \bar{a}_0 + \bar{a}_1 L_x^{-2} + ... \; , 
  \label{eq:MagExtrapolation}
\end{equation}
where eclipses represents higher order terms in
$\frac{1}{L_x}$ (note that \eref{eq:MagExtrapolation} is only a heuristic fit; 
see \onlinecite{Neuberger89} for theoretical predictions). 
One should note that any \emph{independent} growth
of $L_x$ and $L_y$ toward the $L \rightarrow \infty$ limit, can be
interpreted as the existence of an infinitely long cylinder at some
stage. This will collapse the system, essentially, to an inherently 1D
state, for which the behavior of the magnetic moments is essentially
different (see below).

\begin{figure}
  \begin{center}
    \includegraphics[width=0.99\columnwidth]{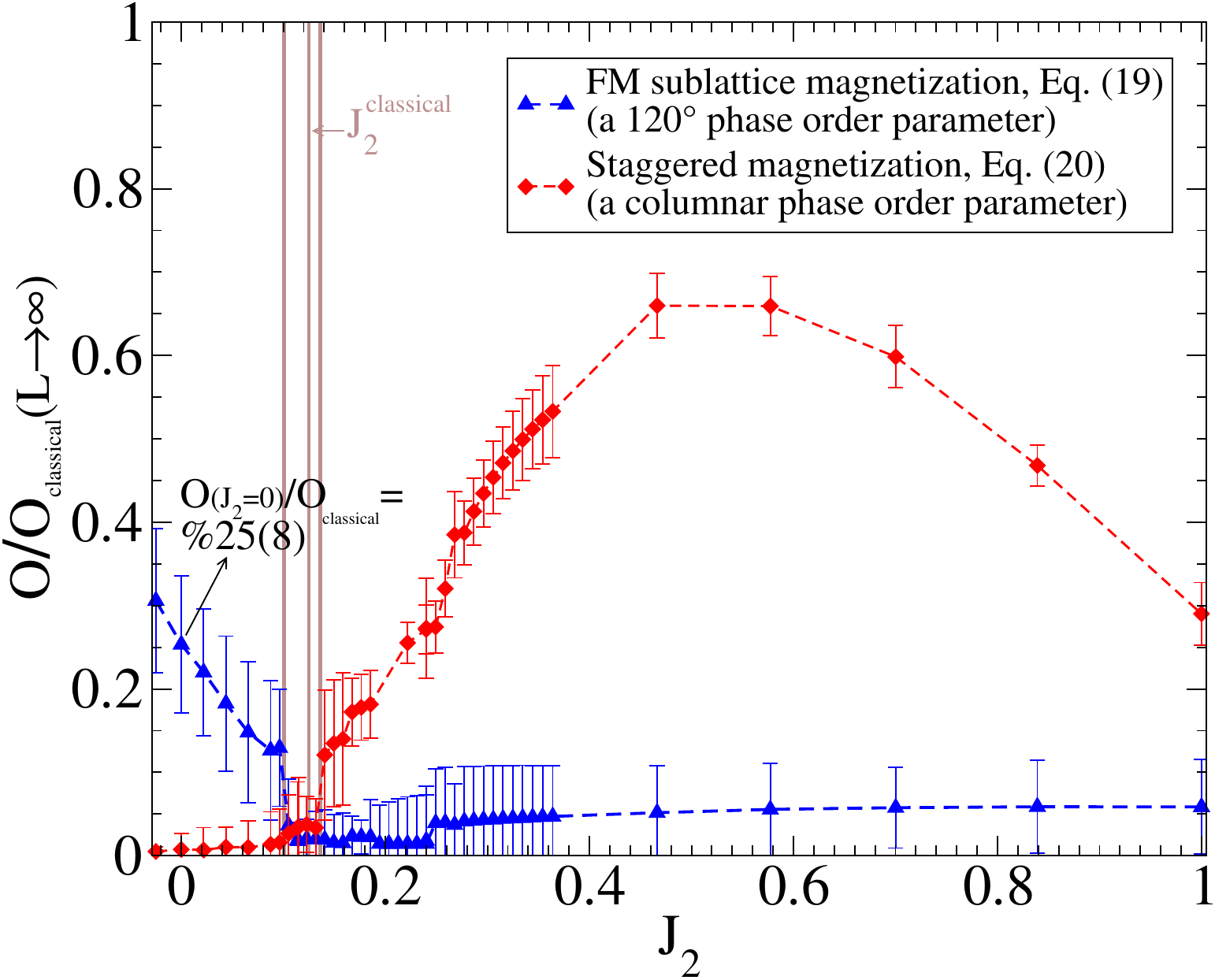}
    \caption{(Color online)
    fDMRG results for the 
    magnetization order parameters of the THM, $O^{\text{FM}}_A$, \eref{eq:FM-SubOP}, and $O^{\text{stag}}$, \eref{eq:StagMag_ch6},
    in the thermodynamic limit of $L\rightarrow\infty$. Each data-point represents a separate 
    extrapolation (and the resulting error) with the method of the fixed-aspect-ratio, \eref{eq:MagExtrapolation}. 
    A variety of cylindrical structures have been used for 
    extrapolation purposes, as listed in \sref{sec:methods}. 
    Brown (outer) stripes are predicted phase boundaries, while the 
    middle stripe is the classical phase transition at $J_2=0.125$~\citep{Jolicoeur89}.
    \label{fig:FiniteMag}}
\end{center}
\end{figure}

The magnetic order parameters that we selected to study the phase
diagram on finite-size cylinders include: the FM sublattice
magnetization, defined arbitrarily on sublattice $A$,
\begin{eqnarray}
  O^{\text{FM}}_A = \frac{2}{\sqrt{N_A(N_A+2)}} 
\sqrt{\left\langle \vektor{S}_A^2 \right\rangle} \;,
  \label{eq:FM-SubOP}  
\end{eqnarray}
where $\vektor{S}_A = \sum_{i \in A} \vektor{S}_i$ is summed over all sites in sublattice $A$,
and $2/\sqrt{N_A(N_A+2)}$ is a normalization factor ($N_A$ is the
total number of sublattice-$A$ spins on the finite lattice).
$O^{\text{FM}}_A$ is a well-defined order parameter for the
$120^{\circ}$ phase.
The \emph{classical} $120^{\circ}$ order will result in the maximum
possible value for the order parameter in the limit of
$L\rightarrow\infty$, i.e.~$O^{\text{FM}}_{A}[\text{classical},
L\rightarrow\infty] = 1$~\cite{SaadatmandThesis}.  The next order
parameter is the staggered magnetization, $\vektor{M}_{\text{stag}}$,
for which the second moment is a well-defined order parameter for the
columnar phase,
\begin{equation}
  O^{\text{stag}} = \frac{1}{L} \sqrt{ \langle \vektor{S}^2_{\text{stag}} \rangle }
  \label{eq:StagMag_ch6} 
\end{equation}
where $\vektor{S}_{\text{stag}} = \vektor{S}_A - \vektor{S}_B$ is the staggered
magnetization for sublattices $A$ and $B$.
The \emph{classical} columnar order will result in
the maximum possible value for the order parameter in the limit of
$L\rightarrow\infty$, i.e.~$O_{\text{stag}}[\text{classical},
L\rightarrow\infty] = 1$\cite{SaadatmandThesis}.

Our results for $O^{\text{FM}}$ and $O^{\text{stag}}$, in the
thermodynamic limit of $L \rightarrow \infty$, are presented in
\fref{fig:FiniteMag}. Individual error-bars are relatively large, but
the overall behavior of the magnetization curves follow the expected
pattern: there exists a small region for $J_2$, where both
$O^{\text{FM}}(L\rightarrow\infty)$ and
$O^{\text{stag}}(L\rightarrow\infty)$ are touching the zero axis
(considering uncertainties), which  provides SL region
boundaries, $J^{\text{low}}_2=0.101(4)$ and
$J^{\text{high}}_2=0.136(4)$.  Next
to the SL phase region, on the left, only
$O^{\text{stag}}(L\rightarrow\infty)$ is touching the zero axis, while
$O^{\text{FM}}(L\rightarrow\infty)$ are increasing for
$J_2\rightarrow-\infty$ (confirming the stabilization of
$120^{\circ}$ order in this region).  On the other hand, next to the SL phase
region on the right, $O^{\text{stag}}(L\rightarrow\infty)$ 
increases rapidly, indicating columnar
order. Interestingly, the value of $O^{\text{FM}}(L\rightarrow\infty)$
($O^{\text{stag}}(L\rightarrow\infty)$) is increasing (decreasing)
again for large $J_2$. This is consistent with the existence 
of a multi-component $120^{\circ}$ order (three copies
of a conventional $120^\circ$ order placed on sublattices;
see~\sref{sec:overview}) in the $J_2 \rightarrow \infty$ limit.

It is worth noting the magnitude of the sublattice magnetization at
$J_2=0$ (NN model). Measurement of variants of a $120^{\circ}$ order
parameter for the NN model has been in the center of
attention~\citep{Jolicoeur89,Bernu94,Li15,Kaneko14,Saadatmand15} to
understand the degree of magnetization reduction (in comparison to
their classical counterparts) in such a frustrated model.  As shown in
\fref{fig:FiniteMag}, we predict
$O^{\text{FM}}[J_2=0] / O^{\text{FM}}[\text{classical}] =
25(8)\%$, which is considerably smaller than approximate results
of $50\%$ by SWT~\citep{Jolicoeur89}, $48\%$ by ED~\citep{Bernu94},
$40\%$ by CCM~\citep{Li15}, and $50\%$ by variational
QMC~\citep{Kaneko14}. 

\section{Correlation Lengths}
\label{sec:CorrLengths}

For infinite cylinders, the gapped or gapless nature of the groundstate can be
understood through the study of the (principal) correlation length,
$\xi$, since the behavior of the magnetic ordering and the scaling
behavior of the static correlation functions are connected
(cf.~\sref{sec:intro}).  Indeed,
the Hastings-Oshikawa-Lieb-Schultz-Mattis 
theorem\cite{Hastings04_PRB,Hastings04_PRL}
relates the size of the energy gap, $\Delta_e$, to $\xi$, for local,
translation-invariant Hamiltonians on even-width cylinders as $\xi \leq
\frac{\mathrm{Const.}}{\Delta_e}$ (i.e.~$\xi^{-1}$ serves as an upper
boundary for the gap size). For the (inherently one-dimensional) MPS
ansatz, the connection between entanglement scaling and the correlation length is
well-understood~\citep{Tagliacozzo08,Stojevic15,McCulloch08}.  In a
critical phase, the correlation length diverges with a signature
\emph{power-law} scaling with the number of states as $\xi(m) =
\tilde{\kappa}_c m^{\tilde{\kappa}}$.  Furthermore, in such states, the entanglement entropy
diverges with a scaling of $S_{EE} \sim \log \xi$.  On the other hand, for
short-range gapped states $\xi$ saturates to a finite value as $m$
is increased, which in the topological spin-liquid state of the THM is
short; of the order of a few lattice spacings.  Interestingly, as we see below,
the correlation length scaling for magnetic
ordering in $SU(2)$-symmetric MPS on infinite cylinders
appears rather differently than the full 2D limit.  Such cylindrical
magnets exhibit some signatures of true LRO (e.g.~in the ES
-- see below), however, due to the explicit preservation of SU(2) and
the dominating 1D physics of the MPS ansatz, the groundstates emerge as
quasi-LRO critical states (note that the correlation length can still
diverge with respect to the cylinder circumference).  Nevertheless, in the iMPS
representation of the wavefunction, the correlation lengths (per
unit-cell) can be conveniently read from the eigenspectrum of the
transfer operator, $\mathbb{T}_I$ (cf.~\sref{sec:methods}):
\begin{equation}
  \frac{\xi_{i}(m)}{L_u} = - \frac{1}{\ln|\eta_{i}(m)| }, ~~ i=2,3,4,...~,
  \label{eq:Xi-def_ch2}  
\end{equation}
where $\eta_i$ are eigenvalues of $\mathbb{T}_I$ (arranged as
$\{|\eta_1| > |\eta_2| > |\eta_3| > ... \}$). $\eta_i$ depends on the number
of states, and are also labeled by an $SU(2)$ spin sector, which is the
symmetry sector of the (block diagonal) transfer operator, and corresponds
to the symmetry of the associated correlation function.
We have discarded $i=1$, as
the largest eigenvalue of $\mathbb{T}_I$ in an orthonormalized basis
always corresponds to $\eta_1=1$ (belonging to the identity eigenmatrix) and
the principal correlation length is the second largest eigenvalue,
$\xi_{2,S} \equiv \xi_{S}$.
For a phase with magnetic ordering,
such as $120^{\circ}$ and columnar order,
the principal correlation
length is expected to belong to the $S\!=\!1$ sector, indicating
that the slowest decaying correlations are in the spin-spin form. For the topological
and algebraic spin liquid phases, we find that the principal
correlation length is in the $S\!=\!0$ sector, indicating that the slowest
decaying correlation is some kind of singlet-singlet correlator
(we have not determined the exact form).
An undesirable effect of the variational convergence of the
groundstate using the iDMRG approach emerges from the constraint of
$SU(2)$ symmetry, whereby spurious
symmetry effects make the wavefunction non-injective (the spectrum of
$\mathbb{T}_I$ contains multiple identity eigenvalues
in each $S$-sector). We have
removed such wavefunctions everywhere except in the immediate vicinity
of the $J_2=0.105(5)$ transition (expectation values are still reliable), 
where the non-injectivity was difficult to avoid (this is likely
due to the closeness of this point to the topological SL region).

\begin{figure}
  \begin{center}
    \includegraphics[width=0.99\columnwidth]{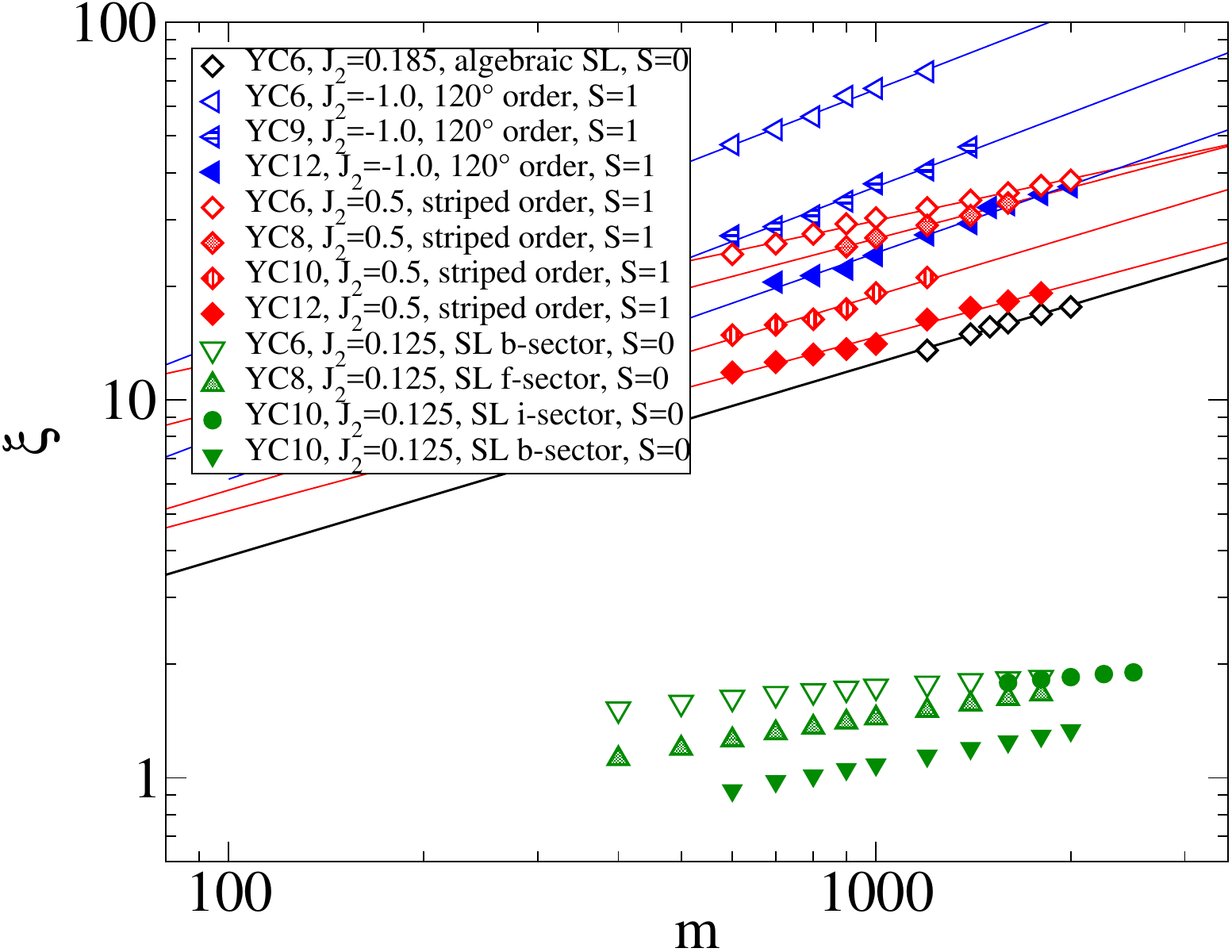}
    \caption{(Color online)
    iDMRG results for the principal correlation lengths (per unit-cell size) 
    versus the number of states, $m$, in a variety of the detected phases and 
    system sizes of the THM on infinite cylinders. Results are labeled with SU(2) 
    quantum numbers, ${S}$. Lines are attempted power-law fits, 
    $\xi = \tilde{\kappa}_c m^{\tilde{\kappa}}$, to quasi-LROs 
    ensuring the existence of a critical phase (see also~\sref{sec:intro}). 
    Green-symbol data are selected from \onlinecite{SaadatmandThesis} to provide 
    comparison between the magnitudes and asymptotic behaviors of $\xi_S(m)$ in gapped and gapless
    phases.
    \label{fig:CorrLength}}
 \end{center}
\end{figure}

We present the correlation length results for the ordered phases in
\fref{fig:CorrLength}, where we compared them against the $\xi_S$
from the topological spin liquid\cite{Saadatmand16}. We immediately
notice that the principal correlation length belongs to the
$S=1$ sector for the magnetic groundstates with $120^{\circ}$ and
columnar ordering, however, it switches to the $S=0$ sector for
all SL states, whether they are quasi-LROs (as in ASLs) or short-range
correlated (as in topological spin liquids). We can see
that both the ASL and the magnetically ordered states have
\emph{power-law} behavior, reflecting their gapless and quantum
critical natures. We emphasize that in the case of the ASL, this
behavior appears to be intrinsic; however for the magnetically ordered
phases the power-law correlations are a consequence of preserving $SU(2)$
symmetry.
In contrast, for the topological spin liquid,
the correlation length is considerably smaller in the size (order of
few lattice spacings) and qualitatively begins to saturate in the
large-$m$ limit, although it is surprisingly difficult to do a rigorous
fit.

\section{Entanglement entropy of quasi-LRO magnets}
\label{sec:EE}

The entanglement entropy is a central quantity in the physics 
of the many-body systems, which
provides a measure of how strongly conjunct subsystems are entangled.
The entropy has proven to be a powerful numerical tool
for characterizing the low-energy spectrum, detection of SSB, and
topological degeneracy of the groundstate (for some examples,
see~\onlinecite{KitaevPreskill06_original,Levin06_original,Eisert10_colloquium,Zhang11,Wang13,Metlitski15}).
Between many different approaches to measure entropy, we employ the method
of Jiang\etal\cite{Jiang12} that calculates the von Neumann
entropy 
along a bipartition cut of the cylinder, as shown in
\fref{fig:TypicalLattice}, since it is computationally convenient to
manipulate in the context of MPS and DMRG algorithms. The bipartite
von Neumann entropy is defined as $S_{EE} = - Tr( \tilde{\rho} \log
\tilde{\rho})$.
In terms of the eigenvalues of $\tilde{\rho}$, i.e.~$\{\lambda_i\}$,
the entropy can be written as $S_{EE} = - \sum_i \lambda_i \log \lambda_i$.
Roughly speaking, $S_{EE}$ counts the number of entangled
pairs on the bipartite boundary.  $S_{EE}$ is a
function of the $(D-1)$-dimensional \textit{area} of the
$D$-dimensional quantum system, i.e.~the boundary size, $L_{cut}$
(note that $L_{cut}=L_y$ for the YC structure). In fact, robust
theoretical
studies~\citep{Hamma05_PRA,Hamma05_PLA,KitaevPreskill06_original,Levin06_original}
proved that for interacting 2D spin systems with only local couplings
and a cut size significantly larger than the correlation length, the
leading term in the entropy scales with the boundary area, $S_{EE} \propto
L_{cut}$, not the system volume, which is known as the \emph{area-law}
(the area-law was originally introduced in the context of the black
holes~\citep{Bekenstein73} and quantum field
theory~\citep{Srednicki93,Plenio05}).
However, for strictly 1D quantum critical states (in the thermodynamic
limit) the condition of the boundary size being considerably larger
than the correlation length \emph{cannot} be met, and the $S_{EE}$
behavior is modified. In this case, the leading term in the entropy relates
to the only length scale of the system, i.e.~the correlation length,
as $S_{EE} \sim \log(L_{eff}) \sim
\log(\xi)$~\citep{Tagliacozzo08,Stojevic15}, where $L_{eff}$ stands
for the effective size of the system.  For the symmetry-broken true
LROs, again, the size of the cut is significantly smaller than the
diverging correlation length and a logarithmic term should be
added to the area-law behavior~\citep{Metlitski15}:
\begin{equation}
  S_{EE} = \beta_0 + \beta_1 L_{cut} + \frac{N_G}{2} \log(L),
  \label{eq:2D-SSB-EE_ch2}
\end{equation}
where $\beta_0$ corresponds to a non-universal constant, which depends
on the system geometry, the topological entanglement
entropy\cite{KitaevPreskill06_original,Levin06_original}, spin
stiffness, and the number, $N_G$, and the velocity of the Nambu-Goldstone
excitations. In addition, $\beta_1$ is another non-universal constant,
which depends on the short-range entanglement in the vicinity of the
cut and a short-distance characteristic cutoff.  For the (quasi-)LRO,
$SU(2)$-symmetric, iMPS groundstates on the infinite cylinders, we
find that the entropy scaling behavior is distinct. 
As discussed in \sref{sec:CorrLengths},
MPS-ansatz symmetry broken magnets appear as quantum critical states 
on the cylinder. Thus, it is expected
that the entropy exhibits a \textit{combination} of the area-law and the
critical behaviors. Our numerical measurements on an $SU(2)$-symmetric,
quasi-LRO groundstate of the $J_1$-$J_2$ THM on the infinite YC
structures confirms such a mixed scaling as of
\begin{equation}
  S_{EE} \simeq a_0(L_y) + a_1(L_y) \log(\xi) \; ,
  \label{eq:iMPS-EE_ch2}
\end{equation}
where
\begin{equation}
  a_1(L_y) = \alpha_0 + \alpha_1 L_y \; .
\label{eq:MPS-EE-coefficient_ch2}
\end{equation}
The behavior of the non-universal constant of $a_0$ proved to be more
challenging to predict, but it can only contain sub-leading
corrections to the area-law term appearing in $a_1$ (see below).

\begin{figure}
  \begin{center}
    \includegraphics[width=0.99\columnwidth]{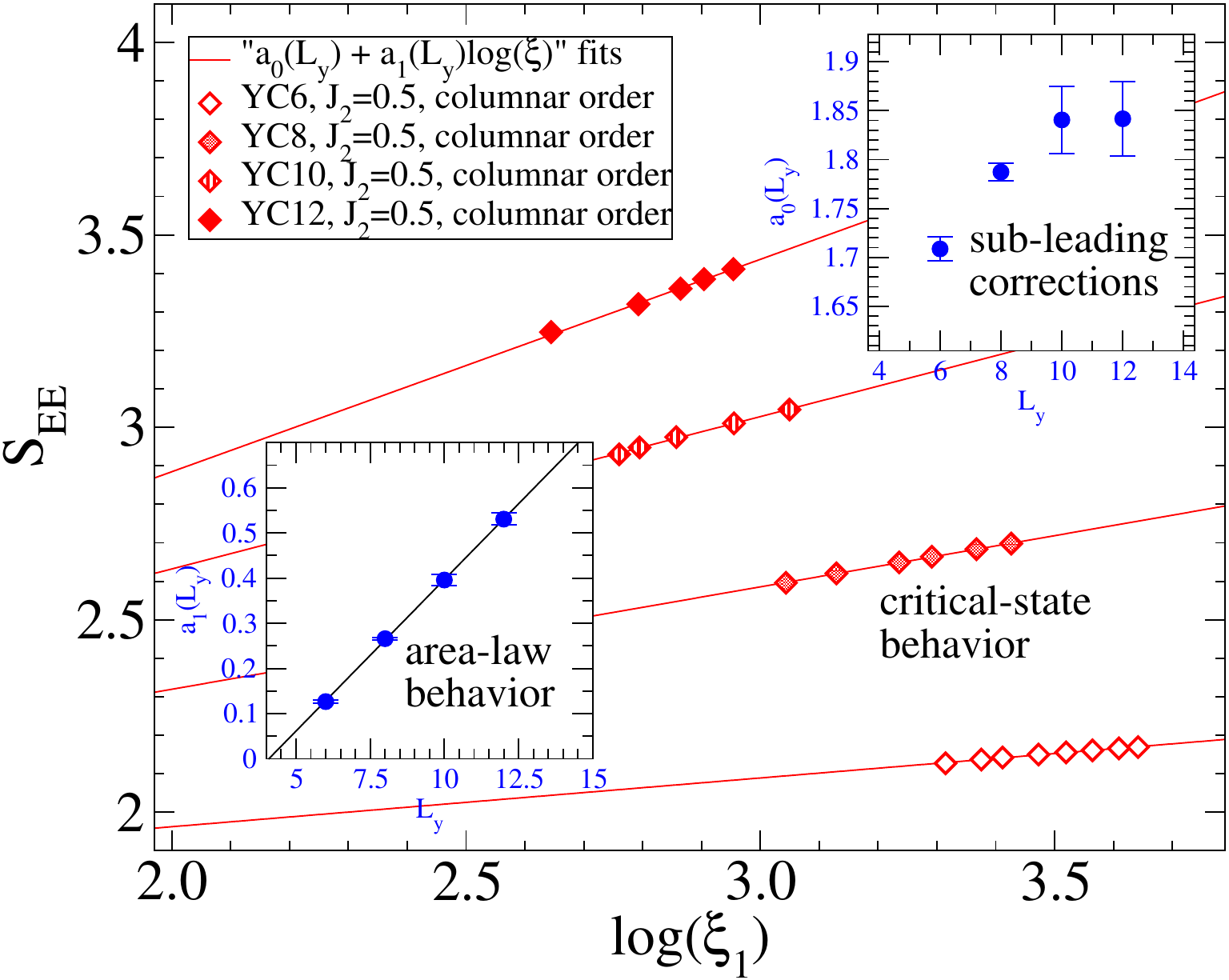}
    \caption{(Color online)
    iDMRG results for the entanglement entropy of the columnar order of the THM at $J_2=0.5$
    on infinite-size YC systems with different system widths. In the main figure, entropies are plotted
    versus $S\!=\!1$ correlation lengths, $\xi_1$, and red lines are attempted fits according to
    \eref{eq:iMPS-EE_ch2}, which is the predicted behavior for the quasi-long-range critical states on infinite cylinders.
    The scaling behaviors of $a_0$ and $a_1$ of \eref{eq:iMPS-EE_ch2} 
    versus the system width are presented in the insets.  
    \label{fig:EE}}
 \end{center}
\end{figure}

In \fref{fig:EE}, we present our entropy measurements for the groundstates
of the THM deep in the \emph{columnar} phase region. Due to
exponential cost of the calculations with the system width we
only obtained a few wavefunctions for different
$L_y$ in the columnar phase. However the results shown in \fref{fig:EE}
confirm the prediction of \eref{eq:iMPS-EE_ch2} and
\eref{eq:MPS-EE-coefficient_ch2}. In the figure, we first fit a line
to the original entropy data and calculate $a_0$ and $a_1$ for each system
size. Clearly, $a_1$-values are consistent with the area-law behavior.
We measured the coefficients of $a_1$ as $\alpha_0=-0.28(1)$ and
$\alpha_1=0.068(1)$. In contrast, there was no obvious fit possible
for $a_0$-values, but their saturating nature for the large-$L_y$
limit is consistent with this term being a sub-leading correction to
the mixed term containing the area-law behavior.

\section{Numerical tools I: cumulants and Binder ratios of the
  magnetization order parameters}
\label{sec:cumulants}

In \sref{sec:methods}, we constructed the theoretical framework for a
method to measure the non-local moments and cumulants
of the magnetic order parameters, in the context of $SU(2)$-symmetric
translation-invariant MPS, where all projection components of the
magnetic order parameter, $\vektor{M}^{[k]}$, vanish by
construction. In this case, the higher moments can play the role of
the order parameter.  It is convenient to connect the moments of the operators to
the $i$th cumulant \emph{per site}, $\kappa_i$, by employing
\begin{equation}
  \la M^n \ra = \sum_{i=1}^{n} B_{n,i}(\kappa_1 L, \kappa_2 L, ... , \kappa_{n-i+1} L), 
  \label{eq:MomentsCumulants_ch2}
\end{equation}
where $B_{n,i}$ are \emph{partial Bell
  polynomials}~\citep{Stuart09_book} and $L$ now stands for the
operator length.
For some examples, we expand
\eref{eq:MomentsCumulants_ch2} to write the relations for the first
few cumulants,
\begin{align}
  \la M \ra &= \kappa_1 L~, \notag \\
  \la M^2 \ra &= \kappa_2 L + \kappa_1^2 L^2~, \notag \\
  \la M^3 \ra &= \kappa_3 L + 3\kappa_2\kappa_1 L^2 + \kappa_1^3 L^3~, \notag \\
  \la M^4 \ra &= \kappa_4 L + ( 4\kappa_3 \kappa_1 + 3\kappa_2^2) L^2
  + 6 \kappa_2 \kappa_1^2 L^3 + \kappa_1^4 L^4~.
  \label{eq:FirstFewMoments_ch2}
\end{align}
The cumulants per site are obtained directly as the asymptotic large
$L$ limit obtained from the summation of the tensor diagrams presented
in \sref{sec:methods}.
For the iMPS ansatz, when the asymptotic limit is taken to derive a
translation-invariant infinite-size system, one should replace the
operator length with the effective system size as $L \rightarrow
L_{eff} \propto \xi$ (see also~\sref{sec:EE}).  Below, we introduced
the magnetic order parameters that are used to measure the cumulants
and characterize the LROs of the THM. We
first construct the MPO forms of the higher moments of a
\emph{staggered magnetization} (the order parameter for
columnar order on cylinders with FM stripes in
$\vektor{a}_{+60^\circ}$-direction),
\begin{equation}
  \vektor{M}_{stag} = \sum_{i=1}^{L_y} (-1)^i \vektor{S}_i, 
  \label{eq:StaggMag_ch6}
\end{equation}
and a \emph{tripartite magnetization} (the order parameter for
the $120^\circ$ phase),
\begin{equation}
  \vektor{M}_{tri} = \sum_{i\epsilon<A,B,C>}^{L_y} ( \vektor{S}_{A_i} + e^{i\frac{4\pi}{3}}\vektor{S}_{B_i}
  + e^{-i\frac{4\pi}{3}}\vektor{S}_{C_i} ), 
  \label{eq:TriMag_ch6}
\end{equation}
on a $L_y$-size unit cell.  Numerical
computation of the moments of such order parameters is a challenging
task due to relatively large dimensions of the resulting MPOs.
Nevertheless, we succeeded to calculate the second cumulant,
$\kappa_2$, and the fourth cumulants, $\kappa_4$, of $M_{stag}$
and $M_{tri}$ (the odd moments vanish due to the $SU(2)$ symmetry)
for a range of the groundstates. We suggest that the most
useful choice of cumulants is
$\kappa_4$, which is connected to the \emph{excess
kurtosis}~\citep{Stuart09_book}, $\gamma_4$, of the block
distribution function associated with the operator $\vektor{M}^{[k]}$:
\begin{equation}
  \gamma_4 = \frac{\kappa_4}{\kappa_2^2 L}~. 
  \label{eq:kurtosis_ch2}
\end{equation}
We emphasize that the above equation is only valid for the $\kappa_1=
\kappa_3=0$ case.  The importance of the fourth cumulant was revealed
by some studies on fourth magnetic moment behavior of 2D Ising
antiferromagnets~\citep{Tsai98,Malakis07}, which established
$\kappa_4$ as an effective tool for pinpointing quantum
critical points. In these studies, the scaling behavior of the fourth
magnetic moment is observed to vary significantly at an Ising
transition (more precisely, $\kappa_4$ changes sign at the critical
point, and changes by many orders of magnitude nearby the critical point).
Another relevant and interesting (dimensionless) quantity is the
Binder cumulant~\citep{Binder81_PRL,Binder81_ZPB,Binder84,Sandvik10},
$U_L = \frac{n_H + 2}{2} ( 1 - \frac{n_H}{n_H + 2} \frac{ \la
  \vektor{M}^4 \ra }{ \la \vektor{M}^2 \ra^2 } ) $, where $n_H$ is the
number of projection spin operators used to construct the order
parameter (e.g.~$n_H=3$ for a vector magnetization). In the vicinity
of a critical point, the Binder cumulant becomes independent of the system
size (lower moments of the order parameter cancel out higher-order
finite-size effects) and can be used to pinpoint the transition.
Previously, we adopted\cite{Saadatmand15} $U_L$ of a (scalar) dimer
order parameter to locate a critical point in the phase diagram of the
THM on three-leg cylinders. However, until now, the scaling behavior
of $U_L$ was less-known for the cases where the order
parameter itself is strictly zero.
In the limit of $L\rightarrow\infty$, as it is clear from
\eref{eq:FirstFewMoments_ch2}, the higher-order corrections in
$\la M^n \ra$ vanish and the conventional method of Binder cumulants
for locating the phase transitions becomes ineffective.  However, the
correlation length, $\xi$, gives us a natural length scale and a
rather precise process to scale a Binder-cumulant-type quantity in the
vicinity of a critical point.
As in the case of the entropy, \sref{sec:EE}, the
key to obtaining the correct scaling of the magnetic moments of iMPS
wavefunctions is to choose $L_{eff} = \tilde{s} \xi$, where
$\tilde{s}$ is any \emph{fixed} scaling constant. For
Binder cumulant, $\tilde{s}$ has no qualitative effect except to change
the value of the critical
binder cumulant, similar to the role of
boundary conditions for the finite-size Binder cumulant.  Therefore,
one can freely choose $\tilde{s}$ to obtain the most numerically stable
fit. When the order parameter is zero by symmetry, so that $\kappa_1 = \kappa_2 = 0$,
the appearance of such
a constant is irrelevant and only the
ratio of the second and fourth cumulants plays a role. By
replacing the explicit relations for $\la \vektor{M}^2 \ra$ and $\la
\vektor{M}^4 \ra$ from \eref{eq:MomentsCumulants_ch2} into $U_L$, 
we propose the ratio (which
we call the ``Binder ratio'' -- see also~\eref{eq:kurtosis_ch2}):
\begin{equation}
  U_r = \frac{\kappa_4}{\kappa_2^2  \xi} \; .
  \label{eq:BinderRatio_ch6}
\end{equation}
We find that numerically this combination of the moments and the correlation
length removes much of the numerical noise that appears in the individual
moments.

\begin{figure}
  \begin{center}
    \includegraphics[width=0.99\columnwidth]{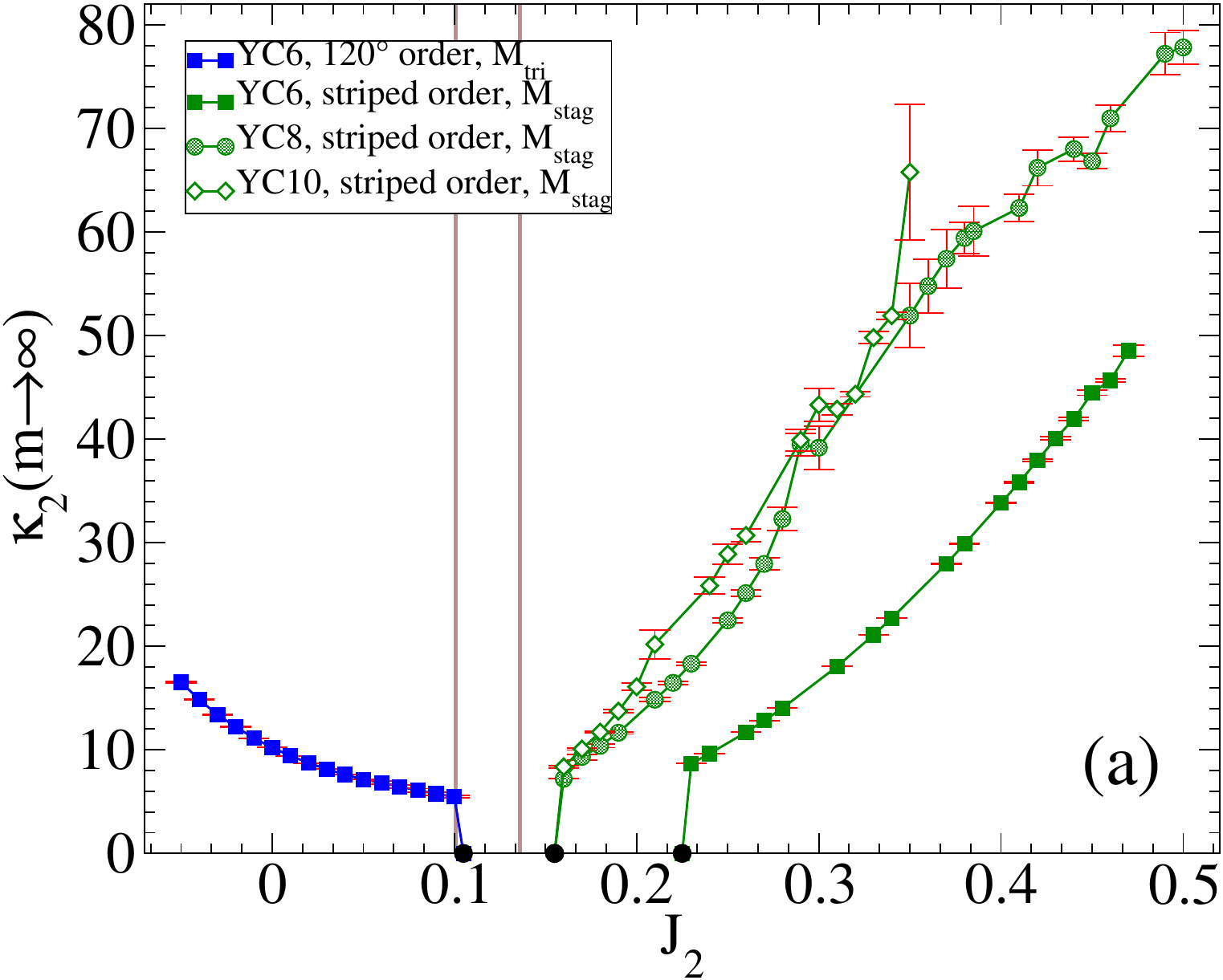}
    \includegraphics[width=0.99\columnwidth]{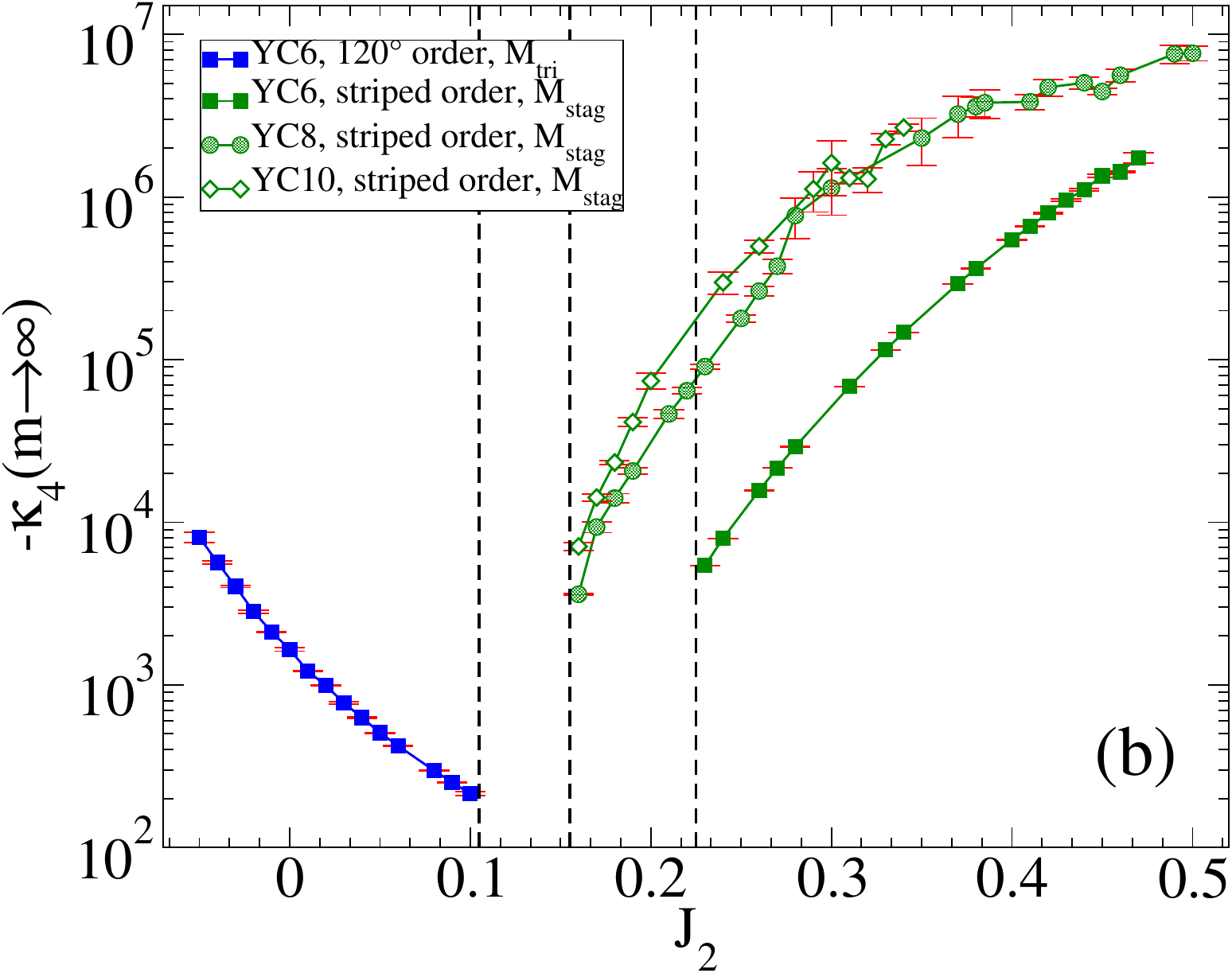}
    \caption{(Color online)
    iDMRG results for the extrapolated (a) second cumulants and
    (b) fourth cumulants 
    of the magnetization order parameters, \eref{eq:StaggMag_ch6} 
    and \eref{eq:TriMag_ch6}, at the thermodynamic limit of $m\rightarrow\infty$, 
    on a variety of phase regions and system widths of the THM.   
    Each colored data-point represents a $\kappa_2(m\rightarrow\infty)$-value (-$\kappa_4(m\rightarrow\infty)$-value), which is
    the result of an extrapolation 	
    according to a power-law fit $\kappa = \breve{a}_0 + \breve{a}_1 e^{-\breve{a}_2 m}$ 
    for  $m\rightarrow\infty$ (see \onlinecite{SaadatmandThesis} 
    for some examples on the individual extrapolations).
    In part (a), brown stripes are fDMRG results for the phase transition obtained
    from direct measurements of the local magnetization, \sref{sec:FiniteMag}. Solid circles, in part (a), 
    and dashed-lines, in part (b),
    mark the borders beyond which an extrapolation was not possible due to the magnetic disorder.
    \label{fig:cumulants}}
 \end{center}
\end{figure}

We present the extrapolated results of $\kappa_2$ and $|\kappa_4|$
for $\vektor{M}_{stag}$ and $\vektor{M}_{tri}$, in the limit of
$m\rightarrow\infty$, in \fref{fig:cumulants}. In the figures, each
data-point is the result of a separate extrapolation of the cumulants
versus $m$.
Upon careful numerical examination of the scaling behaviors of
numerous groundstates in the various phases, we were able to
establish the scaling relation of $|\kappa_{n}| = \breve{a}_0 +
\breve{a}_1 e^{-\breve{a}_2 m},~n=2,4,$ for ordered phase regions and
make sense of the cumulant results in the $m\rightarrow\infty$ limit.
These results show that $\kappa_4$
is comparatively large and negative when there is quasi-long-range magnetic ordering. 
Moreover, $\kappa_2$ is large and
positive for quasi-LROs (see $120^\circ$ and columnar phase regions in \fref{fig:cumulants}).  
This is in contrast to the behavior near phase transitions, and within the
topological and algebraic spin liquids, where we were not able to
find an appropriate analytical fit for the cumulants in the
$m\rightarrow\infty$ limit, as they behave irregularly or quickly decay
to numerically vanishing values. A likely reason for this is that for
a magnetically-ordered, $SU(2)$ $S=0$ groundstate,
the moments $\vektor{M}^{[k]}$ acquire a set of equally-weighted non-zero values
from the limited number of recovered (purely) TOS levels by iDMRG (see
below). In such a case, the distribution function would resemble a
discrete uniform distribution with very large and negative
$\kappa_4$, and large and positive $\kappa_2$. However, for
disordered states with no symmetry breaking in the thermodynamic
limit, the distribution function is expected to resemble the normal
distribution centered around zero magnetization, which has vanishing
$\kappa_4$. For $\kappa_2(m\rightarrow\infty)$, in
\fref{fig:cumulants}(a), we display in bold the boundaries where
we were not able to extrapolate to
$m\rightarrow\infty$. These are quite close to the
phase transitions indicated by fDMRG, \fref{fig:FiniteMag} (except
for the YC6 structure, where we find an additional ASL phase), 
which supports the validity of the iDMRG cumulant method.
The same behavior was observed for $\kappa_4(m\rightarrow\infty)$,
indicated by the black dashed-lines in \fref{fig:cumulants}(b). In
addition, the extremely large (negative) values of
$\kappa_4(m\rightarrow\infty)$ are consistent with our interpretation.

\begin{figure}
  \begin{center}
    \includegraphics[width=0.99\columnwidth]{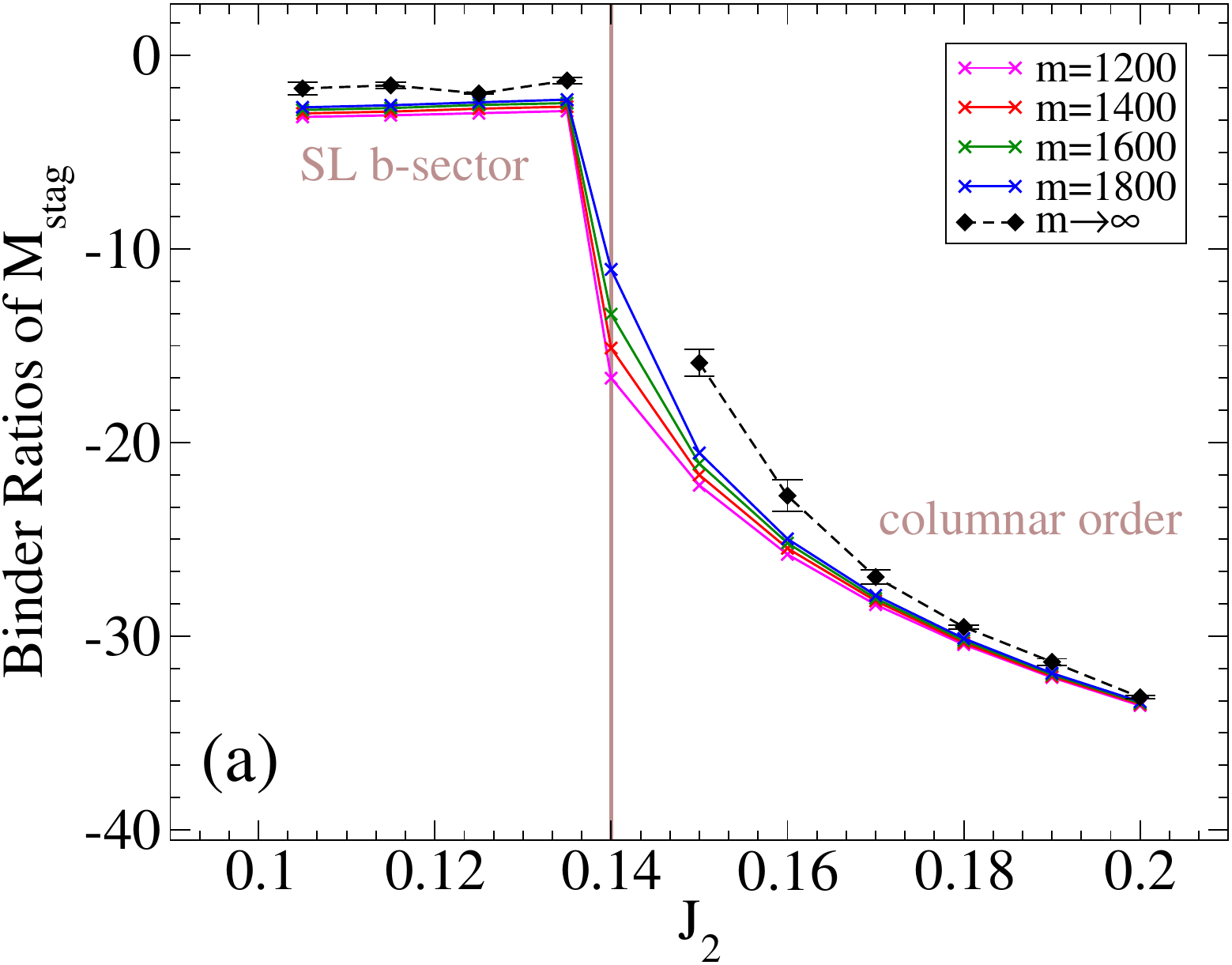}
    \includegraphics[width=0.99\columnwidth]{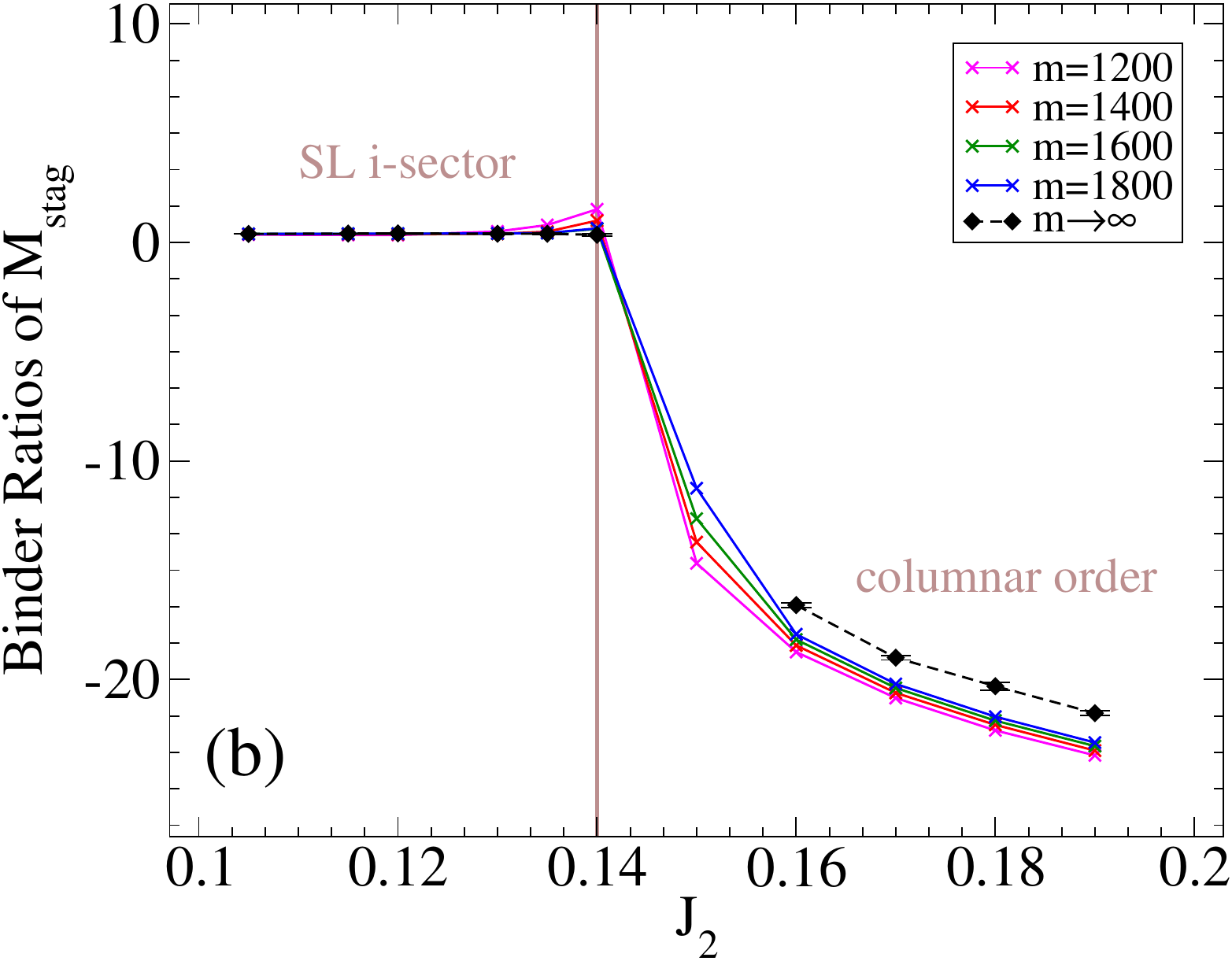}
    \caption{(Color online)
    iDMRG results for the Binder ratios, $U_r(m)$, 
    \eref{eq:BinderRatio_ch6}, of $M_{stag}$, \eref{eq:StaggMag_ch6},
    in the vicinity of the topological SL and the columnar phase regions of the THM 
    on (a) YC10 and (b) YC8 structures. Black diamonds denote
    $U_r(m\rightarrow\infty)$, i.e.~extrapolations of Binder ratios
    according to \eref{eq:Ur-scaling_ch6} to the $m\rightarrow\infty$ 
    limit. Brown stripes are the best estimate for the phase transition
    based on the discontinuity of the dashed line, 
    $U_r(m\rightarrow\infty)$, for the larger system size, part (a).
    \label{fig:YC8&10-TransitionPointPredictions}}
 \end{center}
\end{figure}

\begin{figure}
  \begin{center}
    \includegraphics[width=0.99\columnwidth]{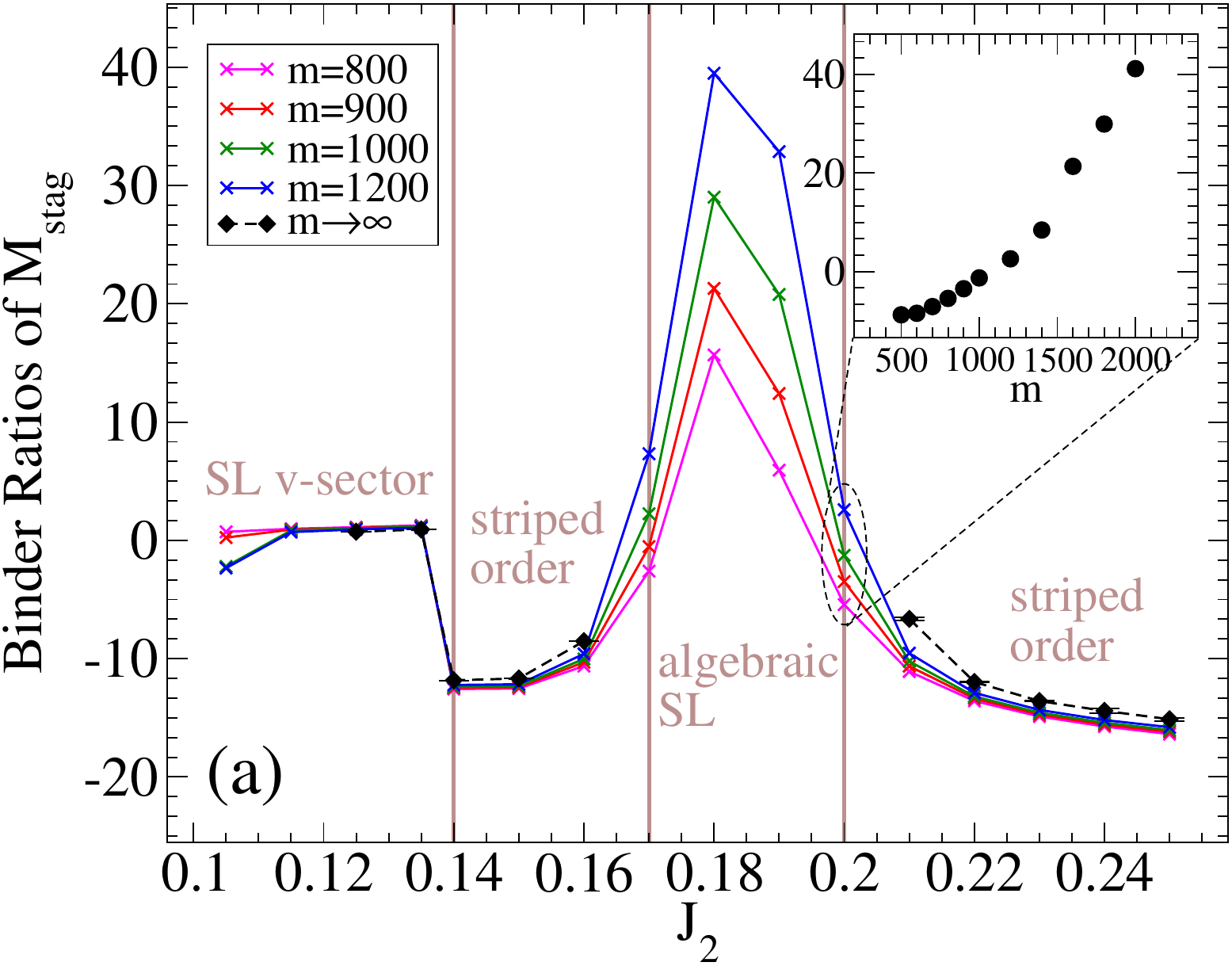}
    \includegraphics[width=0.99\columnwidth]{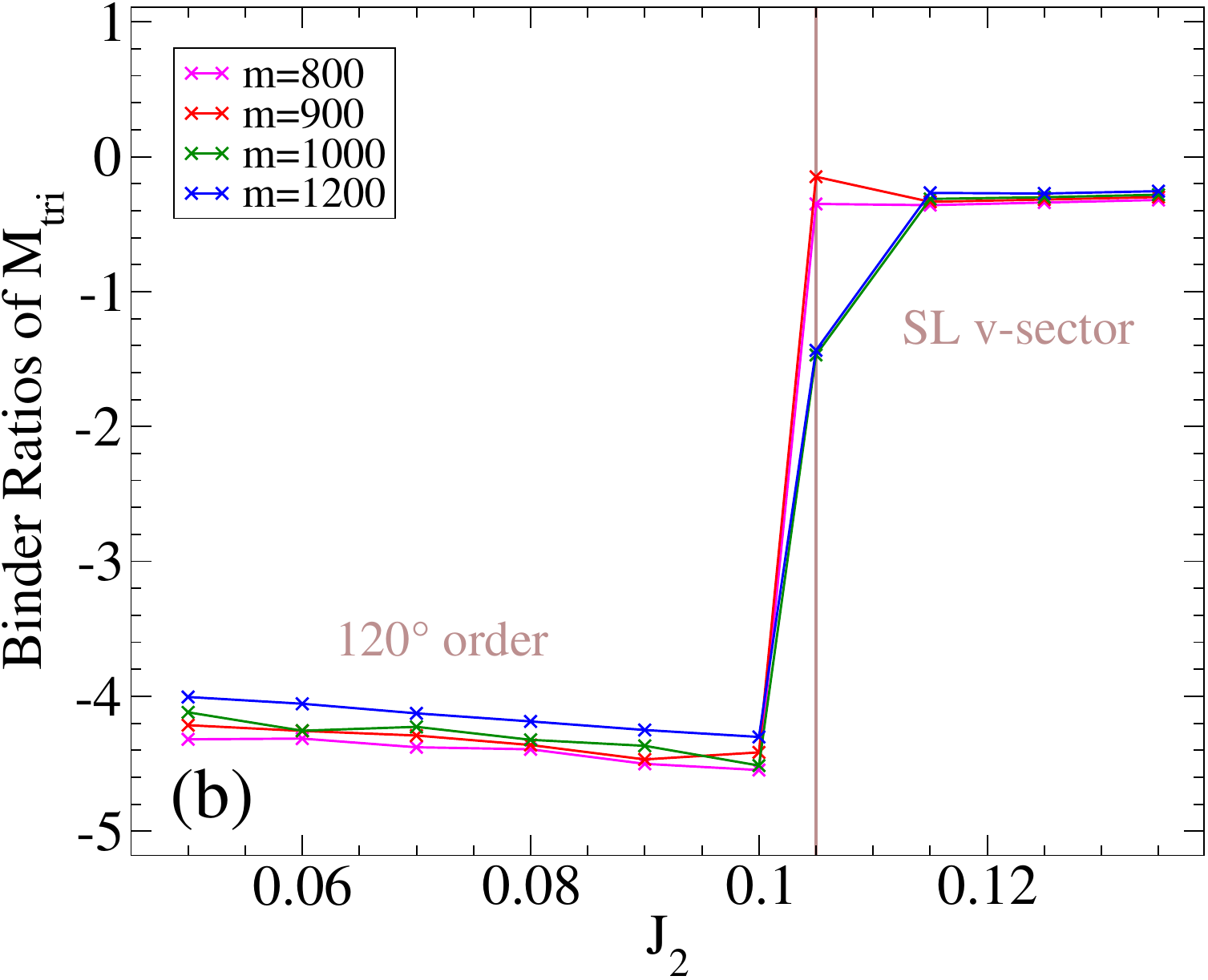}
    \caption{(Color online)
    iDMRG results for the Binder ratios, 
    $U_r(m)$, \eref{eq:BinderRatio_ch6}, of (a) $M_{stag}$, \eref{eq:StaggMag_ch6},  
    and (b) $M_{tri}$, \eref{eq:TriMag_ch6},
    of the THM on YC6 systems. In part (a), black diamonds denote
    $U_r(m\rightarrow\infty)$, i.e.~the
    the extrapolation of the of Binder ratio
    according to \eref{eq:Ur-scaling_ch6} toward the $m\rightarrow\infty$ 
    limit. Furthermore, the inset shows the individual $U_r(M_{stag})$ at $J_2=0.2$.
    Brown stripes are the best estimate for the phase transition
    based on the discontinuities or rapid changes in $U_r$.  
    \label{fig:YC6-TransitionPointPredictions}}
 \end{center}
\end{figure}

\begin{figure}
  \begin{center}
    \includegraphics[width=0.99\columnwidth]{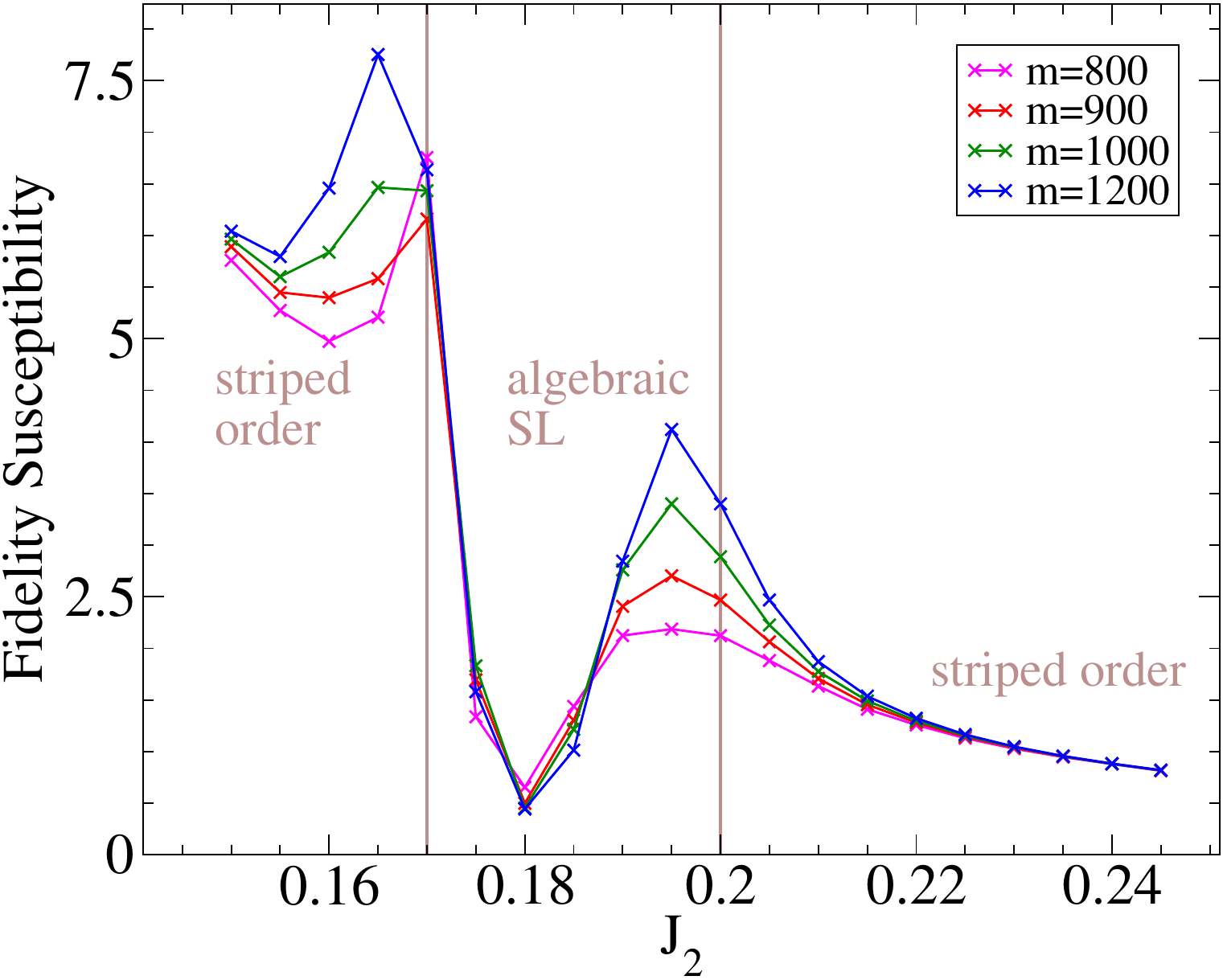}
    \caption{(Color online)
    iDMRG results for the fidelity susceptibility, $\chi_F^{\text{approx}}$, \eref{eq:FidelitySusceptibility},
    of the THM on YC6 systems. Brown stripes are the predicted 
    phase transitions based on \fref{fig:YC6-TransitionPointPredictions}(a) results.
    \label{fig:YC6-FidelitySusceptibility}}
 \end{center}
\end{figure}

Our attempts to pinpoint the phase transitions of the THM on infinite
cylinders, using $U_r$, is presented in
\fref{fig:YC8&10-TransitionPointPredictions} and
\fref{fig:YC6-TransitionPointPredictions}. Based on these results, we
argue that $U_r(m)$, as the ratio between $\kappa_4(m)$ and
$\kappa_2(m)$, scaled with $\xi(m)$, regularly
extrapolates to a finite value in the $m\rightarrow\infty$ limit
everywhere, except close to (or on) a phase transition, or when the
wavefunctions are non-injective (cf.~\sref{sec:CorrLengths}). Careful
numerical examination suggests that the Binder ratios scale with a
saturating behavior similar to the cumulants,
\begin{equation}
  U_r = \breve{b}_0 + \breve{b}_1 e^{-\breve{b}_2 m},
  \label{eq:Ur-scaling_ch6}
\end{equation}
In \fref{fig:YC8&10-TransitionPointPredictions}, we observe that in
the topological SL phase region, $U_r(m)$ has a comparatively
small value, as expected for nonmagnetic phases with
$\gamma_4\rightarrow 0$. In addition, when there is magnetic ordering,
$U_r(m)$ converges to a finite, negative value, while it appears
different $m$-curves tend to group together.  The latter should be due
to the fact that the iMPS magnetic orders are quantum critical states
with an $U_r$ independent from $L_{eff}$.  Furthermore, for the ASL
phase of YC6 structures (see
\fref{fig:YC6-TransitionPointPredictions}(a)), very close to the
expected phase transition points from the short-range correlation
data, \fref{fig:ShortRangeCorr}, and within the entire ASL phase region, $U_r$ diverges
with $m$ (e.g.~see the inset of \fref{fig:YC6-TransitionPointPredictions}(a)),
where it is impossible to extrapolate to a finite $U_r(\infty)$.
In the immediate vicinity of the transition from the $120^\circ$ to topological spin
liquid (cf.~\fref{fig:YC6-TransitionPointPredictions}(b)), it was not
possible to employ \eref{eq:Ur-scaling_ch6} due to unavoidable
non-injectivity of the wavefunctions.  However, we suggest that the
fixed-$m$ results are rather reliable and can be used to estimate a
phase transition.  Overall, we locate critical points of the THM from
the discontinuities of $U_r(m\rightarrow\infty)$-lines (i.e.~where
there is no extrapolation possible) or when there is a significant
kink in fixed-$m$ data. Based on this approach, we estimate the phase
transition points of $J_2 = 0.105(5)$ between the $120^\circ$ and
topological spin liquid states using YC6 results of
\fref{fig:YC6-TransitionPointPredictions}(b), $J_2 = 0.140(5)$ between
topological SL and columnar states using YC10 results of
\fref{fig:YC8&10-TransitionPointPredictions}(a) (YC8 results of
\fref{fig:YC8&10-TransitionPointPredictions}(b) would estimate a
transition very close to this point, so we have based the final
prediction on the larger-width data), and transition points of $J_2 =
0.140(5),0.170(5),0.200(5)$ encapsulating the ASL and columnar states
using YC6 results of \fref{fig:YC6-TransitionPointPredictions}(a).  To
further validate $U_r$ accuracy in estimating the transitions in case
of YC6 structures, we also provide a numerical approximation for the
\emph{fidelity susceptibility}~\citep{You07},
\begin{equation}
  \chi_F^{\text{approx}} = \frac{1 - |\la \psi_0(J_2) | \psi_0(J_2+\delta J_2) \ra|^2}{\delta J_2^2}, 
  \label{eq:FidelitySusceptibility}
\end{equation}
in \fref{fig:YC6-FidelitySusceptibility}, where we set $\delta
J_2=0.05$. The fidelity susceptibility is known to be well-behaved and
small when away from a phase boundary, but can diverge at a
transition. It is clear from the figure that the diverging peaks of
$\chi_F^{\text{approx}}$ (considering their tendency to lean toward
the right) are happening relatively close to the predicted phase
transitions from the Binder ratio results of
\fref{fig:YC6-TransitionPointPredictions}(a).

\section{Numerical tools II: `TOS columns' in the momentum-resolved
  entanglement spectrum}
\label{sec:ES}

The entanglement between the partitions of a quantum system is encoded
in the spectrum of the entanglement Hamiltonian, $H_E =
-\log(\tilde{\rho})$, i.e.~$\{-\log(\lambda_i)\}$, which is known as
the ES and commonly presented using energy-level arrangements
analogous to an energy spectrum.  $\{-\log(\lambda_i)\}$ can be
labeled using any global-symmetry quantum number to extract more
information on the symmetry nature of the state (as long as the
corresponding symmetry is preserved on the bipartite cut). $H_E$
maintains the symmetries of a cylindrical wavefunction, however, they
may exist some symmetries that are \textit{not} explicitly preserved
by the ansatz due to the mapping of the 2D model onto an MPS chain.
Nevertheless, one can still diagonalize such a symmetry operator in
the `auxiliary' basis (i.e.~the basis that diagonalizes $H_E$) to
create a new set of good quantum numbers (see
\onlinecite{Kolley13,Saadatmand16,Saadatmand16,Cincio13} for some
examples).  When the $SU(2)$-symmetry is preserved in the calculation,
the obvious choice for the labels is the spin $S$ quantum number (belonging to
a single partition of the system). We refer to an $H_E$ spectrum that
is plotted against $S$ (where no other label exists) as the
\emph{spin-resolved} ES.  Kolley\etal\citep{Kolley13} showed that the
spin-resolved ES of a magnetically ordered state on finite-length cylinders
shows signatures of 
symmetry-breaking at the thermodynamic limit. This emerges
from a key finding: the
realization~\citep{Metlitski15,Kolley13,Rademaker15} that the
\emph{low-energy} part of the ES of magnetic orders exhibits a
specific type of grouped levels, known as the entanglement-spectrum
TOS (also referred to as ``quasi-degenerate joint states''),
closely resembling the low-lying levels in the energy spectrum, known
as the Anderson TOS levels~\citep{Anderson52,Lhuillier05,Bernu94}
(also referred to as the ``Pisa tower'' structure or the ``thin
spectrum''), which is considered as clear-cut evidence for the
existence of \textit{true} LROs on finite lattices.
Kolley\etal~established that, similar to the energy spectrum, for a
fixed $S$-sector, entanglement-spectrum TOS levels are well-separated
from the denser rest of the spectrum and the lowest energy levels of
the ES, immediately above the TOS levels, are spin-wave states (Nambu-Goldstone
modes).  In this paper, we are interested in exploiting both the $S$
quantum numbers ($SU(2)$ is explicitly preserved in the iDMRG
calculations), and the momenta in the cylinder $Y$-direction, $k$,
i.e.~the complex phase of the eigenvalues of the reduced $T_y$
operator, where $T_y$ is the translation by one site in $Y$-direction;
we can decompose the operator in the same way as the Schmidt
decomposition of the wavefunction~\cite{Schollwock11}, $T_y = T^L_y
\otimes T^R_y$, where $T^L_y$ and $T^R_y$ are the reduced operators and
maintain the unitary property of the original operator. $T_y$ is
\emph{not} preserved exactly in the calculations due to the MPS
mapping on the cylinder, \fref{fig:TypicalLattice}, but it can be
diagonalized straightforwardly\cite{SaadatmandThesis}.  We refer to an
$H_E$ spectrum that is plotted against $k$ and additionally labeled
by $S$, as the momentum-resolved ES, $\{-\log \big(
\lambda_n[k_n,S_n] \big) \}$.  For a system with PBC in $Y$-direction,
dihedral symmetry implies that $T_y^{L_y}=I$.  As a result, the
allowed momentum spacing is as $\Delta k_n = \frac{2\pi n}{L_y}$ for
$n=0, 1, ..., L_y-1$.  We notice that $k_0$, the momentum of the
lowest ES level, is not fixed due to the possibility of inserting a
shift in the expectation value of $T_y$ (one needs to first fix $k_0$,
then measure the rest of the momenta in respect to it; physically only 
$\Delta k_n$ matters here -- see also~\onlinecite{SaadatmandThesis,Michel10}).
The study of momentum-resolved forms of the ES is now finding a place
in the literature of the low-dimensional quantum magnets. Another key
breakthrough was the realization of that such ES can be used to fully
classify anyonic sectors of chiral~\citep{Cincio13} and
$\mathbb{Z}_2$-gauge~\citep{Zaletel15}
topological orders on infinite
cylinders~\citep{Saadatmand16}. Below, we argue that the symmetry-breaking can be
recognized and characterized using the momentum-resolved ES, which
shows the symmetry properties even more robustly than the 
spin-resolved ES.

\begin{figure}
  \begin{center}
    \includegraphics[width=0.49\columnwidth]{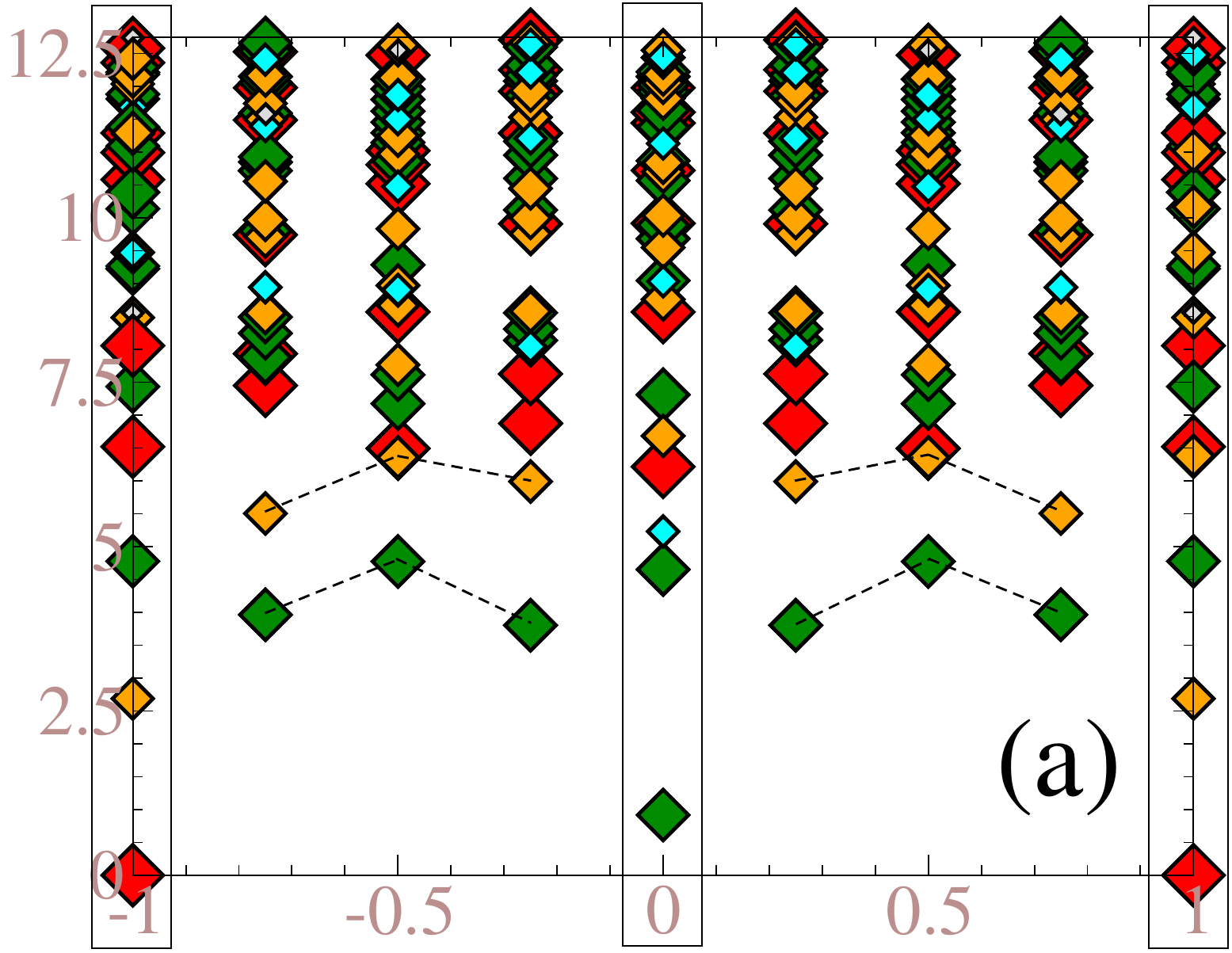}
    \includegraphics[width=0.49\columnwidth]{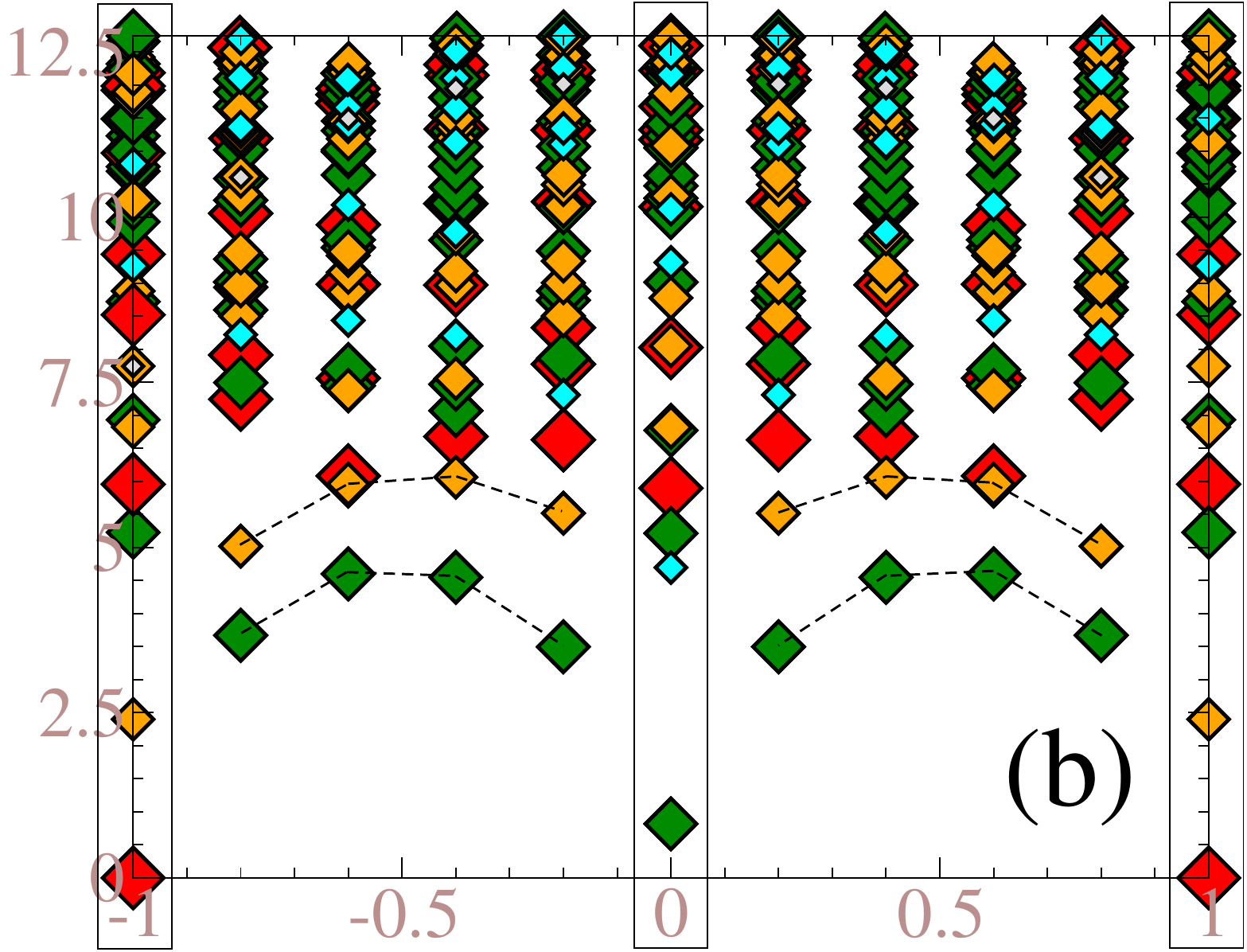} \\
    \includegraphics[width=0.99\columnwidth]{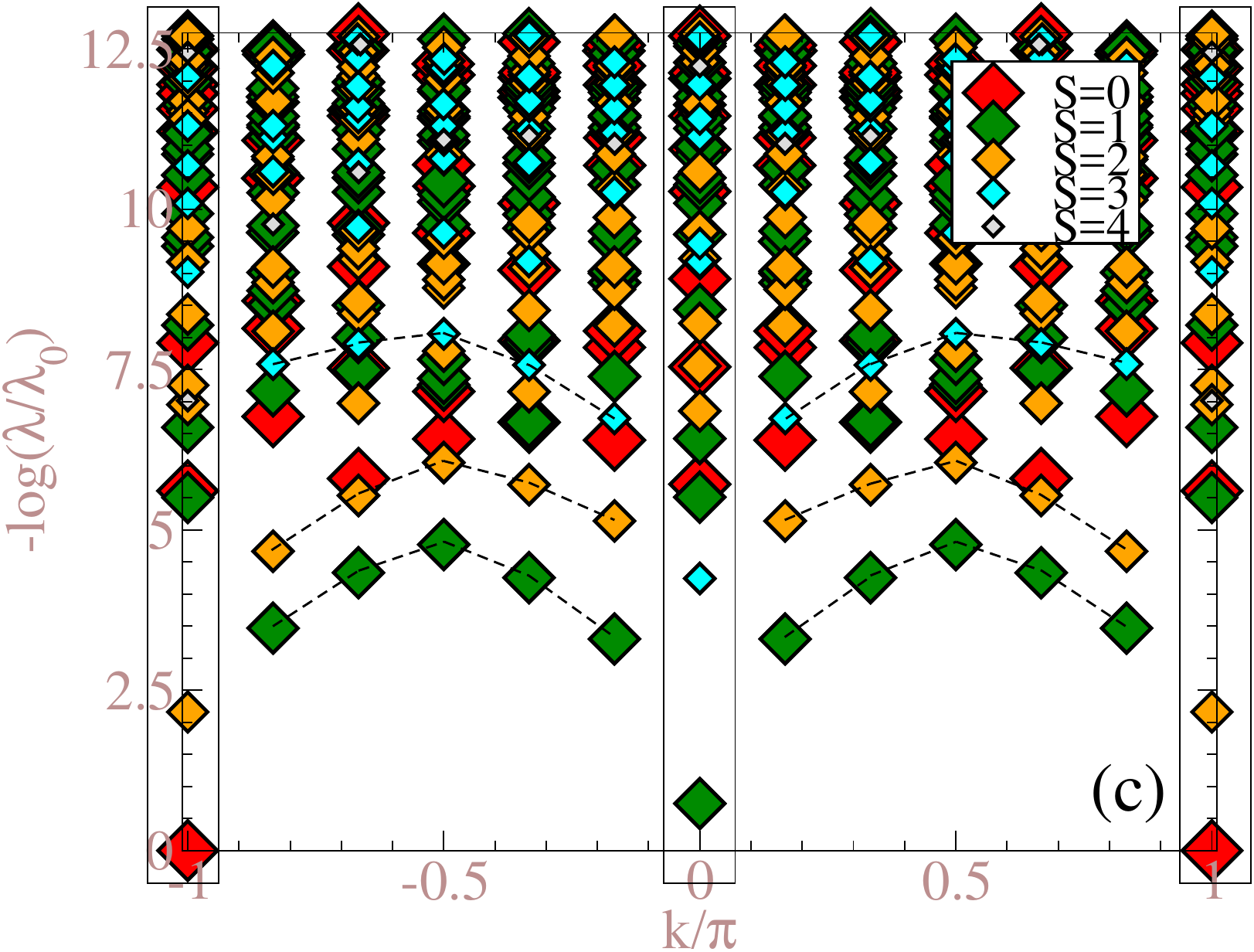}
    \caption{(Color online)
    iDMRG momentum-resolved ES of the columnar order, $J_2=0.5$, for 
    (a) YC8, (b) YC10 and (c) YC12 structures of the THM versus
    $Y$-direction momenta (the reference momentum is fixed as 
    $k^{\text{TOS}}[\lambda_0]=\pm\pi$). Boxes emphasize TOS 
    columns at the unique momenta of $k_\nu^{\text{TOS}} = 0, \pi$. Dashed-lines 
    are guides to the eyes and connect the Nambu-Goldstone modes
    of the ES for the first few levels on the top of the TOS levels.
    \label{fig:ES-columnar}}
 \end{center}
\end{figure}

Upon careful examination of the momentum-resolved ES of the magnetic
orders in the THM on infinite cylinders and noticing the underlying
symmetries of the sublattices, we find that 
the spectrum contains exactly $N_s$ (number of the
groundstate sublattices) column-like structures, which are
the low-lying component TOS levels, independent of the system width.
We shall refer to these particular patterns as `TOS columns'. The
appearance of TOS columns is due to that, as previously discussed, the
TOS levels are clear features in the low-lying ES. These columns
also have a momentum structure. Consider an
ideal magnetic order that consists of $N_s$ fully FM sublattices,
represented as $\{ \vektor{\tilde{S}}_1, \vektor{\tilde{S}}_2, ...,
\vektor{\tilde{S}}_{N_s} \}$ ($L_y = 0\!\mod\!N_s$) in a big-S
notation of the spins.  The $SU(2)$-symmetric groundstate is, of course,
the $S_{\text{total}}\!=\!0$-singlet, constructed by adding all spins,
$|| \vektor{\tilde{S}}_1, \vektor{\tilde{S}}_2, ...,
\vektor{\tilde{S}}_{N_s} ; 0 \ra$ in a reduced dimension basis
notation (see for example \onlinecite{Biedenharn09_book}).
Importantly, this is the true groundstate of the effective Hamiltonian
of $H_{\text{eff}} \propto \frac{1}{\sqrt{L}}
\vektor{S}_{\text{total}}^2$ describing purely the TOS
levels~\cite{Kolley13}. The only non-trivial sets of unitary symmetry
operations that are allowed to act on the
$S_{\text{total}}\!=\!0$-singlet and leave a Heisenberg-type
Hamiltonian between the sublattices unchanged (sublattices should be
still arranged on the physical lattice), can be written as the cyclic
translations of sublattices, $T_\nu$, where $\nu$ is the number of
sublattices that will be shifted (for example to the right). One can
then write
\begin{align}
  T_{\nu \!=\! N_s} &|| \vektor{S}_1, \vektor{S}_2, \vektor{S}_3, ..., \vektor{S}_{N_s} ; 0 \ra = \notag \\
  T_{\nu=1}^{N_s}   &|| \vektor{S}_1, \vektor{S}_2, \vektor{S}_3, ..., \vektor{S}_{N_s} ; 0 \ra = \notag \\
  &|| \vektor{S}_1, \vektor{S}_2, \vektor{S}_3, ..., \vektor{S}_{N_s}
  ; 0 \ra
  \label{eq:T-sigma_ch6}
\end{align}
There are obviously only $N_s$ distinct values that $\nu$ can take,
including the identity operator. \eref{eq:T-sigma_ch6} already implies that the TOS levels
can only acquire lattice momenta of $k^{\text{TOS}}_\nu = \frac{2\pi
  \nu}{N_s}$ for $\nu=0, 1, ..., N_s-1$, between the equal or greater
group of general ES momenta, $k_n$. The only complication emerges
from the distribution pattern of $n'$ TOS-levels between $N_s$ momenta
for a fixed $S$-sector. To clarify this, let us focus on the more
general case of $n'> N_s$ and choose the momentum of the lowest ES
level to be $k^{\text{TOS}}_0[S\!=\!0] = 0$, presumably, corresponding
to the action of $I$ on the sublattices (chosen differently in
\fref{fig:ES-columnar}). Trivially, all other $(n'-1)$-levels should
arrange symmetrically in respect to $k^{\text{TOS}}_0[S\!=\!0]$ (there
is no relative net momentum). So, they can either, altogether, fill
the zero-momentum state on top of $k^{\text{TOS}}_0[S=0]$ or occupy
$\pm k_\nu$ ($\nu\neq0$) states around it. The former is \emph{not}
possible, due to the fact that $T_\nu$ ($\nu\neq0$) and $I$ posses a
distinct set of eigenvalues and therefore produce different momenta
(this can be easily observed by writing the bipartite Schmidt
decomposition of the $S_{\text{total}}\!=\!0$-singlet state and switch
to the basis of fixed-$S$ states for $L$ or $R$ partition to reveal
distinct eigenspectra of $T_\nu$ and $I$). In addition, we notice that
some states appearing in a TOS column are \emph{not} essentially TOS
levels. This is partly due to the fact that the non-TOS levels are
also allowed to fill $k^{\text{TOS}}_\nu$ states, and partly because
in an MPS representation, there is always a \emph{fixed} number of
states kept and consequently, only the first few TOS levels of
$H_{\text{eff}}$ will be recovered. Nevertheless, such initial states
(having a clear gap to the higher levels) certainly follow the TOS
level counting as governed by the degree of symmetry-breaking in the
thermodynamic limit.  I.e.~for a state that \emph{fully} breaks
$SU(2)$-symmetry (e.g.~the $120^{\circ}$ order), there are
$N_S^{\text{TOS}}=(2S+1)$ levels grouped together, and for a state
that \emph{partially} breaks the $SU(2)$-symmetry down to $U(1)$ (e.g.~the
columnar order), there is only $N_S^{\text{TOS}}=1$ level per each
fixed $S$-sector (\emph{not} counting the degeneracy that comes from
the $SU(2)$ quantum numbers themselves; the overall degeneracy of the ES
levels is always $(2S+1) N_S^{\text{TOS}}$ -- see
\onlinecite{SaadatmandThesis,Kolley13} for more details).  We discover
another striking feature in the momentum-resolved ES of
symmetry-broken phases, however, this time for the states between the
TOS columns: the first few Nambu-Goldstone modes exhibit \textit{sine-like}
dispersion patterns (as in the energy spectrum), if $L_y$ chosen to be
large enough.

\begin{figure}
  \begin{center}
    \includegraphics[width=0.49\columnwidth]{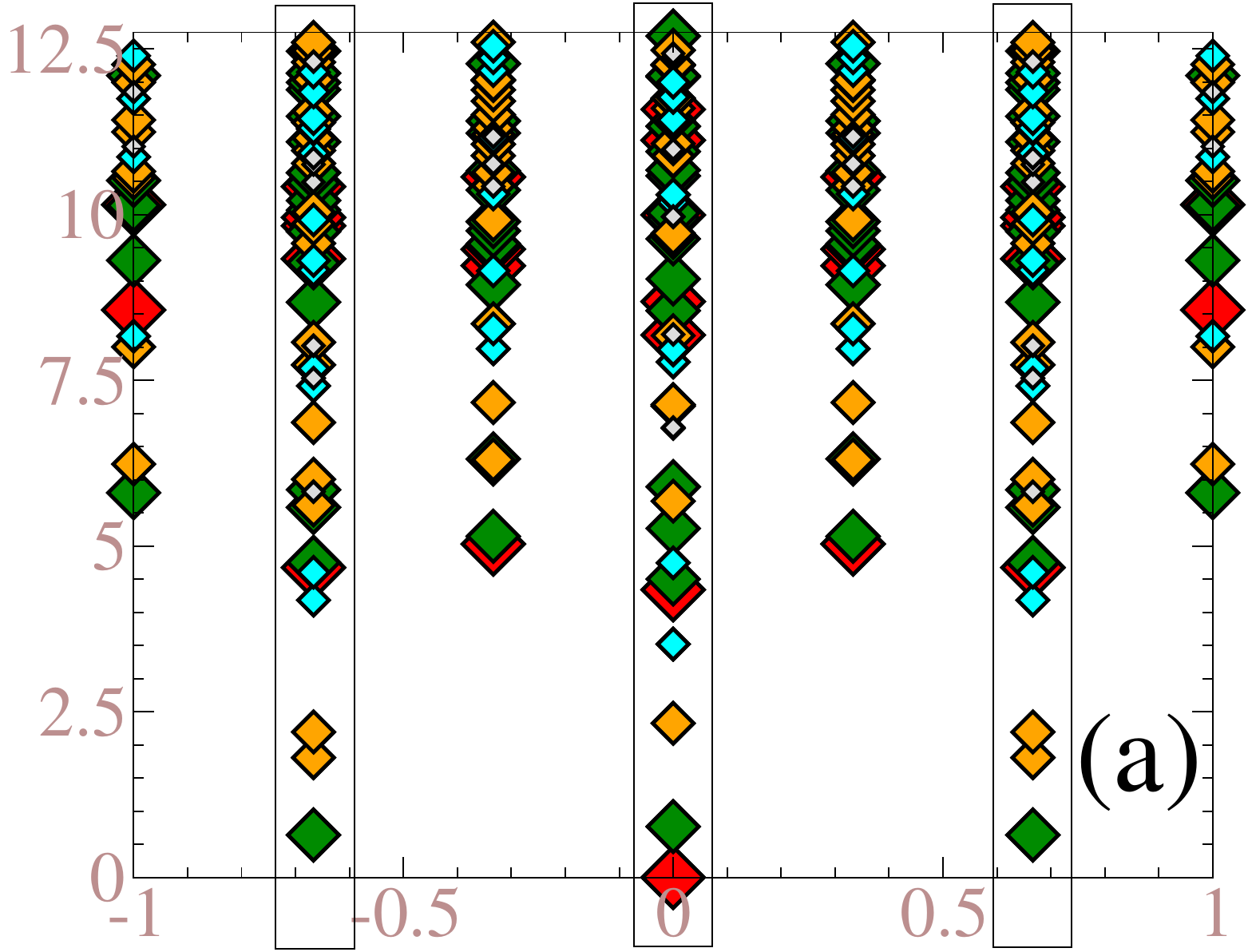}
    \includegraphics[width=0.49\columnwidth]{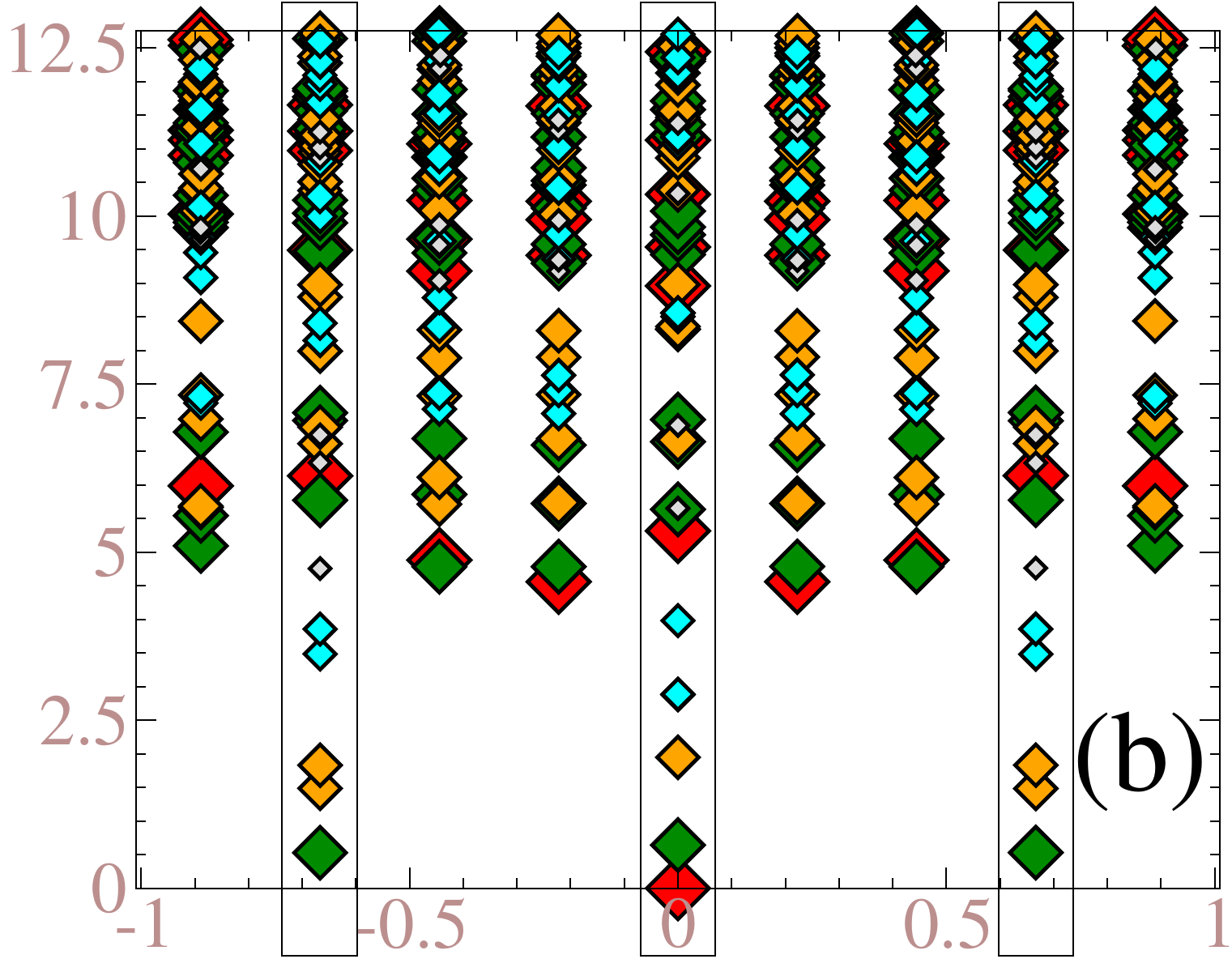} \\
    \includegraphics[width=0.99\columnwidth]{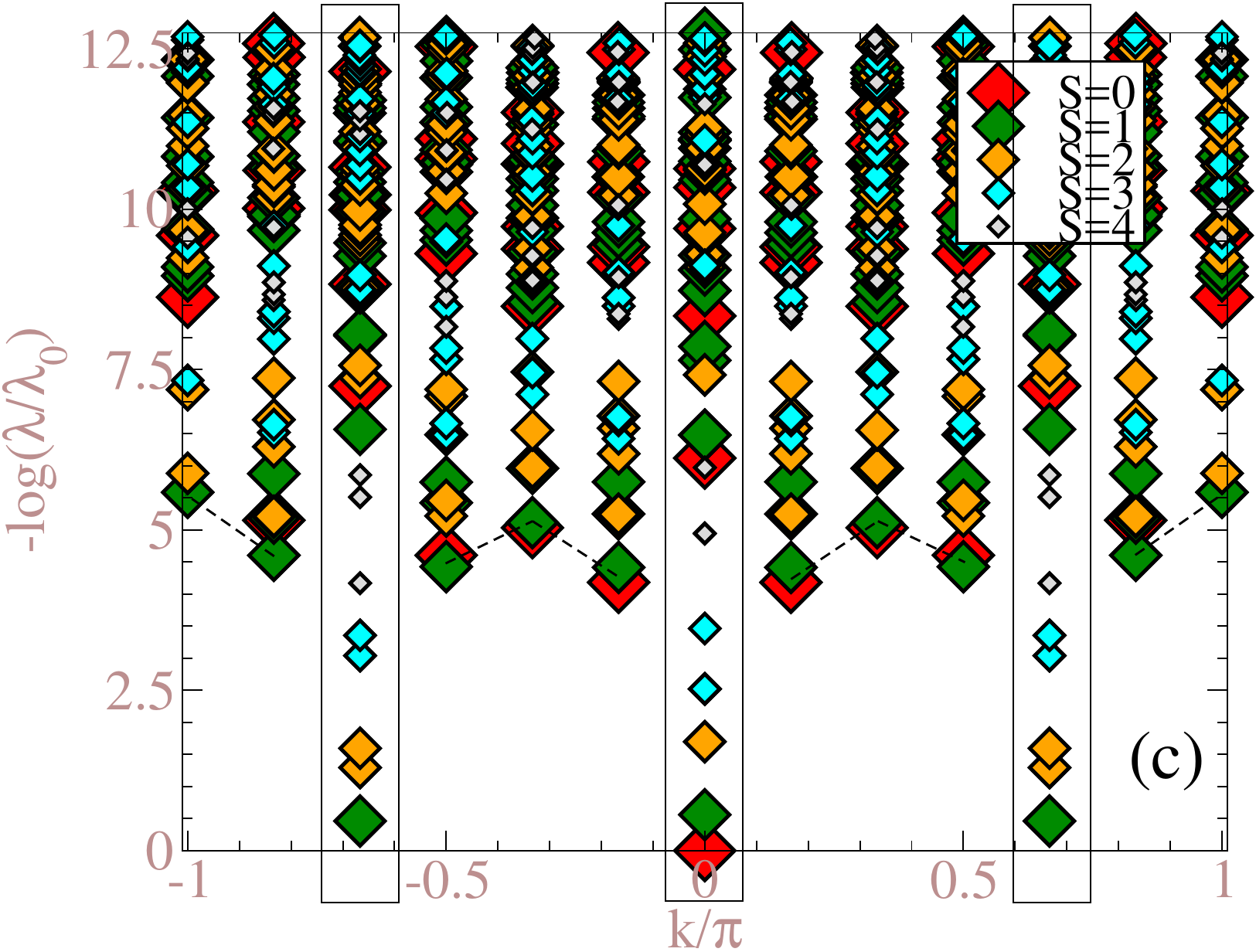}
    \caption{(Color online) 
    iDMRG momentum-resolved ES of the $120^{\circ}$ order, $J_2=-1.0$, for 
    (a) YC6, (b) YC9, and (c) YC12 structures of the THM versus
    $Y$-direction momenta (the reference momentum is fixed to $k^{\text{TOS}}[\lambda_0]=0$). Boxes emphasize TOS 
    columns at the unique momenta of $k_\nu^{\text{TOS}} = -\frac{2\pi}{3}, 0, \frac{2\pi}{3}$. In part (c), 
    dashed-lines are guides to the eyes and connect the Nambu-Goldstone modes
    of the ES for the first few levels on the top of the TOS levels.
    \label{fig:ES-120}}
 \end{center}
\end{figure}

In \fref{fig:ES-120}, we present the momentum-resolved ES of the
$120^{\circ}$ order on different width of the YC structure (for more
visibility, we have limited the display of the ES levels to
$S_{max}=4$ in all ES figures of this section).  The presence of
\emph{three} characterizing TOS columns is clear for
all system widths, consistent with the theory for a $N_s\!=\!3$-state.
The low-lying levels inside the TOS columns (purely TOS levels) have a
clear gap to the higher levels, which qualitatively observed to
converge to a finite value, linearly with $\frac{1}{L_y}$, at the
thermodynamic limit\cite{SaadatmandThesis}. The number of low-lying
levels in the TOS columns agree with the full $SU(2)$-symmetry breaking
in the thermodynamic limit. That is $N_S^{\text{TOS}}=(2S+1)$ for all
$S=0,1,2,3,4$, as previously observed by Kolley\etal\cite{Kolley13}.
For low-lying Nambu-Goldstone modes between the TOS columns, we suggest the
triangular-shape dispersion patterns of \fref{fig:ES-120}(c) are signs
for the formation of sine-like structures, however, due to relatively
small size of $L_y$, the $k_n$-resolution does not suffice to
discern more details.

In \fref{fig:ES-columnar}, we present the momentum-resolved ES of the
columnar order for different widths of the YC structure.  The presence
of \emph{two} characterizing TOS columns (note that
$k^{\text{TOS}}[\lambda_0]=\pm\pi$-columns are the same) is clear for all system
widths, as predicted by the theory for a $N_s\!=\!2$-state. As before,
the low-lying levels inside the TOS columns have a clear gap to the
higher levels and observed to converge to a finite value, linearly
with $\frac{1}{L_y}$, at the thermodynamic limit\cite{SaadatmandThesis}.
The partial breaking of $SU(2)$ to $U(1)$ symmetry can be confirmed by the
level counting of $N_S^{\text{TOS}}=1$ for low-lying
$S=0,1,2,3,4$-levels in the TOS columns. A sine-like dispersion pattern
for the low-lying levels between the TOS columns is apparent, at
least, for the larger $L_y=12$ system, \fref{fig:ES-columnar}(c).

\begin{figure}
  \begin{center}
    \includegraphics[width=0.99\columnwidth]{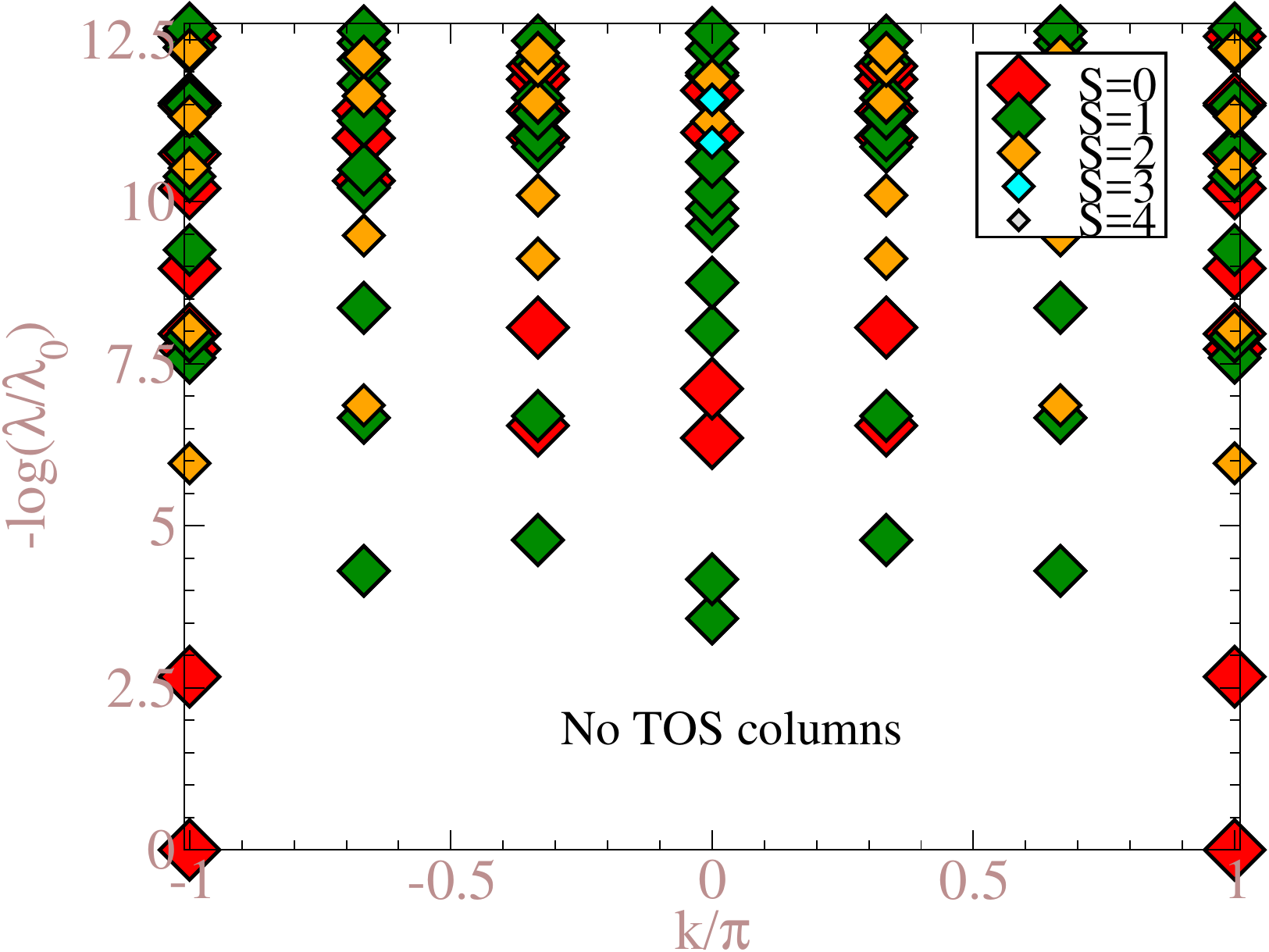}
    \caption{(Color online)
    iDMRG momentum-resolved ES of an ASL state, $J_2=0.185$, for 
    the YC6 structure of the THM versus
    $Y$-direction momenta (the reference momentum is fixed as 
    $k[\lambda_0]=\pm\pi$).
    \label{fig:ES-AlgebraicSL}}
 \end{center}
\end{figure}

In \fref{fig:ES-AlgebraicSL}, we present the momentum-resolved ES of
an ASL state on a $L_y=6$ cylinder.  Clearly, there is no signature
for the presence of TOS columns, which suggests the nonmagnetic nature
of the phase. In addition, we observe \emph{no} non-trivial degeneracy
of low-lying ES levels. So, there exist \emph{no} fractionalization of
symmetries to identify SPT and/or some intrinsic topological ordering
with anyonic excitations (see also \onlinecite{Saadatmand16}).

\section{Time-reversal symmetry-breaking and the robustness of the
  topological phase against the chirality}
\label{sec:chirality}

The existence of the time-reversal symmetry is a key feature of
$H_{J_2}$, \eref{eq:J1J2-Ham}. A chiral groundstate spontaneously
breaks time-reversal, $\tau$, and parity reflection, $P$, symmetry,
but respects the combined $P\tau$-symmetry.  After consistent numerical
observations of a nonmagnetic phase in the $J_1$-$J_2$ THM phase
diagram (cf.~\sref{sec:intro} and \ref{sec:overview}), the natural
question is, whether the new state stabilizes due to SSB of
$\tau$, which would result in a CSL. For a scenario in which the true
groundstate in the SL phase region is $\text{Z}_2$ topological
ordered (advocated by DMRG results\cite{Zhu15,Hu15,Saadatmand16}), we
already investigated\cite{Saadatmand16} the chirality of anyonic
sectors in detail, using direct measurement of the $\tau$-operator
expectation values and calculating a scalar chiral order parameter,
\begin{equation}
  O_\chi = \frac{1}{L_u} \sum_{\la i,j,k \ra} (\vektor{S}_{i} \times \vektor{S}_{j}) \cdot \vektor{S}_{k} \; ,
  \label{eq:ChiralOP} 
\end{equation}
where $\la i,j,k \ra$ represent a NN triangular plaquette and the sum
goes over the wavefunction unit cell. We discovered that the
topological sectors are all $\tau$-symmetric as the $O_\chi$-values
observed to be small and decreasing rapidly to numerically vanishing
magnitudes at the thermodynamic limit of $m \rightarrow \infty$
(furthermore, \sector{b} and \sector{f}-sector are fractionalizing time-reversal symmetry). However,
Hu\etal\cite{Hu15} determined the \sector{i}-sector groundstate as
strongly prone to the chirality by adding directional
($\vektor{a}_{\pm60^\circ}$-axis) anisotropy to the
Hamiltonian. This is, in part, leading another question of our
interest: is the SL phase robust against perturbing $H_{J_2}$ with a
term that explicitly breaks the $\tau$-symmetry and forms a chiral
long-range order? To answer this question, one can study the
$J_1$-$J_2$-$J_\chi$ model,
\begin{equation}
  H_\chi = H_{J_2} + 
  J_\chi \sum_{\la i,j,k \ra} ({\bf S}_i \times {\bf S}_j) \cdot {\bf S}_k \; ,
  \label{eq:HamiltonianChiral_ch6}
\end{equation}
where $\langle i,j,k \rangle$ indicates the sum over all NN triangular
plaquettes in a Hamiltonian unit cell. The phase diagram of $H_\chi$
is previously studied using variational QMC~\citep{Hu16} and
ED~\citep{Wietek16} techniques, however, no clear result has
emerged on the nature of the $J_\chi \rightarrow 0$ limit. To shed
some lights on this matter, in this section we study the response of
the YC8-\sector{i} groundstates\cite{Saadatmand16} to the chiral field by
adiabatically adding a $J_\chi$-term to $H_{J_2}$, as in
\eref{eq:HamiltonianChiral_ch6}, and finding new groundstates using
the $SU(2)$-symmetric iMPS and iDMRG methods.

\begin{figure}
  \begin{center}
    \includegraphics[width=0.99\columnwidth]{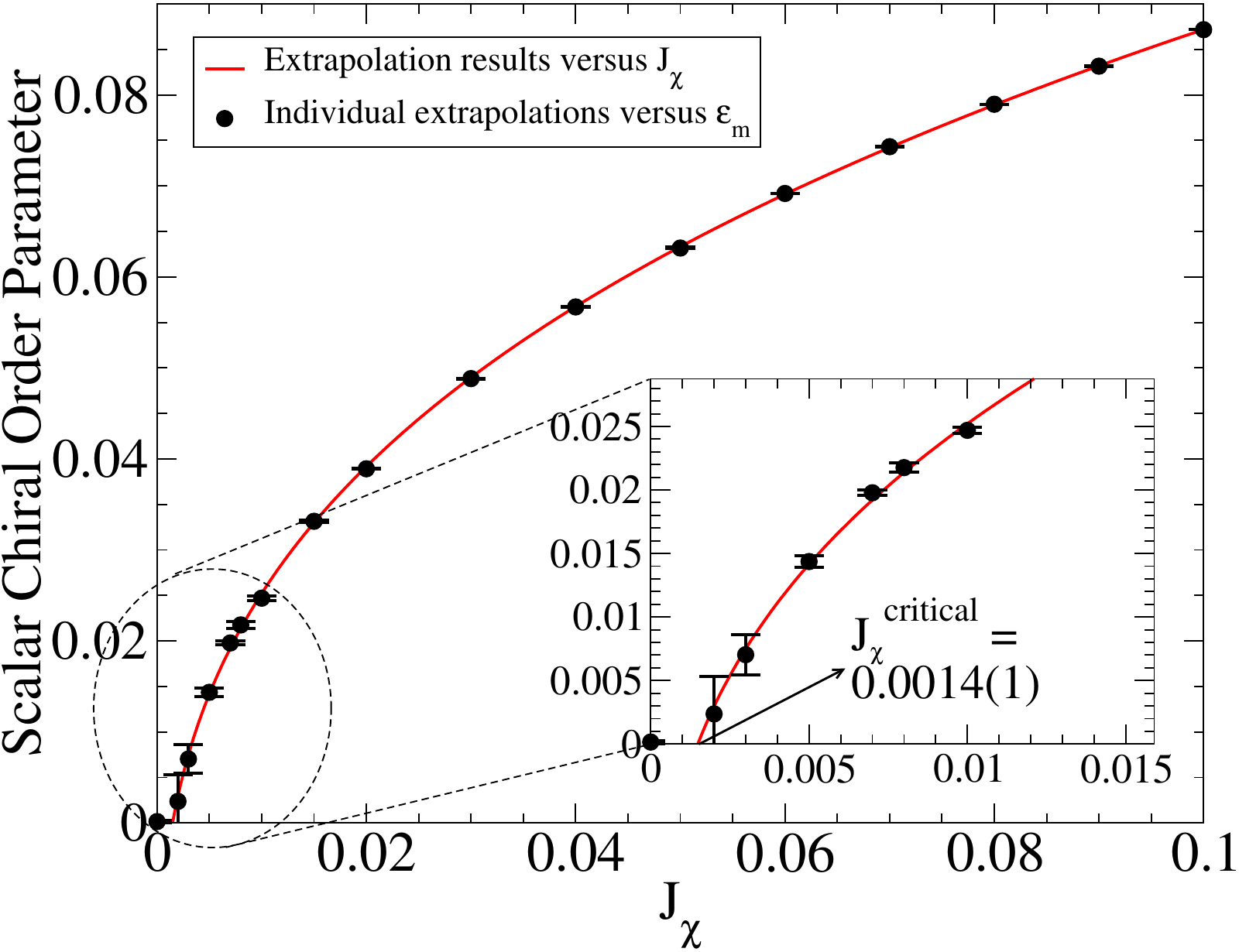}
    \caption{(Color online) 
    iDMRG results for the scalar chiral order parameter, $O_\chi$, \eref{eq:ChiralOP}, versus $J_\chi$ for the groundstates 
    of $H_\chi$, \eref{eq:HamiltonianChiral_ch6}, constructed from a YC8-\sector{i} sector. 
    Each data-point represents a $O_\chi[m\rightarrow\infty,J_\chi]$, 
    which is the result of a separate extrapolation 
    on individual $O_\chi$ versus $\varepsilon_m$ toward the 
    thermodynamic limit of $\varepsilon_m\rightarrow0$ ($m\rightarrow\infty$). The red line
    is our attempted fit of $\tilde{b}_0 + \tilde{b}_1 J_\chi^{\tilde{b}_2}$ to 
    the black circles, excluding the first two $J_\chi$ points (where the chirality
    is zero within the error-bars), which is used to estimate the phase transition when 
    $O_\chi[m\rightarrow\infty,J_\chi^{\text{critical}}]=0$. A zoom-in plot is presented in the inset, as a guide to the eyes.
    \label{fig:ChiralOP}}
 \end{center}
\end{figure}

We present our results for the extrapolated $O_\chi$ in the
thermodynamic limit of $m\rightarrow\infty$ in \fref{fig:ChiralOP}. We
notice that, within our resolution, upon varying $J_\chi$, there is at
least one (significant) point exposed to nonzero chiral perturbations,
but has negligible $O_\chi(m\rightarrow\infty)$ within the error-bars.
This means that the topological SL phase is robust against chirality and
one needs to provide $\tau$-symmetry-breaking terms larger than a
finite-value, namely $J_\chi^{\text{critical}}$, to impose a chiral
groundstate.  To further predict this small
$J_\chi^{\text{critical}}$, we applied a fit of $\tilde{b}_0 +
\tilde{b}_1 J_\chi^{\tilde{b}_2}$ to the data and find
$J_\chi^{\text{critical}} = (-\frac{b_0}{b_1})^\frac{1}{b_2} =
0.0014(1)$.  These results also suggest the existence of a
\emph{second order} phase transition toward the CSL phase. This is
consistent with the suggestion from Wietek and
L\"{a}uchli\cite{Wietek16}, and may clarify the results of
Hu\etal\cite{Hu16}, where it is unclear if $O_\chi$
would be zero or not in the $J_\chi\rightarrow0$ limit.

\section{Conclusion}
\label{sec:conclusion}

We have presented comprehensive results for the phase diagram of the
$J_1$-$J_2$ Heisenberg model on triangular lattices, using infinite-length
YC structures. Using the Binder ratio of the magnetization order parameter,
$U_r$, \eref{eq:Ur-scaling_ch6}, and TOS columns of the momentum-resolved ES,
we have obtained phase boundaries and characterized the nature of the symmetry
breaking magnetic order. We found that the Binder ratio reliably detects 
phase boundaries between magnetically ordered states, even when using $SU(2)$ symmetry,
where the order parameter itself is zero by construction.
We identified the $120^\circ$-ordered groundstate as a three-sublattice LRO
with full $SU(2)$-symmetry-breaking in the thermodynamic limit; the
columnar-ordered groundstate as a two-sublattice LRO with partial
$SU(2)$-symmetry-breaking at the thermodynamic limit, and confirm the
nonmagnetic nature of the SL states on infinite cylinders of widths up
to 12 sites. In addition, we have discovered the stabilization of a
new ASL phase, with power-law correlation lengths, for width-6
infinite cylinders. We have pinpointed the phase transitions between
the infinite cylinder's groundstates of the THM, precisely, using the Binder
ratios. The transitions are relatively close to the phase boundaries
found from the direct measurements of the local order parameters using
fDMRG on $L_y=3,4,5,6$-cylinders, and short-range correlations and
fidelity susceptibility phase diagrams from iDMRG calculations. In
addition, for the columnar order, we have numerically proved that the
entropies consistently obey $S_{EE} = a_0(L_y) + (\alpha_0 + \alpha_1 L_y)
\log(\xi)$, a mixture of the area-law and the quantum critical
behavior, as expected for the magnetic phases built by the inherently
one-dimensional $SU(2)$-symmetric iMPS ansatz. To the best of our knowledge, a set of
numerical tools to efficiently distinguish and classify LROs were
previously absent in the $SU(2)$-symmetric iDMRG literature.
Considering the advantages of $SU(2)$-symmetric calculations, we suggest
that the proposed methods can be applied widely to detect symmetry broken
states using the iMPS.

Finally, to unravel the true nature of time-reversal symmetry-breaking
in the topological SL, we have investigated the
robustness of YC8-\sector{i} sector under perturbing $H_{J_2}$ with a
chiral term, \eref{eq:HamiltonianChiral_ch6} (it was previously
suggested~\citep{Hu15} that YC8-\sector{i} states are 
prone to become chiral under applying bond anisotropies to the
Hamiltonian). The results of the scalar chiral order parameter,
$O_\chi(m\rightarrow\infty)$, versus $J_\chi$ can be fitted using
$\tilde{b}_0 + \tilde{b}_1 J_\chi^{\tilde{b}_2}$ with high
accuracy and shows the existence of a continuous phase transition
to the CSL phase at small, but non-zero, $J_\chi^{\text{critical}}
= 0.0014(1)$. Therefore, for finite-width cylinders the topological
state of the THM are time-reversal symmetric, and not a chiral topological liquid.

\begin{acknowledgments}
  
  The authors would like to thank Jason Pillay for useful discussions.
  This work has been supported by the Australian Research Council
  (ARC) Centre of Excellence for Engineered Quantum Systems, grant
  CE110001013. I.P.M.~also acknowledges support from the ARC Future
  Fellowships scheme, FT140100625.
  
  \emph{Notes added.} -- After completing this work, a 
  related paper~\citep{Gong17} appeared in which the authors study the phase diagram
  of the $J_1$-$J_2$-$J_\chi$ model, \eref{eq:HamiltonianChiral_ch6}, on finite-$L_x$ 
  cylinders using the $SU(2)$-symmetric fDMRG algorithm. In agreement to 
  \sref{sec:chirality} results, Gong\etal~find a smooth phase transition from 
  $J_1$-$J_2$ SL to a CSL at a small but finite chiral coupling strength 
  ($J_\chi^{\text{critical}} \approx 0.02$ for $J_2=0.1$).
  
  In addition, \fref{fig:ChiralOP} caption should specify $J_2=0.125$.

\end{acknowledgments}



\cleardoublepage

\end{document}